\documentclass{vldb}
\usepackage{times}
\usepackage{framed}
\usepackage[normalem]{ulem}

\usepackage[space]{grffile}

\usepackage[,
            hyperfigures=false,
            hyperindex=false,
            pdfborder={ 0 0 0 },
            pdftitle={},
            pdfauthor={},
            pdfkeywords={}]{hyperref}

\usepackage{amsmath,amssymb,epsfig}
\usepackage{graphicx}
\usepackage{balance}
\usepackage{verbatim}
\usepackage{color}
\usepackage{caption}
\usepackage{subcaption}
\usepackage[capitalize]{cleveref}
\usepackage{url}
\usepackage{balance}
\usepackage{paralist}
\usepackage{booktabs}
\usepackage{siunitx}
\usepackage[table]{xcolor}
\usepackage[scaled]{inconsolata}
\usepackage[T1]{fontenc}

\usepackage{tabularx}
\newcolumntype{b}{X}
\newcolumntype{s}{>{\hsize=.1\hsize}X}
\usepackage{multirow}

\usepackage{appendix}

\usepackage{enumitem}
\setlist[enumerate]{itemsep=0mm}

\newcommand{\sampar}[1]{\vspace{3pt}\noindent{\bf #1}}

\usepackage[ruled,vlined]{algorithm2e}

\graphicspath{{./}}

\newif\ifsubmission
\submissionfalse

\ifsubmission
\newcommand{\todo}[1]{}
\newcommand{\amol}[1]{}
\newcommand{\agp}[1]{}
\newcommand{\srm}[1]{}
\newcommand{\aje}[1]{}
\newcommand{\michael}[1]{}
\newcommand{\david}[1]{}
\newcommand{\com}[1]{}

\newcommand{\cut}[1]{}

\else
\newcommand{\com}[1]{\smallskip\noindent\textsf{\small\textcolor{red}{#1}}}

\newcommand{\todo}[1]{\textcolor{red}{\bf [TODO: #1]}}
\newcommand{\amol}[1]{\noindent{\textcolor{blue}{[Amol: #1]}}}
\newcommand{\agp}[1]{\noindent{\textcolor{blue}{[Aditya: #1]}}}
\newcommand{\srm}[1]{\noindent{\textcolor{red}{[Sam: #1]}}}
\newcommand{\aje}[1]{\noindent{\textcolor{red}{[Aaron: #1]}}}
\newcommand{\michael}[1]{\noindent{\textcolor{red}{[Michael: #1]}}}
\newcommand{\david}[1]{\noindent{\textcolor{red}{[David: #1]}}}

\newcommand{\cut}[1]{}
\fi

\newcommand{\papertext}[1]{}
\newcommand{\techreport}[1]{#1}

\newenvironment{itemize*}{
    \vspace*{-0.05in}
    \begin{itemize}
        \setlength{\itemsep}{0pt}
        \setlength{\parsep}{3pt}
        \setlength{\topsep}{3pt}
        \setlength{\partopsep}{0pt}
        \setlength{\leftmargin}{1.95em}
        \setlength{\labelwidth}{1.5em}
        \setlength{\labelsep}{0.5em}
    }{\end{itemize}
    \vspace*{-0.05in}
}
\newenvironment{enumerate*}{
    \vspace*{-0.05in}
    \begin{enumerate}
        \setlength{\itemsep}{0pt}
        \setlength{\parsep}{3pt}
        \setlength{\topsep}{3pt}
        \setlength{\partopsep}{0pt}
        \setlength{\leftmargin}{2em}
        \setlength{\labelwidth}{1.5em}
        \setlength{\labelsep}{0.5em}
    }{\end{enumerate}
    \vspace*{-0.05in}
}

\makeatletter
\def\@copyrightspace{\relax}
\makeatother

\definecolor{comment-red}{rgb}{0.56,0,0}

\newcommand{\ta}[1]{\vspace{-3pt}\begin{framed}\vspace{-5pt}\noindent\textit{\underline{Takeaway:} #1}\vspace{-5pt}\end{framed}\vspace{-3pt}}

\newenvironment{denselist}{
    \begin{list}{\small{$\bullet$}}%
    {\setlength{\itemsep}{0ex} \setlength{\topsep}{0ex}
    \setlength{\parsep}{0pt} \setlength{\itemindent}{0pt}
    \setlength{\leftmargin}{1.5em}
    \setlength{\partopsep}{0pt}}}%
    {\end{list}}

\begin{document}


\title{Understanding Workers, Developing Effective Tasks,  \\ and Enhancing Marketplace Dynamics: \\ 
{\Large A Study of a Large Crowdsourcing Marketplace}}
\subtitle{\papertext{[Experiments and Analysis Paper]}\techreport{[Extended Technical Report]}}

\setcounter{secnumdepth}{2}

\author{ Ayush Jain$^\dagger$, Akash Das Sarma$^\star$, Aditya Parameswaran$^\dagger$, Jennifer Widom$^\star$ \\ 
$^\dagger$University of Illinois \ \ \ \ $^\star$Stanford University}


\maketitle

\begin{abstract}
We conduct an experimental analysis of a dataset comprising over 27 million microtasks performed by over 70,000 workers issued to a large crowdsourcing marketplace between 2012-2016. Using this data---never before analyzed in an academic context---we shed light on three crucial aspects of crowdsourcing: (1) Task design --- helping requesters understand what constitutes an effective task, and how to go about designing one; (2) Marketplace dynamics --- helping marketplace administrators and designers understand the interaction between tasks and workers, and the corresponding marketplace load; and (3) Worker behavior --- understanding worker attention spans, lifetimes, and general behavior, for the improvement of the crowdsourcing ecosystem as a whole.    
\end{abstract}

\section{Introduction}

Despite the excitement 
surrounding artificial intelligence and 
the ubiquitous need for large volumes of 
manually labeled training data, 
the past few years have been a relatively tumultuous period 
for the crowdsourcing industry.
There has been a recent spate of {\em mergers}, e.g., \cite{odesk-elance-merger},
{\em rebrandings}, e.g., \cite{mobileworks-leadgenius, odesk-upwork},
{\em slowdowns}, e.g., \cite{mturk-downturn},
and moves towards {\em private crowds}~\cite{crowd-book}.
For the future of crowdsourcing marketplaces, 
it is therefore both important and timely
to step back and study how these marketplaces
are performing, how the requesters are making and can make best
use of these marketplaces, and how workers are participating in these
marketplaces---in order to {\em develop more efficient marketplaces,
understand the workers' viewpoint and make their experience less tedious,
and design more effective tasks from the requester standpoint.}
\techreport{Achieving all these goals would support and sustain 
the {\em ``trifecta'' of key participants}
that keeps crowdsourcing marketplaces ticking---the marketplace administrators,
the crowd workers, and the task requesters. }

At the same time, developing a better understanding of 
how crowdsourcing marketplaces function can help
us design crowdsourced data processing algorithms and systems that
are more efficient, in terms of latency, cost, and quality.
Indeed, crowdsourced data processing is performed 
at scale at many tech companies, 
with tens of millions of dollars spent every year~\cite{crowd-book}, so the efficiency
improvements can lead to substantial savings for these companies. 
In this vein, there have been a number of papers on both optimized algorithms, e.g.,~\cite{so-who-won, crowdscreen, DBLP:conf/icdt/DavidsonKMR13, searching, DBLP:conf/sigmod/AmsterdamerGMS13},
and systems, e.g.,~\cite{qurkCIDR, crowddb, cdas, DBLP:conf/www/BozzonBC12, deco-survey}, all from the database community, and such findings can have an impact in the design of all of these 
algorithms and systems.

Unfortunately, due to the proprietary nature of crowdsourcing marketplace
data, it is hard for academics 
to perform such analyses and identify pain points and solutions. 
Fortunately for us, one of the more forward thinking crowdsourcing
marketplaces made 
a substantial portion of its data from 2012 to date available to us:
this includes data ranging from worker answers to specific questions and 
response times, all the way to the HTML that encodes the user interface for a 
specific question.

This data allows us to answer some of the most important 
open questions in microtask crowdsourcing: what constitutes an ``effective'' task,
how can we improve marketplaces, 
and how can we enhance workers' interactions. 
In this paper, using this data, we study the following key questions:
\begin{denselist}
\item Marketplace dynamics: helping marketplace administrators understand the interaction between tasks and workers, and the corresponding marketplace load; e.g., questions like: 
(a) how much does the load on the marketplace vary over time, 
and is there a mismatch between the number of workers and the number of tasks available, 
(b) what is the typical frequency and distribution of tasks that are repeatedly issued, 
(c) what types of tasks are requesters most interested in, 
and what sorts of data do these tasks operate on?
\item Task design: helping requesters understand what constitutes an effective task, 
and how to go about designing one; e.g., 
questions like: 
what factors impact (a) the accuracy of the responses; 
(b) the time taken for the task to be completed; or (c) the time taken for the task to be picked up? 
Do examples and images help? Does the length or complexity of the task hurt?
\item Worker behavior: understanding worker attention spans, lifetimes, and general behavior; e.g., questions like (a) where do workers come from, 
(b) do workers from different sources show different characteristics, 
such as accuracies and response times, 
(c) how engaged are the workers within the marketplace, and relative to each other, 
and (d) how do their workloads vary?
\end{denselist}

\noindent 
The only paper that has performed an extensive analysis of crowdsourcing marketplace
data is the recent paper by Difallah et al.~\cite{DBLP:conf/www/DifallahCDIC15}.
This paper analyzed the data obtained via crawling a public crowdsourcing marketplace (in this case Mechanical Turk).
Unfortunately, this publicly visible data 
provides a restricted view
of how the marketplace is functioning, since the worker responses, 
demographics and characteristics of the workers, and the speed at which
these responses are provided are all unavailable.
As a result, unlike the present paper, that paper only considers
a restricted aspect of crowdsourcing marketplaces, specifically,
the price dynamics of the marketplace (indeed, their title reflects this as well)---for instance, 
demand and supply modeling, 
developing models for predicting throughput, and
analyzing the topics and countries preferred by requesters.
\techreport{That paper did not analyze the  
the full ``trifecta'' of participants that constitute a crowdsourcing marketplace.}
Even for marketplace dynamics, 
to fully distinguish the results of the present paper 
from that paper, we exclude any experiments or analyses
that overlap with the experiments performed in that paper. 
We describe this and other related work in Section~\ref{sec:related}. 

This experiments and analysis paper is organized as follows:
\begin{denselist}
\item {\bf Dataset description and enrichment.} In Section~\ref{sec:dataset},
we describe what our dataset
consists of, and the high-level goals of our analysis.
In Section~\ref{sec:operational} through \ref{sec:available-data}
we provide more details about the marketplace mechanics, the scale and timespan of 
the dataset, and the attributes provided. 
 We also enrich the dataset by manually labeling tasks 
ourselves on various features of interest,
described in Section~\ref{sec:data-enrich},
e.g., what type of data does the task operate on,
what sort of input mechanism does the task use
to get opinions from workers.

\item {\bf Marketplace insights.} 
In Section~\ref{sec:marketplace}, 
we address questions on the 
(a) marketplace load --- {\em task arrivals} (Section~\ref{sec:marketplace-bursty}), {\em worker availability} (Section~\ref{sec:marketplace-worker-availability}), and {\em task distribution} \techreport{(Section~\ref{sec:heavy-hitter})}\papertext{(in our technical report~\cite{techreport})}, with the aim of helping improve marketplace design, 
and (b) the types of tasks, {\em goals, human operators and data types}, and correlations between them (Section~\ref{sec:label-analyses}), with the aim of characterizing the spectrum of crowd work.
\item {\bf Task design improvements.} In Section~\ref{sec:task_analyses}, we (a) characterize and quantify metrics governing the ``effectiveness'' of tasks (Section~\ref{sec:metrics}), (b) identify features affecting task effectiveness and detail how they influence the different metrics (Sections~\ref{sec:feature-num-words} through~\ref{sec:feature-image}), (c) perform a classification analysis in Section~\ref{sec:prediction} wherein we look at the problem of predicting the various ``effectiveness'' metric values of a task based on simple features, and (d) provide final, summarized recommendations on how requesters can improve their tasks' designs to optimize for these metrics (Section~\ref{sec:task-summary}).
\item {\bf Worker understanding.} In Section~\ref{sec:worker}, we analyze and provide insights into 
the worker behavior. We compare characteristics of different worker demographics and sources---provided by different crowdsourcing marketplaces; as we will find, the specific marketplace whose data we work with solicits workers from many sources (Section~\ref{sec:worker-demo}). We also provide insights into worker involvement and task loads taken on by workers (Section~\ref{sec:worker-load}), and characterize and analyze worker {\em engagement} (Section~\ref{sec:worker-engagement}).
\end{denselist}
\section{Dataset Details and Goals}\label{sec:dataset}

We now introduce some terms that will help us operate in a  
marketplace-agnostic manner. 
The unit of work undertaken by a single worker is defined to be a {\em task}.
A task is typically listed in its entirety on one webpage, 
and may contain multiple short {\em questions}.
For example, a task may involve flagging whether each image in a set of ten images is inappropriate; so this task contains ten questions.
Each question in a task operates on an {\em item}; in our example, this item is an image.
These tasks are issued by {\em requesters}.
Often, requesters issue multiple tasks in parallel 
so that they can be attempted by different
workers at once.  
We call this set of tasks a {\em batch}.
Requesters often use multiple batches to issue identical units of work---for example, 
a requester may issue a batch of 100 ``image flagging'' tasks one day, operating on a set of items, 
and then another batch of 500 ``image flagging'' tasks after a week, on a different set of items.
We overload the term {\em task} to also refer to these 
identical units of work issued across time and batches, 
independent of the individual items being operated upon,
in addition to a single instance of work.
The usage of the term task will be clear from the context; if it is not clear, we will refer
to the latter as a {\em task instance}.


\subsection{Operational Details\label{sec:operational}}

Due to confidentiality and intellectual property reasons,
we are required to preserve the anonymity of 
the commercial crowdsourcing marketplace we operate on,
who have nevertheless been generous enough to provide
access to their data for research purposes.
To offset the lack of transparency due to the anonymity,
we discuss some of the crucial operational aspects
of the marketplace, that will allow us to understand how the marketplace
functions, and generalize from these insights to other similar marketplaces.

The marketplace we operate on acts as an aggregator or an intermediary
for many different sources of crowd labor. 
For example, this marketplace uses
Mechanical Turk~\cite{mturk}, Clixsense~\cite{clixsense}, and NeoDev~\cite{neodev}, 
all as sources of workers, as well as an internal worker pool.
For task assignment, i.e., assigning tasks to workers, 
the marketplace makes use of both push and pull mechanisms. 
The typical setting is via {\em pull}, where the workers
can choose and complete tasks that interest them.
In a some sources of workers that we will discuss later on,
tasks are {\em push}ed to workers by the marketplace. 
For example, Clixsense injects paid surveys into webpages so that
individuals browsing are attracted to and work on specific tasks.
In either case, the marketplace allows requesters to specify various
parameters, such as a minimum accuracy for workers who are allowed to work
on the given tasks, any geographic constraints, any constraints on the sources of crowd labor,
the minimum amount of time that a worker
must spend on the task, the maximum number of tasks in a batch a given worker
can attempt, and an answer distribution threshold (i.e., the threshold of skew
on the answers provided by the workers below which a worker is no longer allowed
to work on tasks from the requester).
The marketplace monitors these parameters and prevents workers
from working on specific tasks if they no longer meet the desired criteria.
\papertext{We provide additional details of our data collection process in our technical report.}
\techreport{
The marketplace also provides optional templates for HITs for  some common standard tasks, such as {\em Sentiment Analysis}, {\em Search Relevance}, and {\em Content Moderation}, as well as for tools such as one for image annotation. Usage of these standard templates leads to some uniformity in interfaces, and also gives potential for the improvement of task design simultaneously across requesters.

This marketplace categorizes certain workers as ``skilled'' contributors, who are given access to more advanced tasks, higher payments, and are also sometimes responsible for meta-tasks such as generating test questions, flagging broken tasks, and checking work by other contributors. Our research highlights that having a pool of engaged and active workers is just as, if not more important than having access to a large workforce. It might be interesting to explore an incentive program for the ``active'' or ``experienced'' workers as well.

This marketplace also provides a number of additional features. Notable among them is its module for machine learning and AI. This allows requesters without any machine learning background to generate and evaluate models on training data with easy computation and visualizations of metrics such as accuracy, and confusion matrix. 
}

\subsection{Origin of the Dataset\label{sec:data-details}} 
Our dataset consists of tasks issued on the marketplace
from 2012--2016. 
Unfortunately, we do not have access to all data about all tasks. 
There are about 58,000 batches in total, of which
we have access to complete data for a sample of about 12,000 batches,
and minimal data about the remaining, consisting only of the title of the task
and the creation date.
Almost 51,000, or 88\% of the 58,000 of batches have some representatives 
in our 12,000 batch sample---thus, the sample is missing about 10\% of the tasks. 
(That is, there are identical tasks in the 12,000 batch sample.)
From the task perspective, there are about 6600 distinct
tasks in total, spread across 58,000 batches, 
of which our sample contains 5000, i.e., 76\% of all  distinct tasks.
Thus, while not complete, our sample is a significant and representative portion
of the entire dataset of tasks.
{\em We will largely operate on this 12,000 batch sample, consisting of 27M task instances,
a substantial number.}
Figure~\papertext{\ref{fig:sampling-week-cluster}}\techreport{\ref{fig:sampling}} compares the number of distinct tasks sampled 
versus the total number of distinct tasks that were issued to the marketplace across different weeks, 
We observe that in general we have a significant fraction of tasks from each week.

\techreport{
\begin{figure*}[!t]
\centering
\begin{subfigure}[b]{0.33\textwidth}
\centering
\includegraphics[width=\textwidth]{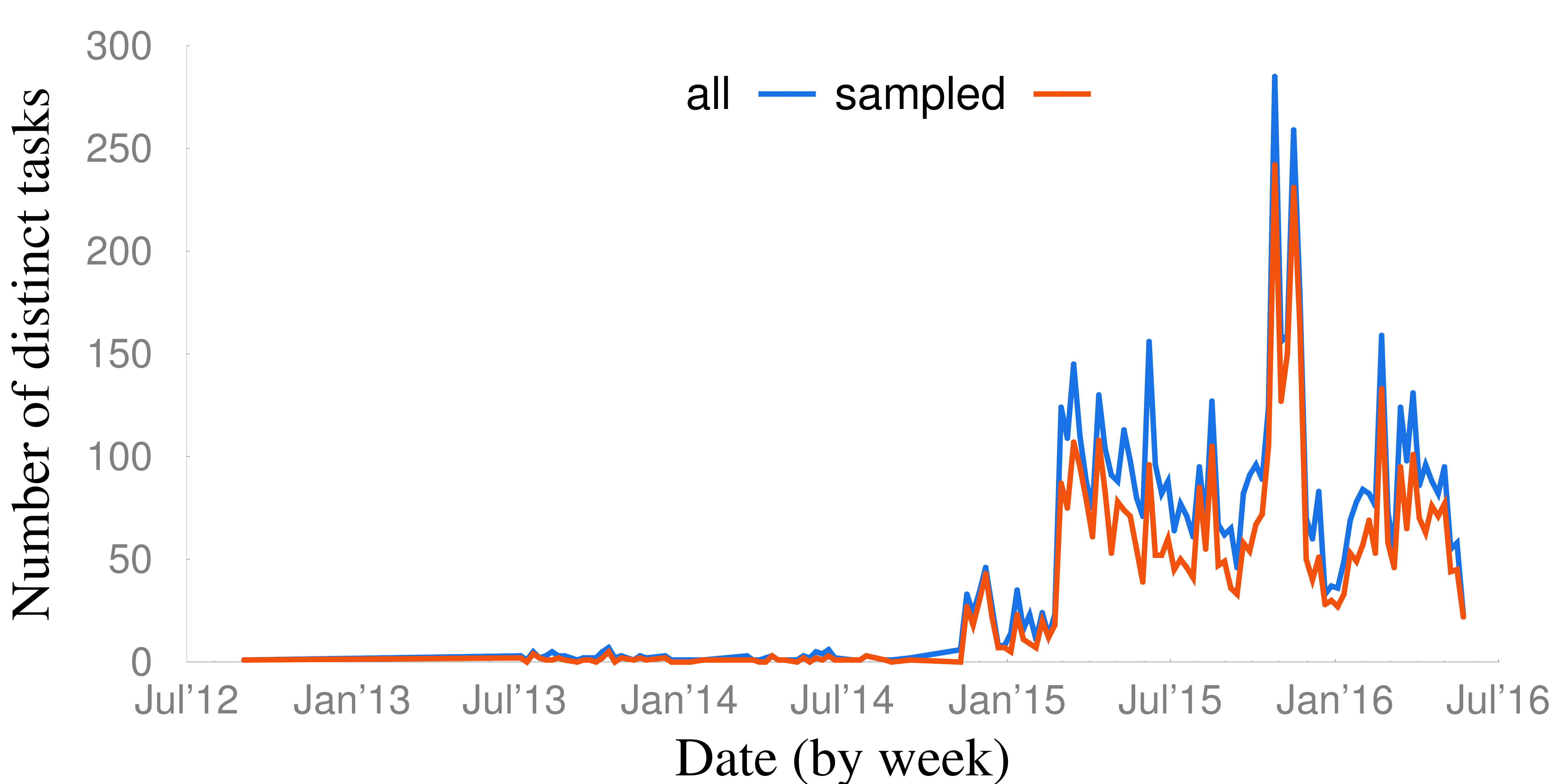}
\caption{\label{fig:sampling-week-cluster}Entire duration}
\end{subfigure}%
\begin{subfigure}[b]{0.33\textwidth}
\centering
\includegraphics[width=\textwidth]{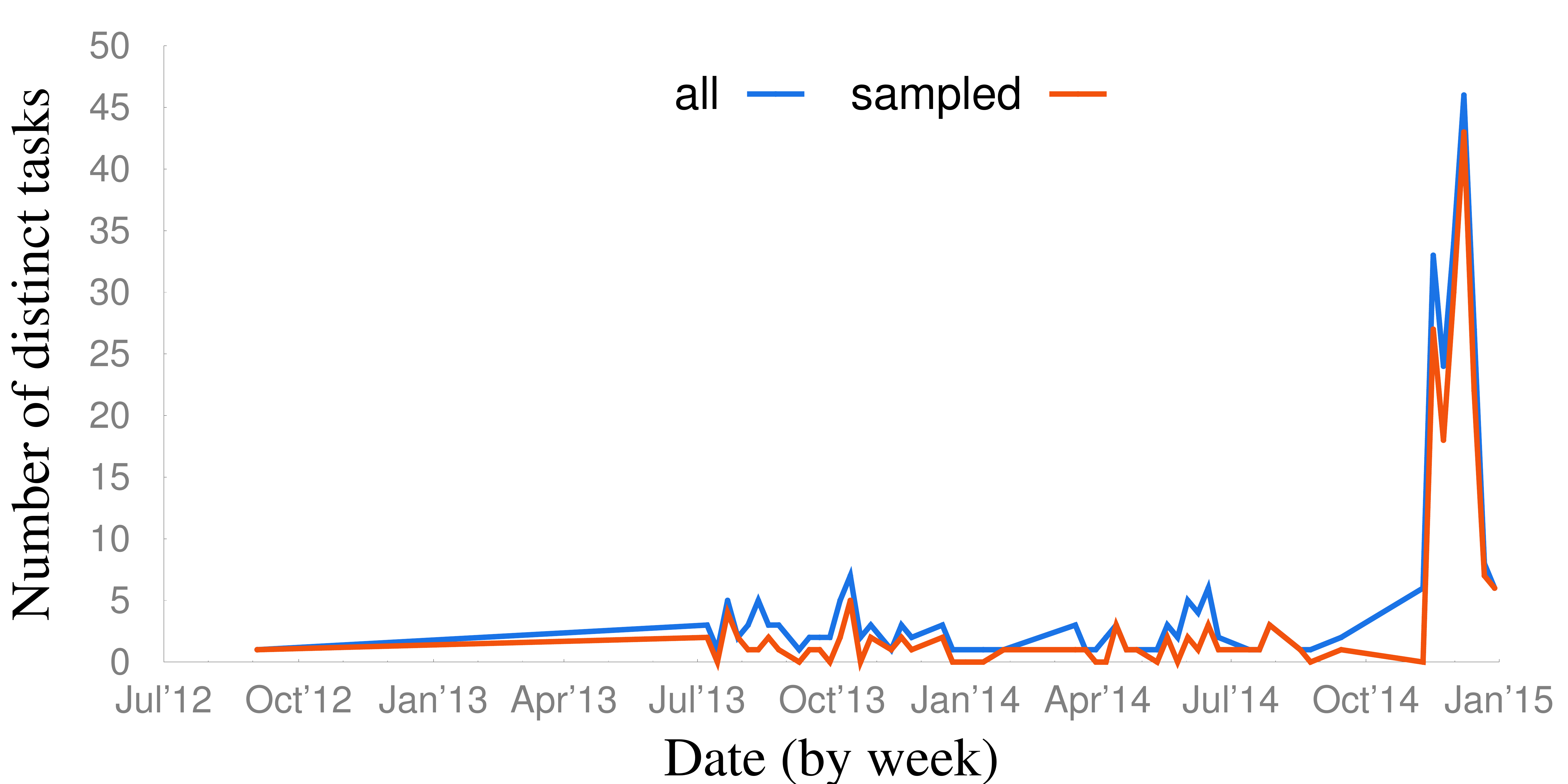}
\caption{\label{fig:sampling-week-cluster-post}Pre Jan 2015}
\end{subfigure}
\begin{subfigure}[b]{0.33\textwidth}
\centering
\includegraphics[width=\textwidth]{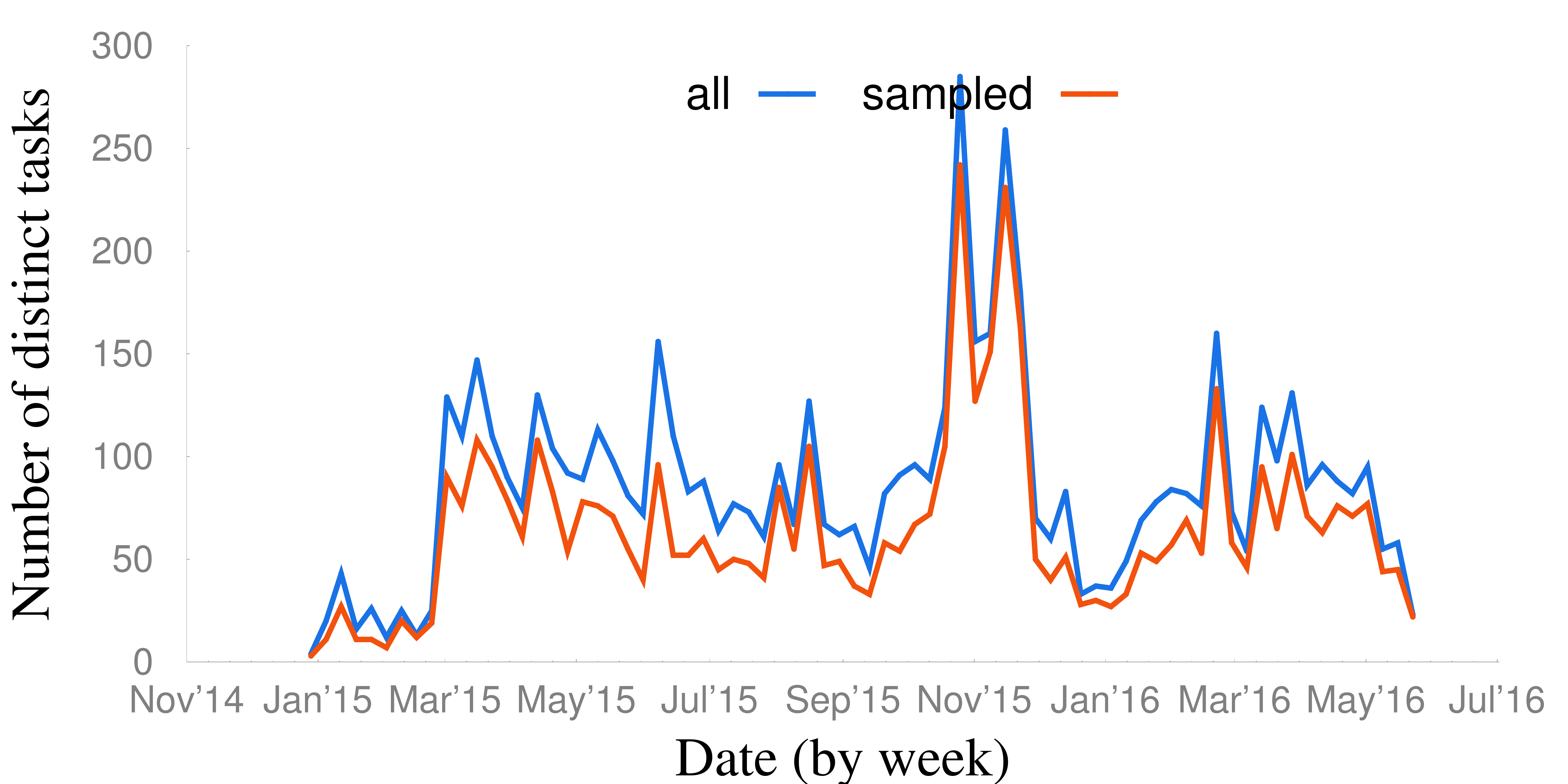}
\caption{\label{fig:sampling-week-cluster-post}Post Jan 2015}
\end{subfigure}
\caption{\label{fig:sampling}Number of tasks sampled (by week)}
\end{figure*}

}

\subsection{Dataset Attributes}\label{sec:available-data}
The dataset is provided to us at the batch level. 
For each batch in our sample, we have metadata containing a short textual 
description of the batch (typically one sentence), 
as well as the source HTML code to one sample task instance 
in the batch.
In addition, the marketplace also provides a comprehensive set of
metadata for each task instance within the batch,
containing
\begin{denselist}
\item Worker attributes such as \texttt{worker ID}, location (\texttt{country}, \texttt{region}, \texttt{city}, \texttt{IP}), and \texttt{source} (recall that this marketplace recruits workers from different sources);
\item Item attributes such as \texttt{item ID}; and
\item Task instance attributes such as \texttt{task ID},
\texttt{start time}, \texttt{end time}, \texttt{trust score}, and \texttt{worker response}.
\end{denselist}
As we can see in this list, the marketplace assigns workers a \texttt{trust score}
for every task instance that they work on. 
This trust score reflects the accuracy of these workers on
test tasks the answers to whose questions is known.
The marketplace administers these test tasks before
workers begin working on ``real'' tasks. 
\techreport{Unfortunately, we were not provided these test tasks, and only 
have the {\tt trust score} as a proxy for the true accuracy of the worker for that specific type of task.
In addition to the {\tt trust score}, we also have information 
about worker identities, and other attributes of the 
items being operated on, and the start and end times for each task instance.}

At the same time there are some important 
attributes that are not visible to us from this dataset.
For instance, we do not have requester IDs, 
but we can use the sample task HTML to  
{\em infer} whether two separate batches have the same type of task, 
and therefore were probably issued by a single requester.
Nor do we have have ``ground truth'' answers for questions in the 
tasks performed by workers.
However, as we will describe subsequently, we find
other proxies to be able to estimate the accuracies of workers or tasks. 
Finally, we do not have data regarding the payments 
associated with different tasks and batches.

\subsection{How did we enrich the data?}\label{sec:data-enrich}

The raw data available for each batch, as described above,
is by itself quite useful in exploring high-level marketplace statistics 
such as the number of tasks and workers over time, the geographic distribution of workers, 
typical task durations, and 
worker lifetimes and attention spans. 
\techreport{That said, this raw data is insufficient to address many of the important
issues we wish to study. 
For example, we cannot automatically identify whether a task operates on web data
or on images, and whether or not it contains examples, or free-form text response boxes.}
To augment this data even further, we {\em enrich} the dataset by
inferring or collecting additional data.
We generate three additional types of task attribute data:
\begin{denselist}
\item Manual labels---we also manually annotate each 
batch on the basis of their {\em task goal}, e.g., entity resolution, sentiment analysis, 
{\em operator type}, e.g., rating, sorting, labeling, 
and the {\em data type} in the task interface, e.g., text, image, social media,
discussed further in Section~\ref{sec:label-analyses}.
\item Design parameters---we extract and store {\em features} 
from the sample HTML source 
as well as other raw attributes of the tasks 
that reflect design decisions made by the requesters. 
For example, we check 
whether a task contains 
instructions, examples, text-boxes and images---we discuss
these further in Section~\ref{sec:task_analyses}. 
\item Performance metrics---we compute and store 
different metrics to characterize the latency, cost and error of
 tasks to help us perform 
quantitative analyses on the ``effectiveness'' of a task's design,
discussed further in Section \ref{sec:metrics}.
\end{denselist}

\techreport{
\subsection{What are the Goals of Our Analysis?}\label{sec:goals}
As previously mentioned, our main goals (at a broad level) 
are to quantitatively address the questions of 
(1) Marketplace dynamics --- helping marketplace administrators and owners 
understand the interaction between tasks and workers, and the corresponding marketplace load; 
(2) Task design --- helping requesters understand what 
constitutes an effective task, and how to go about designing one; and 
(3) Worker behavior --- understanding worker attention spans, lifetimes, and general behavior,
 for the improvement of the crowdsourcing ecosystem as a whole. While the first two goals directly impact requesters and marketplace administrators, we believe they will also help indirectly improve the general worker experience in terms of availability of desirable tasks, and a reduction in the laboriousness of performing tasks.
We now discuss our goals in a little more detail, by breaking each of them down into sub-goals and describing the experiments we perform to answer them.

\noindent
{\bf Marketplace dynamics.}
To understand the marketplace, the first goal of our analysis is to examine general statistics that help us estimate the scale of crowdsourcing operations within the marketplace. This first order analysis is useful for marketplace administrators, helping them estimate the resources required to manage this scale of operations, and identify key limitations. Thereafter, we look at the availability and flow of tasks or workers on the marketplace --- specifically, we check if the marketplace witnesses sudden bursts or a steady stream of activity. This analysis gives us concrete take-aways that can help future marketplace design better load-balancing strategies. Lastly, we analyze the manual labels assigned to tasks to look at the types of tasks that have become popular within the marketplace as a means towards a better
characterization of the spectrum of  crowd work as a whole.

\noindent
{\bf Task Design.}
To help improve task design, we must first be able to characterize the effectiveness of tasks both qualitatively and quantitatively. The three well known aspects used to talk about the effectiveness of a crowdsourced task are (a) latency, (b) accuracy and (c) cost. Consequently, the first step in our analysis is to identify performance metrics that measure these aspects. The next step is to study the impact of varying different design parameters on each of these metrics. This analysis, when performed on a dataset as large as ours, allows us to make data-informed recommendations to requesters looking to design tasks that that are answered accurately by workers with low latency and at low cost.

\noindent
{\bf Worker analyses.}
A worker-centric view of the marketplace can help us in understanding the workers' experience and make changes appropriately to make it easier for them to work. In this respect, we first look at the various labor sources that provide workers that perform work for the marketplace. The load, resource and quality distribution across these sources can point researchers in devising appropriate load-balancing strategies, and can point practitioners towards the ideal source(s) for crowd work. 
Next, we examine the geographical distribution of workers --- this gives us vital information about the active time-zones of the workforce and can help marketplace administrators in ensuring a constant response rate on the marketplace. Finally, we also study the end lifetimes and attention spans of the workers to figure out what fraction of the workforce are regularly active on the marketplace, and how much time is typically spent on the marketplace by workers on a single day.
}
\section{Marketplace Analyses}\label{sec:marketplace}
In this section, we aim to gain insights into the high level, aggregate workings of the marketplace. 
First, we examine some basic statistics of the marketplace, to understand the worker supply and task demand interactions. Specifically, we look at (a) task instance arrival distribution (Section~\ref{sec:marketplace-bursty}), (b) worker availability (Section~\ref{sec:marketplace-worker-availability}), and, \techreport{(c) marketplace load, or contribution from ``heavy\--hitter'' tasks, occupying a bulk of the tasks in the marketplace.}\papertext{in our extended technical report~\cite{techreport}, we additionally examine the contribution of
``heavy-hitter'' tasks relative to other tasks that are ``one-off'' in our marketplace.}
Then, in Section~\ref{sec:label-analyses}, we explore the types of tasks observed in our dataset, to better understand the questions and data types of interest for requesters. 
We also look for correlations across these labels to understand what types of tasks occur together. 
\techreport{Finally, we look at trends in the complexity of tasks over time to gain additional insights into the evolution of the marketplace (Section~\ref{sec:open-ended}).}
\papertext{In our technical report~\cite{techreport}, 
we additionally explore the complexity of tasks
issued over time, and argue that the fraction of ``easy'' tasks has gone down while
the fraction of ``hard'' ones has gone up.}
\subsection{Are tasks uniform or bursty over time?}\label{sec:marketplace-bursty}
\begin{figure*}[!t]
\centering
\begin{subfigure}[b]{0.33\textwidth}
\centering
\includegraphics[width=\textwidth]{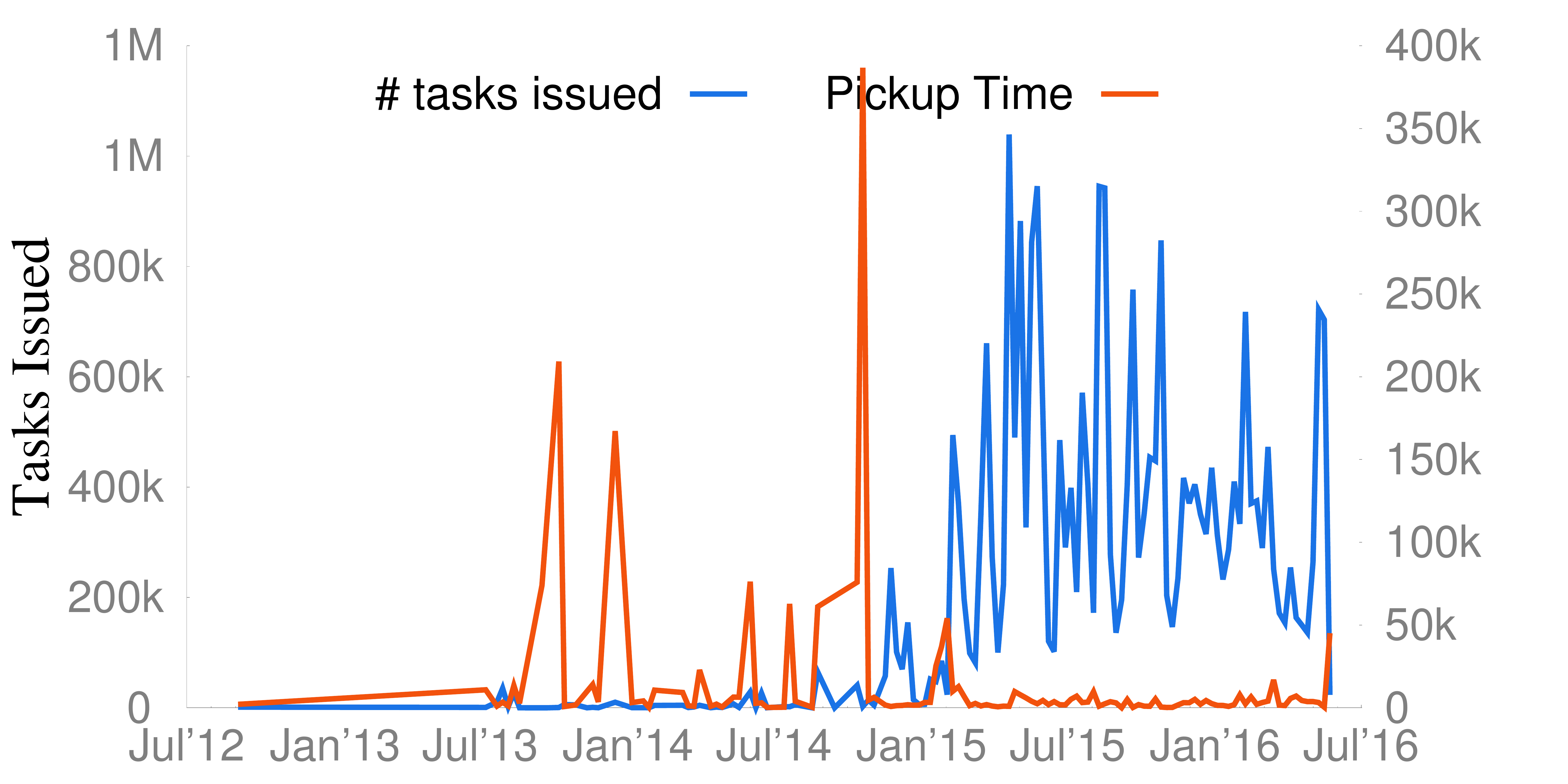}
\caption{\label{fig:task_arrival_monthly}Task instance arrival vs Median pickup times}
\end{subfigure}%
\hfill
\begin{subfigure}[b]{0.66\textwidth}
\centering
\includegraphics[width=0.5\textwidth]{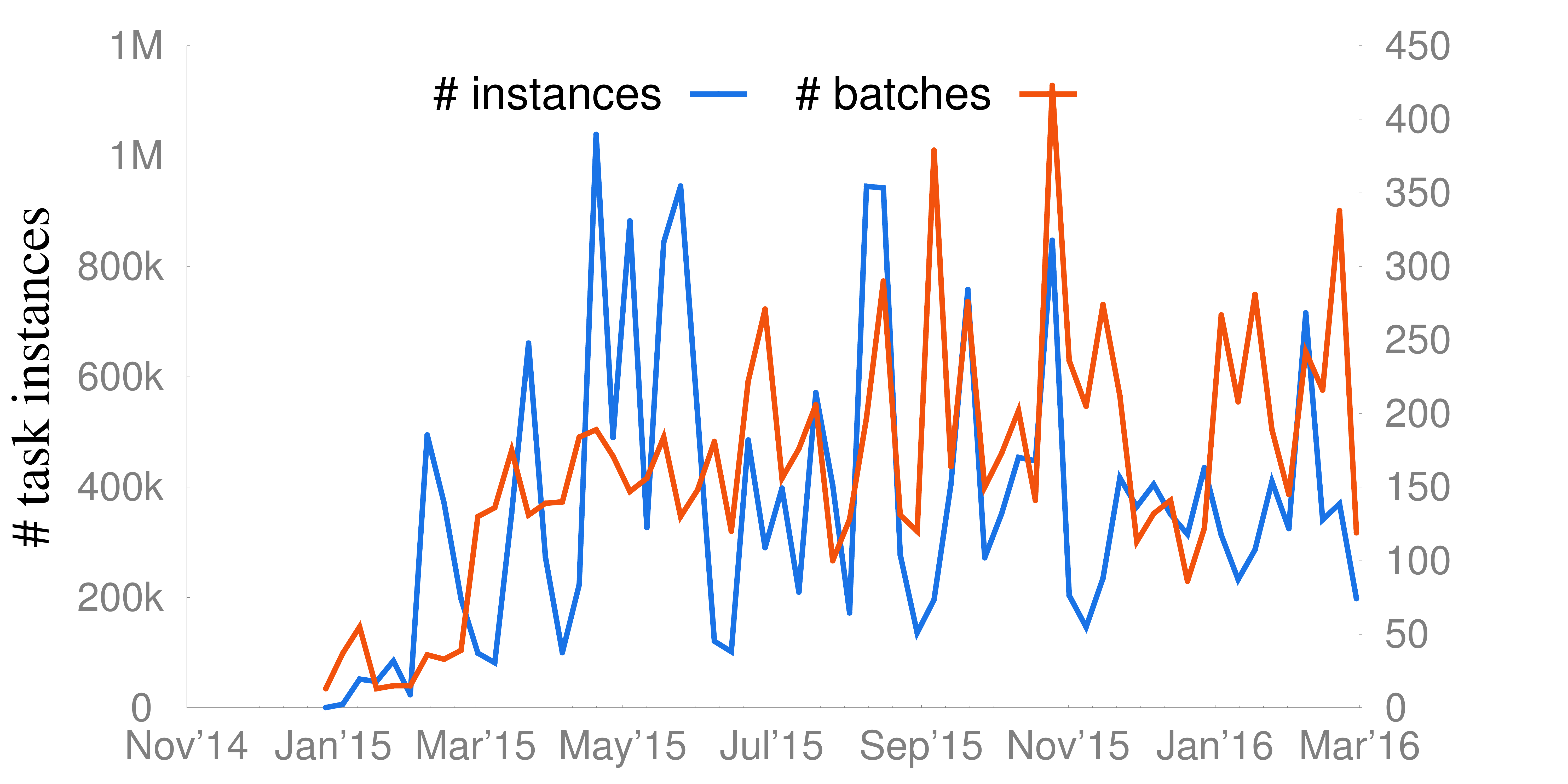}%
\includegraphics[width=0.5\textwidth]{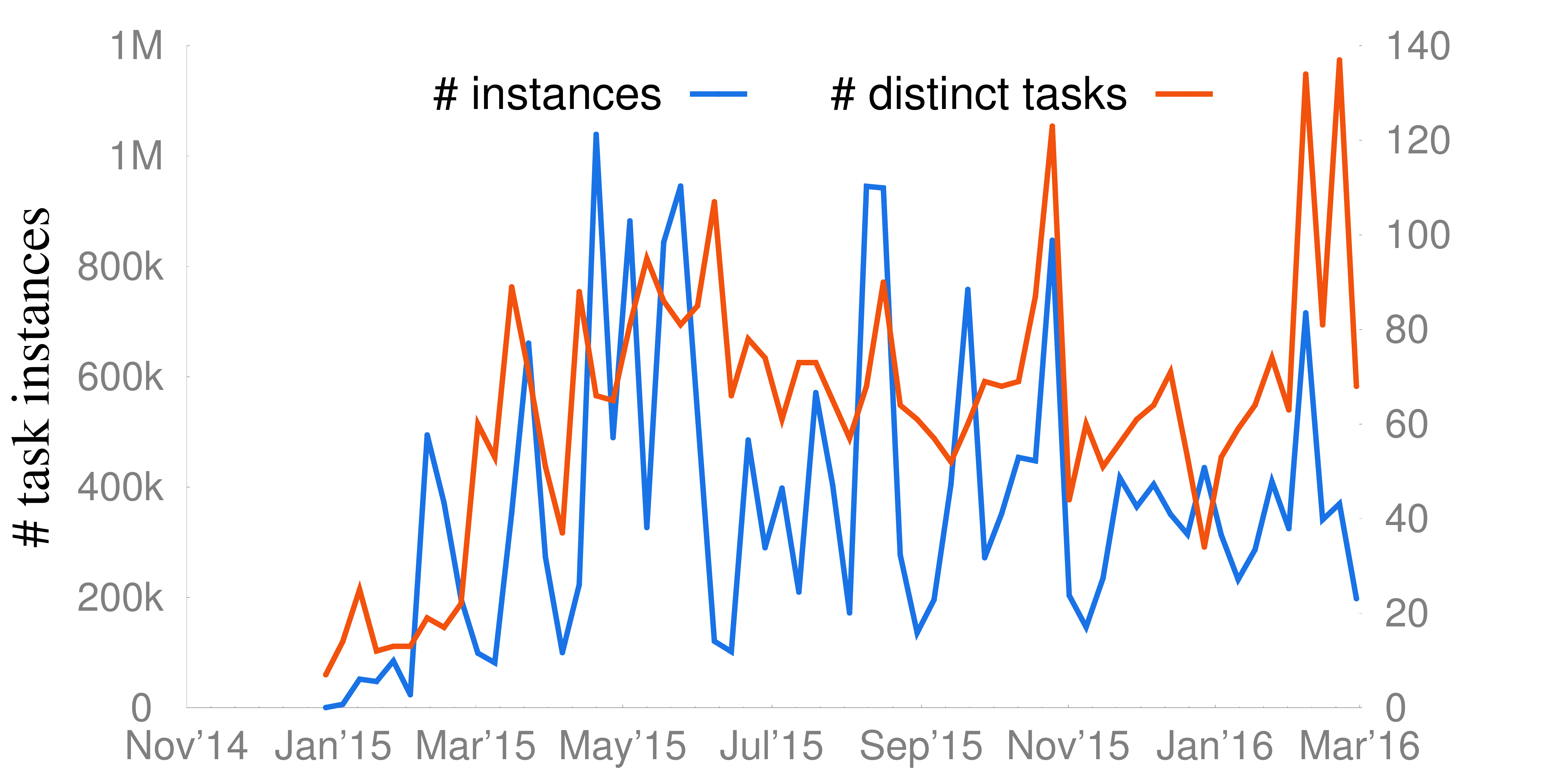}
\caption{\label{fig:arrival_overlay_batch_cluster}Task instance arrival (post Jan 2015) vs (1) batch arrival, (2) distinct task arrival}
\end{subfigure}
\caption{\label{fig:arrivals_overlay}Task Arrivals by week}
\end{figure*}

We first study
the rate at which task instances arrive into the marketplace, 
and the rate at which they are completed. 
\techreport{Note that the load on the marketplace is governed by the number of task instances, which is the fundamental unit of work visible to workers, rather than the number of batches; 
batches can be arbitrarily small or large.}
We plot the number of task instances arriving and being completed each week in Figure~\ref{fig:task_arrival_monthly} in blue. 
First, note that the task arrival plot is relatively sparse until Jan 2015, 
which is presumably when the marketplace started attracting more requesters.
Second, after June 2014, there are some very prominent peaks, 
on top of regular activity each week. 
This suggests that while task instances arrive fairly regularly, there are periods of {\em burstiness}.
Considering the period from Jan 2015 onwards, the median of the number of task instances issued in a day on the marketplace is about 30,000. 
In comparison, on its busiest day, more than 900,000 task instances 
were issued, a $30\times$ increase over normal levels. 
Similarly, the number of task instances 
issued on the lightest day is $0.0004\times$ smaller than the median. 
This raises the question: 
where does the high variation in the number of task instances 
derive from---do the number of batches of tasks issued fluctuate a lot, or do the
number of distinct tasks issued themselves fluctuate a lot? 
For this analysis, we overlay the number of task instances issued on the marketplace 
with the number of batches
and the number of distinct tasks issued for the period post January 2015 in Figure~\ref{fig:arrival_overlay_batch_cluster}.
For both these measures, we find that the fluctuation is similar to the 
fluctuation in the number of task arrivals, indicating
that both factors contribute to the high variation in the market load.

\techreport{
Note that a common explanation for why crowdsourcing is used in companies
is the ability to shrink or grow labor pools on demand~\cite{crowd-book};
this finding seems to suggest that even marketplaces 
need to be able to shrink and grow labor pools based on demand.
For this marketplace, having access to both push and pull mechanisms provides great flexibility. Not only can they route the harder tasks to their more skilled, ``on-demand'' workers, they can also use this push mechanism to reduce latencies for requesters and clear backlogged of tasks. On the other hand, the presence of a large freelance workforce implies that for the majority of tasks they do not have to rely on the more expensive skilled workers, and can therefore be conservative in their use of resources in maintaining a pool of these internal ``super-workers''.
This has huge implications for individuals who rely on crowdsourcing
as a sole source of income: depending on the week, they may not have enough
tasks that suit their interest or expertise. 

Striking a good balance between the two task routing mechanisms and worker pools is crucial to ensuring that all three parties are satisfied: (1) the marketplace is able to clear pending tasks without a building backlog, while maintaining requisite levels of accuracy and cost, (2) requesters receive quality responses for easy and hard tasks, and do not see high latencies in responses, and (3) workers have access to as much work as they can handle, as well as tasks that can cater to their varied interests and expertise levels.
}


Besides the bursty nature of task instance arrivals across weeks,
the marketplace also witnesses periods of low task arrivals on the weekends---the number of instances posted on a weekday is up to $2\times$ the number of instances posted
on Saturdays or Sundays on average.
Further, the average number of instances posted at the start of the week is the highest, 
following which the number decreases over the week.
\techreport{We plot this in Figure~\ref{fig:tasks_weekdays}.}
\papertext{This chart can be found in our technical report~\cite{techreport}.}


\techreport{
\begin{figure}[h]
\centering
\includegraphics[width=0.7\linewidth]{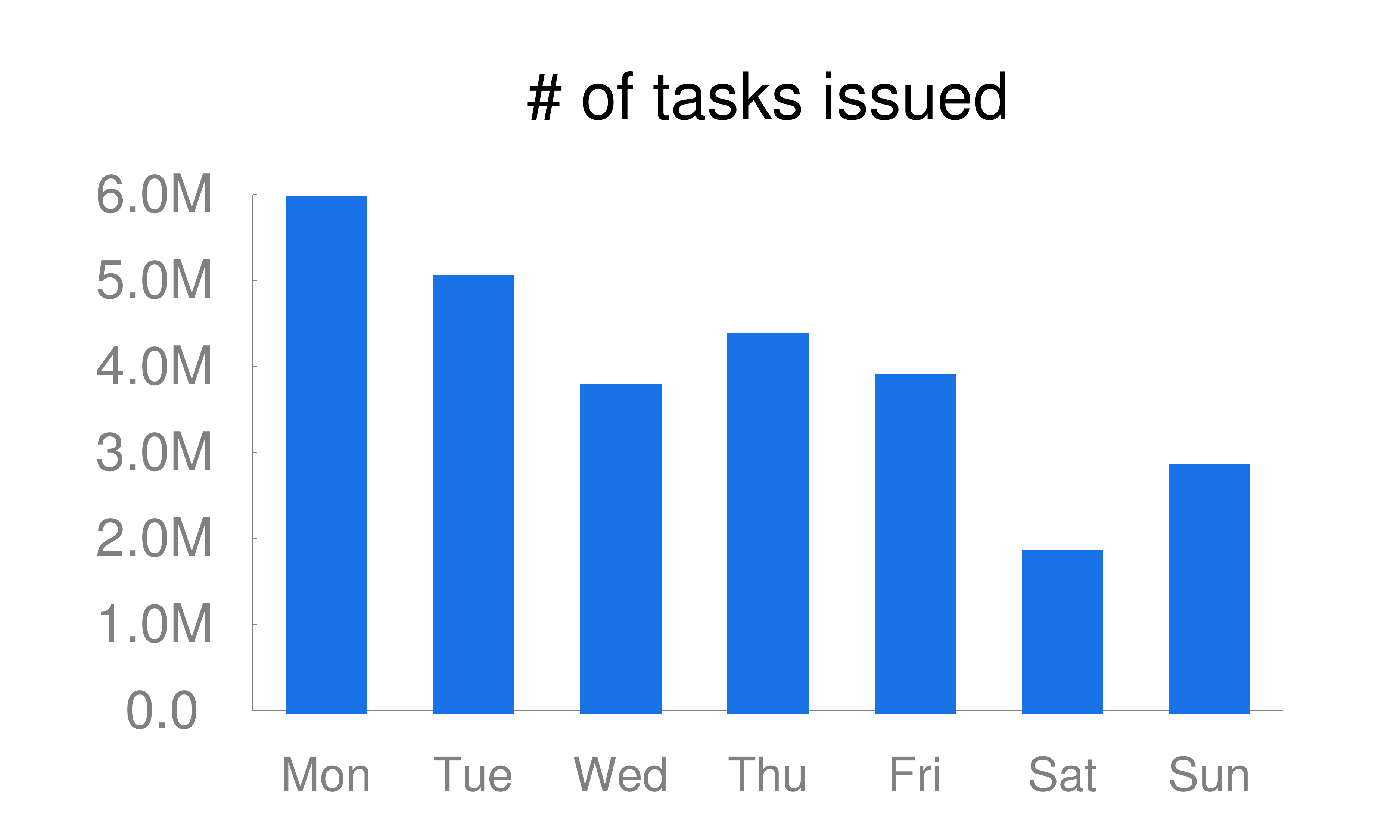}
\caption{\label{fig:tasks_weekdays}Distribution of tasks over days of the week}
\end{figure}
}

\techreport{\ta{Marketplaces witness wide variation in the number of tasks issued, with daily number of issued instances varying between $0.0004\times$, to up to $30\times$ the median load (30,000 instances).}}

\subsection{How does the availability and participation of workers vary?}\label{sec:marketplace-worker-availability}
\begin{figure}[h]
\centering
\vspace{-10pt}
\includegraphics[width=0.8\linewidth]{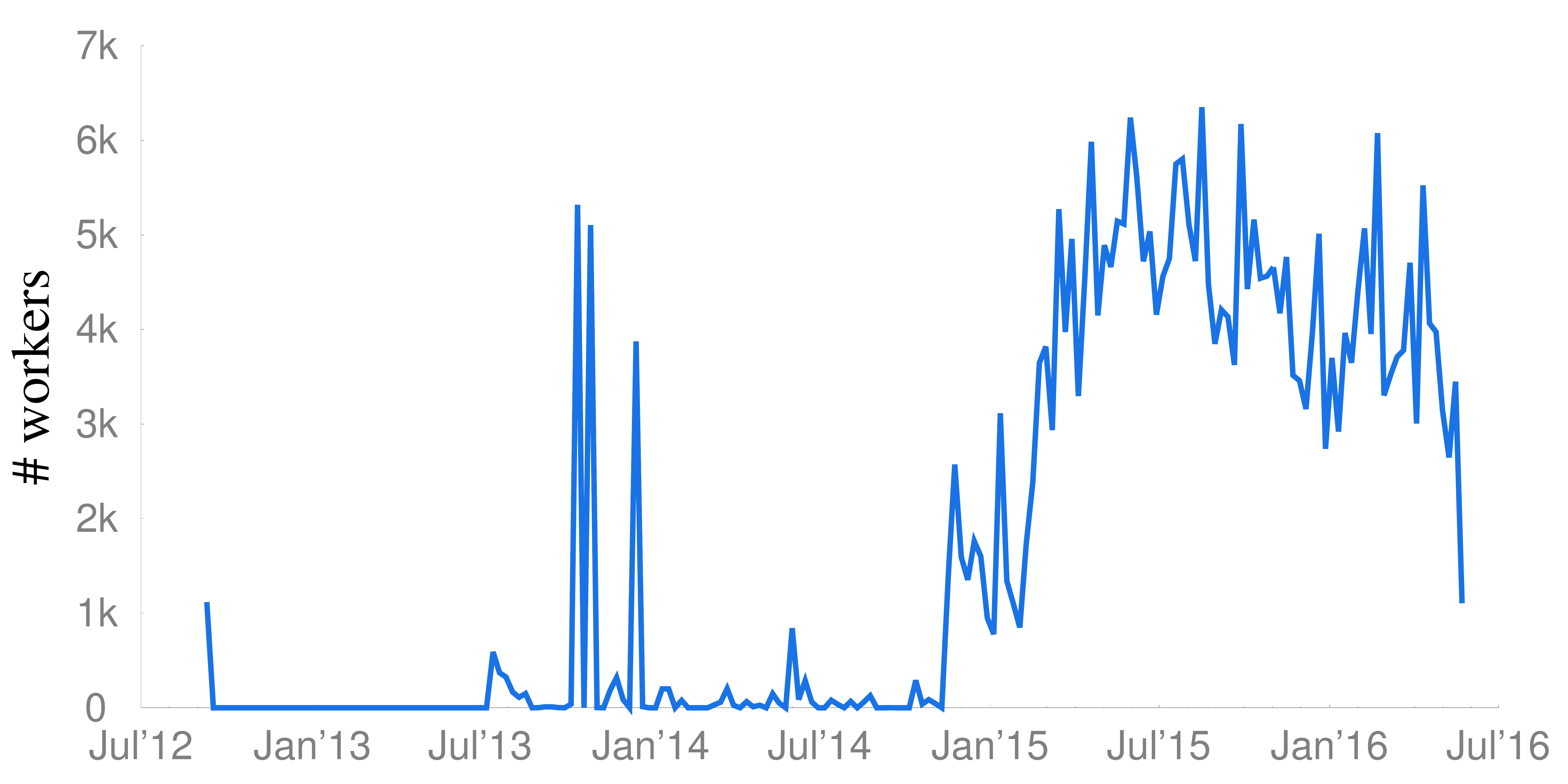}
\vspace{-5pt}
\caption{\label{fig:worker_arrival}Number of workers performing tasks}
\vspace{-10pt}
\end{figure}

\sampar{Worker Availability.}
As described earlier, the specific marketplace we work with 
attracts workers from a collection of labor sources. 
In this manner, it is able to keep up with the spikes in demand. 
We investigate the sources the marketplace draws from in Section~\ref{sec:worker}.
In this section, we focus on studying the number of active workers across different
weeks: Figure~\ref{fig:worker_arrival} depicts this statistic.

Unlike Figure~\ref{fig:task_arrival_monthly} that had a huge variation in the number 
issued task instances, especially after 2015, 
Figure~\ref{fig:worker_arrival} does not show this level of variation.
Thus, somewhat surprisingly, even though there are huge changes 
in the number of available task instances, 
roughly the same number of workers are able to ``service'' a greater number of
requests. This indicates that there is a limitation more in the supply
of task instances than availability of workers.


\techreport{
\ta{Despite the huge variation in the number of available tasks, roughly the
same number of workers (with small variations) 
are able to service all of these tasks.}
}

\begin{figure*}[!t]
\centering
\begin{subfigure}[b]{0.33\textwidth}
\centering
\includegraphics[width=\textwidth]{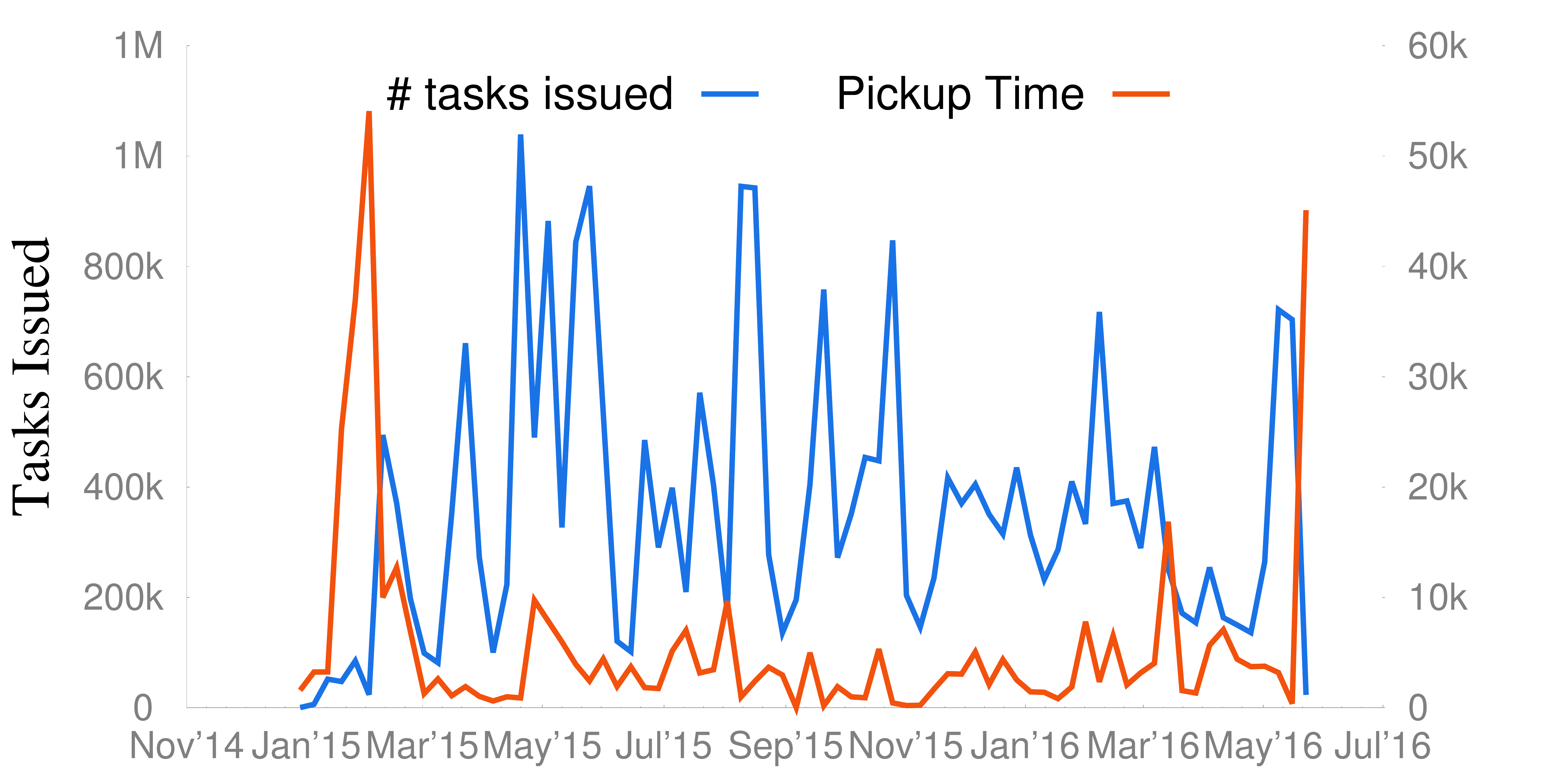}
\caption{\label{fig:task_arrival_weekly_post}Task Arrivals vs Median Pickup Time}
\end{subfigure}%
\hfill
\begin{subfigure}[b]{0.66\textwidth}
\centering
\includegraphics[width=0.5\textwidth]{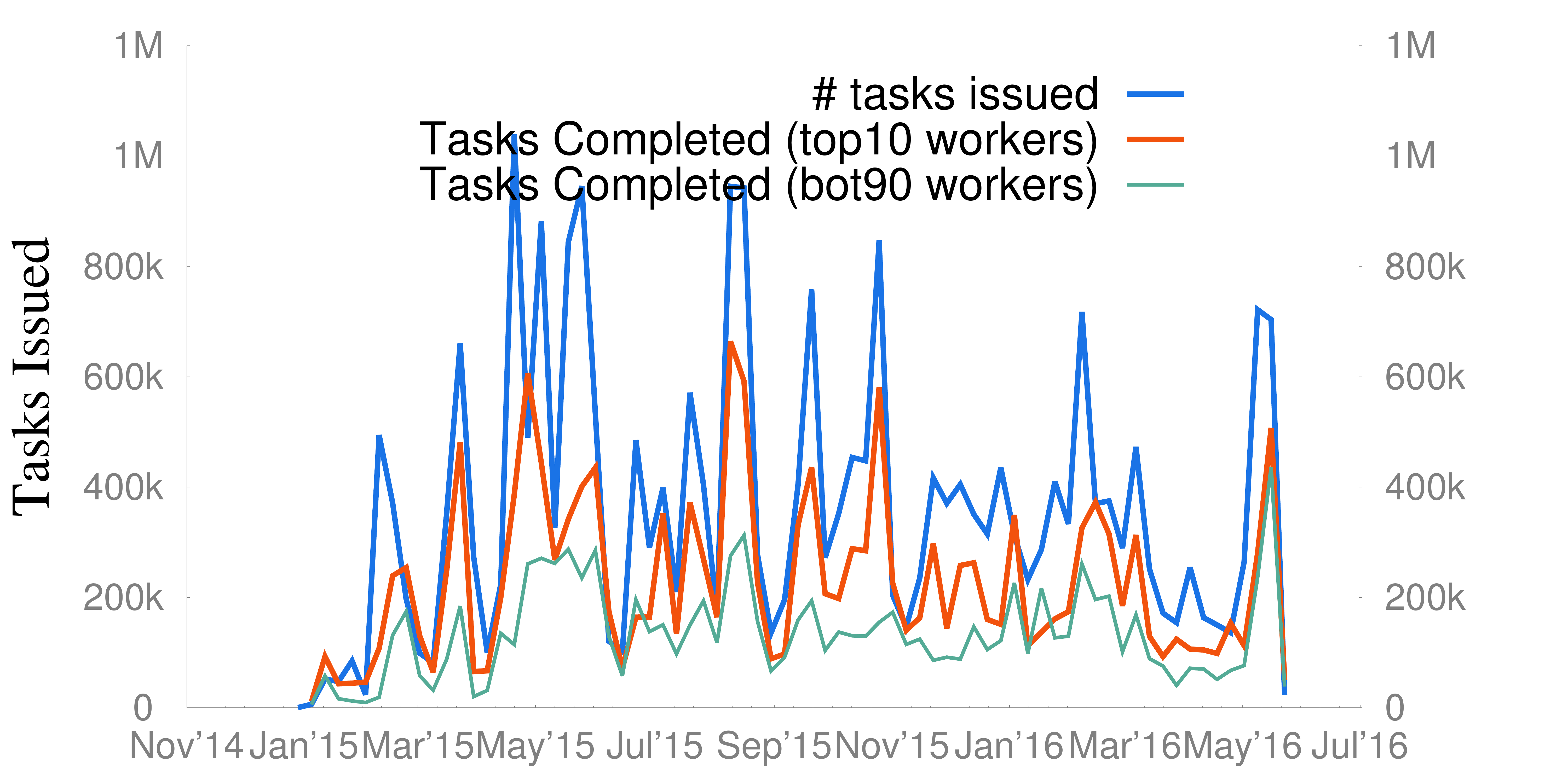}%
\includegraphics[width=0.5\textwidth]{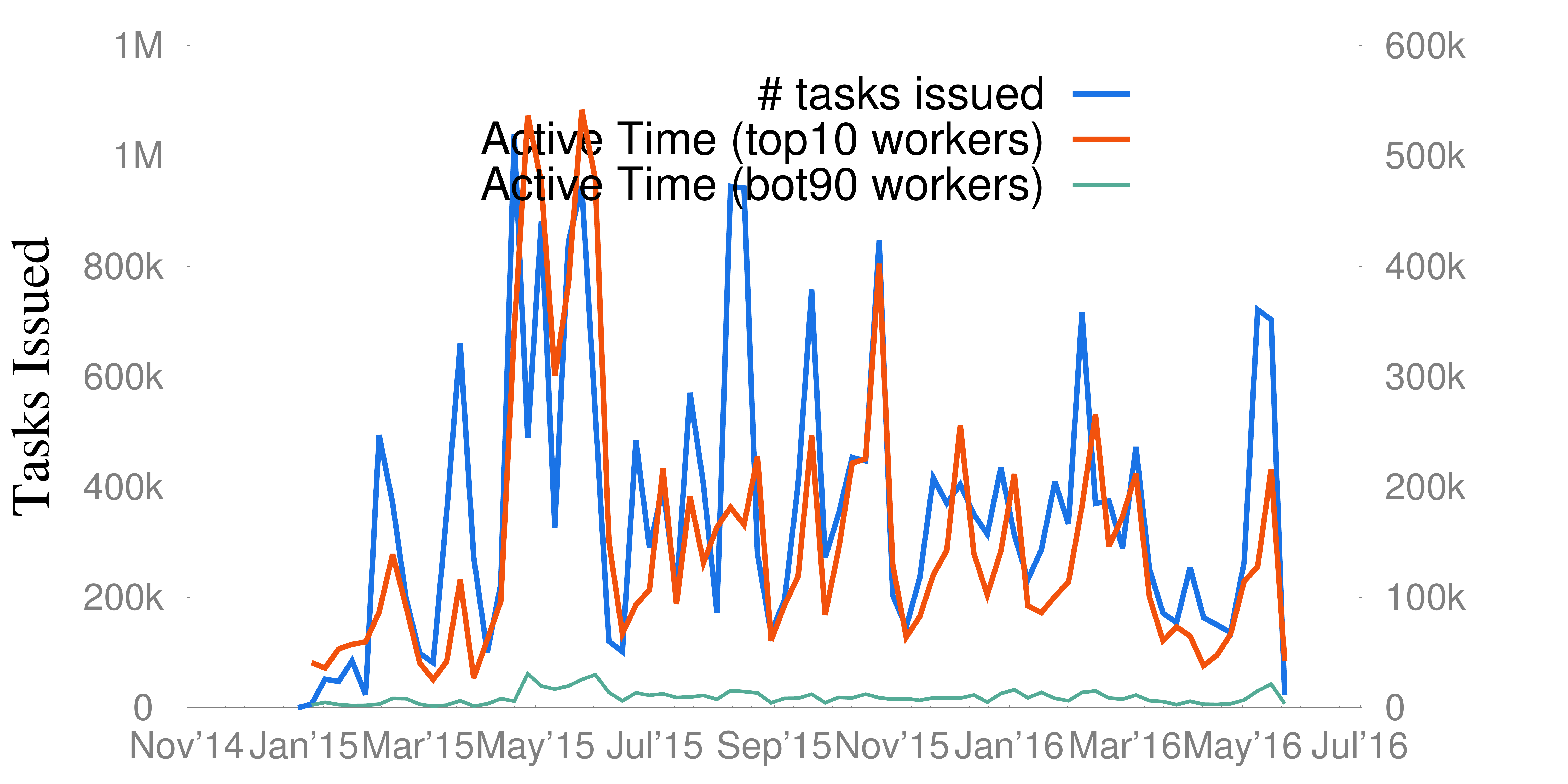}
\caption{\label{fig:task_arrival_top_bot_split}Engagement of {\em top-10\%} and {\em bot-90\%} workers: (1) \# Tasks (2) Active Time}
\end{subfigure}
\caption{\label{fig:arrival_overlay_split}Task Arrivals by Week (Post Jan 2015)}
\end{figure*}

\sampar{Worker Latencies, Idleness, and Task-Distribution.}
We now attempt to explain how roughly the same number of workers
are able to accommodate for the variation in the number of tasks 
on the platform. 
Our first observation is that the median latency in task instances getting 
picked up by workers, noted as {\em pickup time} (and defined formally later in Section~\ref{sec:metrics}) in 
Figure~\ref{fig:task_arrival_monthly}, and depicted in red,
shows that during periods of high load, the marketplace tends to move faster. 
We also zoom in to the high activity period after January 2015 
in Figure~\ref{fig:task_arrival_weekly_post} to further highlight this trend.
One possible explanation for this observation is that when more 
task instances are available, a larger number of workers are attracted to the marketplace
or recruited via a push-mechanism---leading to lower latencies. 
Another possibility (supported by our discussion below for Figure~\ref{fig:task_arrival_top_bot_split}) 
is that with a higher availability of tasks, workers are spending 
a lot more active time on the platform, and hence are more likely to pick up 
new tasks as soon as they are available.

Next, we look into how the workload is being distributed 
across the worker-pool. 
In Figure~\ref{fig:task_arrival_top_bot_split}, 
we plot the number of tasks completed by the top-10\% (in red color)
and the bottom-90\% (in green color) of workers in each week and compare it 
to the total number of tasks issued. 
We observe that while the bottom-90\% also take on a lot more 
tasks during periods of high load, it is the top-10\% that handles most of the flux, 
and is consistently performing a lot more tasks than the remaining 90\%. 
Similarly, examining the same plot for average amount of {\em active time} 
spent by workers on the platform in Figure~\ref{fig:task_arrival_top_bot_split} also 
shows that the top-10\% are indeed spending a lot more time on average per 
week to handle the varying task load as compared to the bottom-90\%.
This observation indicates that while having a large workforce certainly helps, 
it is crucial to focus on worker interest and engagement---attracting more
``active'' workers can allow marketplaces to handle fluctuating workloads better.
We also examined the workload handled by workers 
from different labor sources to verify whether the majority of this 
variation is assigned to the marketplace's internal or external workers. 
We observed that the internal workers account 
for a very small fraction of tasks\papertext{---we defer a full account 
of our results to~\cite{techreport}}.
\techreport{task arrival overlay with internal and external}

\techreport{
\subsection{What is the distribution of work across different distinct tasks?}\label{sec:heavy-hitter}
Next, we wanted to study whether there are a small number of tasks that
dominate the marketplace (e.g., repeatedly labeling different items, issued by a single requester). 
To study this, we first clustered the batches in our dataset
based on metadata from the extracted HTML source
corresponding to the tasks (see Section~\ref{sec:data-enrich}),
and tuned the threshold of a match to ensure that the tasks that on inspection
look very similar and have similar purposes are actually clustered together. 
We shall henceforth refer to these clusters of similar batches 
corresponding to a distinct task, as simply {\em clusters}. 
We denote the number of batches in a cluster by {\em cluster size}.
Then, in Figure~\ref{fig:cluster_dist_batches}, we plot the distribution of 
the number of clusters that have different cluster sizes (both on log scale). 
For example, there were 5 clusters with size larger than 100, indicating that
there were 5 distinct tasks (each lumped into their own clusters)
that were issued across at least 100 batches each. 
As can be seen in the chart, 
there seem to be a large number of tasks that are ``one-off'' with a small number ($<10$) of 
batches:
these tasks, being one-off, cannot benefit from much fine-tuning of the 
interface prior to issuing the task to the marketplace.
On the other hand, there are a small number of ``heavy hitters'': 
more than 10 tasks had cluster sizes of over 100, indicating that 
these tasks had been issued across 100s of batches.
Notice that even within a batch the number of tasks may be large: we study that in
the next plot in Figure~\ref{fig:cluster_dist_tasks}. We see a wide variation in the number of tasks issued - while 204 clusters have less than 10 tasks issued, 3 clusters have more than 1M tasks each. Furthermore, these ``bulky'' clusters have issued close to 80k tasks/batch, so even slight improvements in the design of these batches can lead to rich dividends for the requester. Across this chart, the median number of tasks per cluster is 400. 
\begin{figure}[h]
\centering
\includegraphics[width=0.7\linewidth]{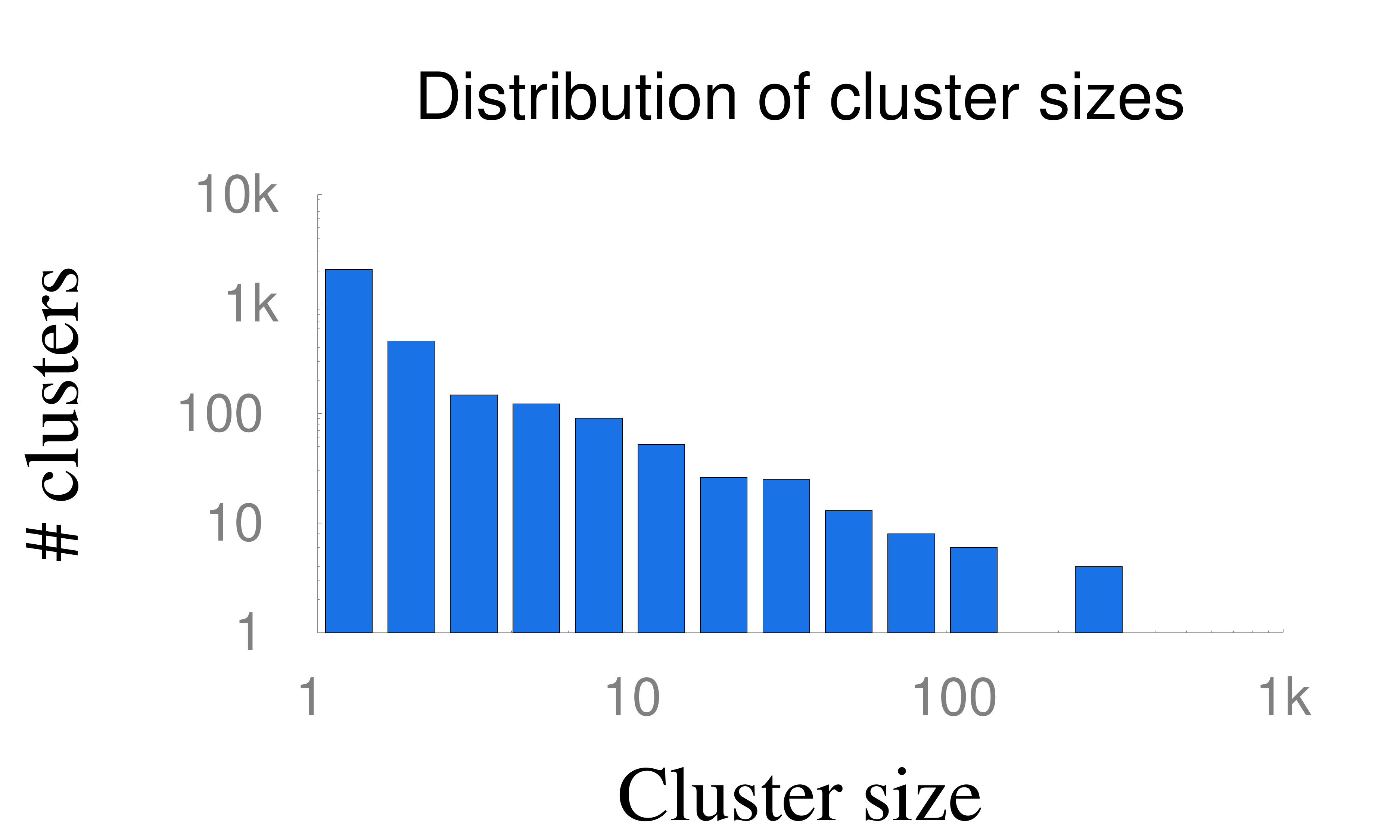}
\caption{\label{fig:cluster_dist_batches} \# of batches in clusters}
\end{figure}
\begin{figure}[h]
\centering
\includegraphics[width=0.7\linewidth]{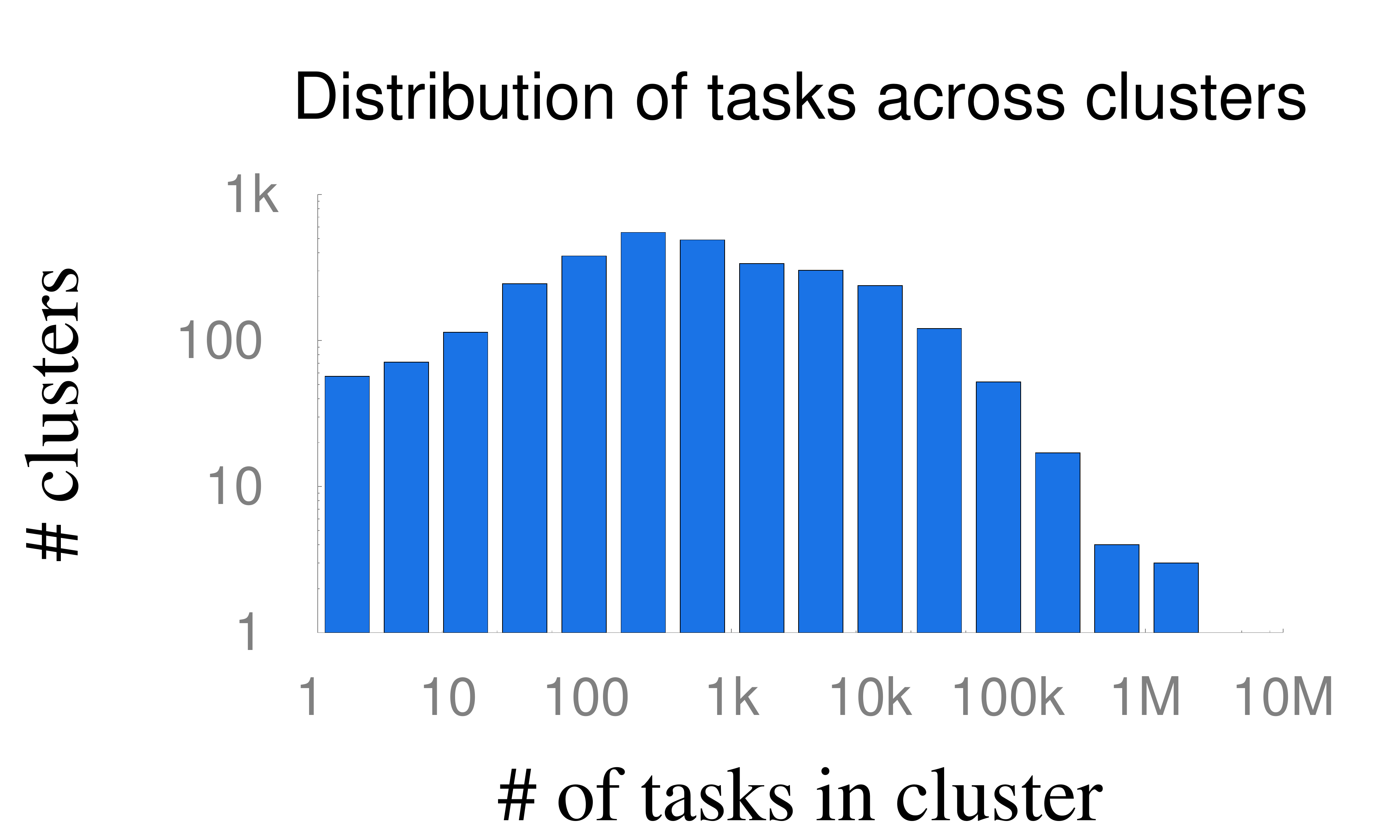}
\caption{\label{fig:cluster_dist_tasks} \# of tasks in clusters}
\end{figure}
Next, we drill down into the top 10 tasks which had over 100 batches, 
the so-called ``heavy hitters''.
In Figure~\ref{fig:heavy_hitter_dist} we plot the cumulative number of tasks issued over time,
one line corresponding to each heavy hitter distinct task.
As can be seen in the figure, 
these tasks show both uniform and bursty availabilities. As an example, the task corresponding to the purple line has only been active in July 2015  while the task corresponding to cyan line has had batches issued regularly over 11 months from May 2015 to April 2016. 

\ta{A huge fraction of tasks and batches come from a few clusters, so fine-tuning towards those clusters can lead to rich
dividends.
The heavy hitter task types have a rapid increase to a  steady stream of activity followed by a complete shutdown, after which that task type is never issued again.}
}

\techreport{

\begin{figure}[h]
\centering
\includegraphics[width=\linewidth]{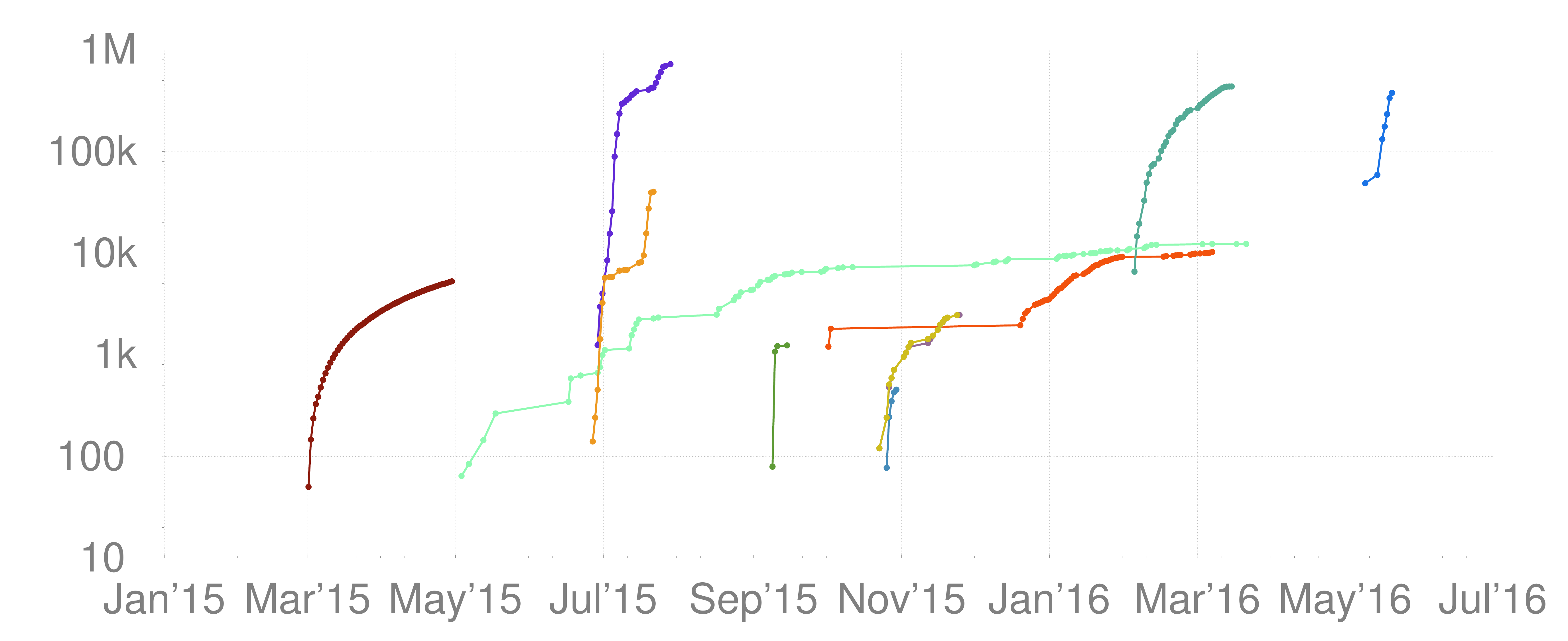}
\caption{\label{fig:heavy_hitter_dist}Heavy Hitter Distribution}
\end{figure}


}

\subsection{What kinds of tasks do we see?}\label{sec:label-analyses}
We now study our enriched task-labels from Section~\ref{sec:data-enrich}
in order to characterize the spectrum of crowd work on in the marketplace. 
Such an analysis can be very useful, for example, to develop a workload of crowdsourcing, 
and to better understand the task types that are most important for further research.

\smallskip
\noindent
{\bf Label Categories.}
We label each task under four broad categories\footnote{\scriptsize Labeling was performed independently by two of the authors, following which the differences were resolved via discussion.}. 
Tasks have one or more label under each category. 
Our mechanism to label tasks is to first cluster batches 
together based on similarity of
constituent tasks, and then we label one task corresponding to 
each cluster, since all tasks within each cluster have identical characteristics. 
The goal of our clustering is to capture the separation between distinct tasks, which is not known to us. As labeling is a labor-intensive process, 
we currently have labels available for about 10,000 
out of the total 12,000 batches ($\approx 83\%$) and 
24 million out of the total 27 million task instances ($\approx 89\%$). 
These batches fall into about $\sim$3,200 clusters.
\begin{denselist}
\item Task \textsf{Goal}: 
Here, we separate tasks based on their end goal. 
We find that most tasks can be characterized as having one (or more) of the following 7 goals\footnote{\scriptsize Tasks that had uncommon or unclear goals and did not fall into one of these classes, were automatically classified as \textsf{Other} or \textsf{Unsure} respectively. This holds for the other categories besides goals as well.}: 
(1) \textsf{Entity Resolution (ER)}, for instance, identifying if two webpages talk about the same business, or if two social media profiles correspond to one single person, 
(2) \textsf{Human Behavior (HB)}, including psychology studies, surveys and demographics, and identifying political leanings, 
(3) \textsf{Search Relevance Estimation (SR)}, 
(4) \textsf{Quality Assurance (QA)}, including spam identification, content moderation, and data cleaning, 
(5) \textsf{Sentiment Analysis (SA)}, 
(6) \textsf{Language Understanding (LU)}, including parsing, NLP, and extracting grammatical elements, and (7) \textsf{Transcription (T)}, including captions for audio and video, and extracting structured information from images.
\item Task \textsf{Operator}: In this category, we label tasks based on the human-operators, or underlying data processing building blocks used by requesters to achieve tasks' goals. 
We observe primarily 10 different operators: 
(1) \textsf{Filter (Filt)}, i.e., separating items into different classes or answering boolean questions, 
(2) \textsf{Rate (Rate)}, i.e., rating an item on an ordinal scale 
(3) \textsf{Sort (Sort)}, 
(4) \textsf{Count}, 
(5) \textsf{Label or Tag (Tag)}, 
(6) \textsf{Gather (Gat)}, i.e., provide information that isn't directly present in the data, for instance by searching the web, 
(7) \textsf{Extract (Ext)}, i.e., convert implicit information already present in provided data into another form, such as extracting text within an image. 
(8) \textsf{Generate (Gen)}, i.e., generate additional information by using inferences drawn from given data, using worker judgement and intelligence, such as writing captions or descriptions for images,
(9) \textsf{Localize (Loc)}, i.e., draw, mark, identify, or bound specific segments of given data and perform some action on individual segments, e.g., draw bounding boxes to identify humans in images, 
and 
(10) \textsf{External Link (Exter)}, i.e., visit an external webpage and perform an action there, e.g., fill out a survey form, or play a game.
\item \textsf{Data} Type: We also separate tasks based on the type of data that is used. 
The same goals and operators can be applied on multiple data types.
All tasks contain a combination of the following 7 types of data: 
(1) \textsf{Text}, 
(2) \textsf{Image}, 
(3) \textsf{Audio}, (4) \textsf{Video}, (5) \textsf{Maps}, (6) \textsf{Social Media}, and 
(7) \textsf{Webpage}.
\end{denselist}

\begin{figure*}[!htbp]
\centering
\vspace{-10pt}
\begin{subfigure}[b]{0.32\textwidth}
\includegraphics[scale=0.45]{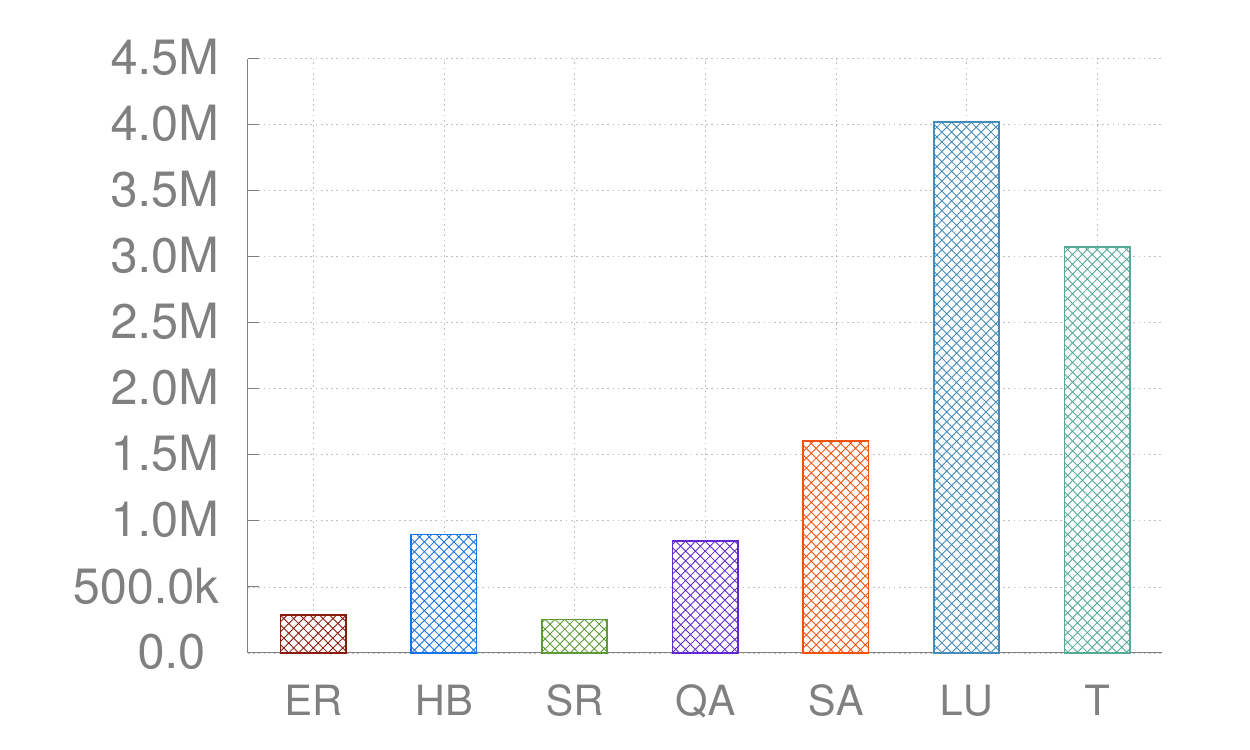}
\vspace{-15pt}
\caption{\label{fig:label_freq_task_goals}Popular Task Goals}
\end{subfigure}
\begin{subfigure}[b]{0.32\textwidth}
\includegraphics[scale=0.45]{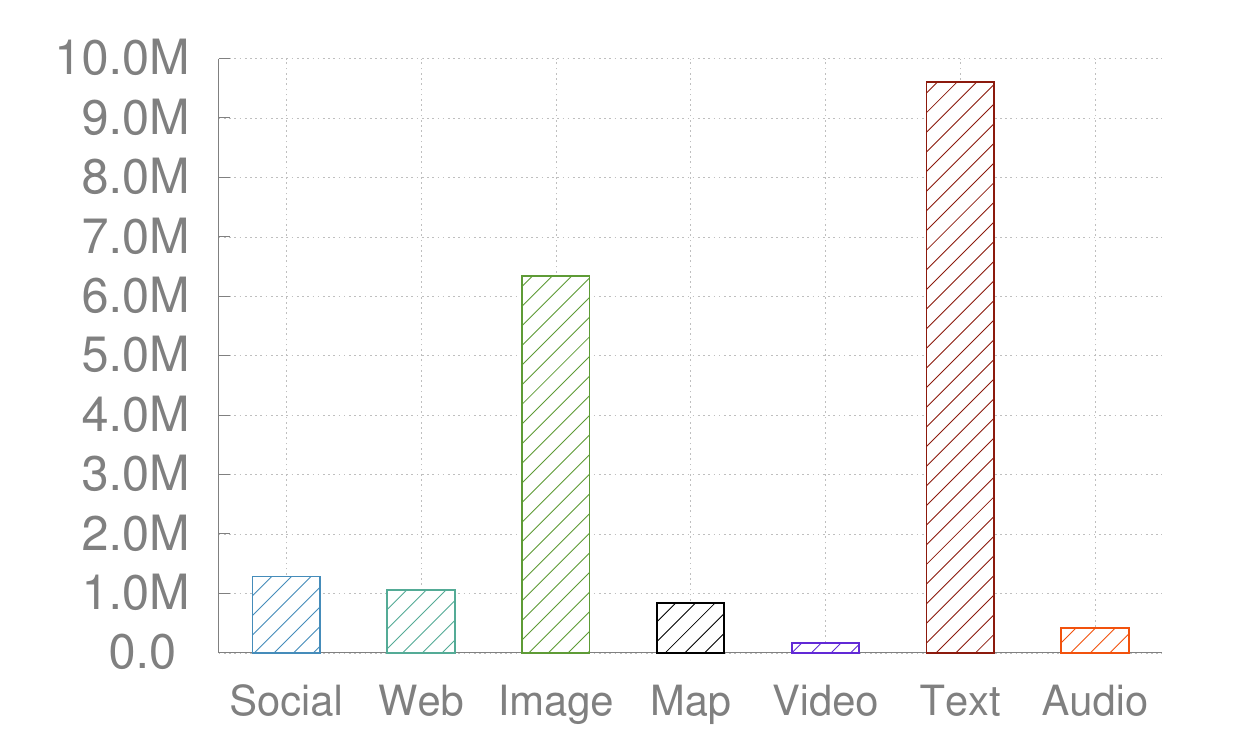}
\vspace{-15pt}
\caption{\label{fig:label_freq_data_type}Popular data types}
\end{subfigure}%
\begin{subfigure}[b]{0.32\textwidth}
\includegraphics[scale=0.45]{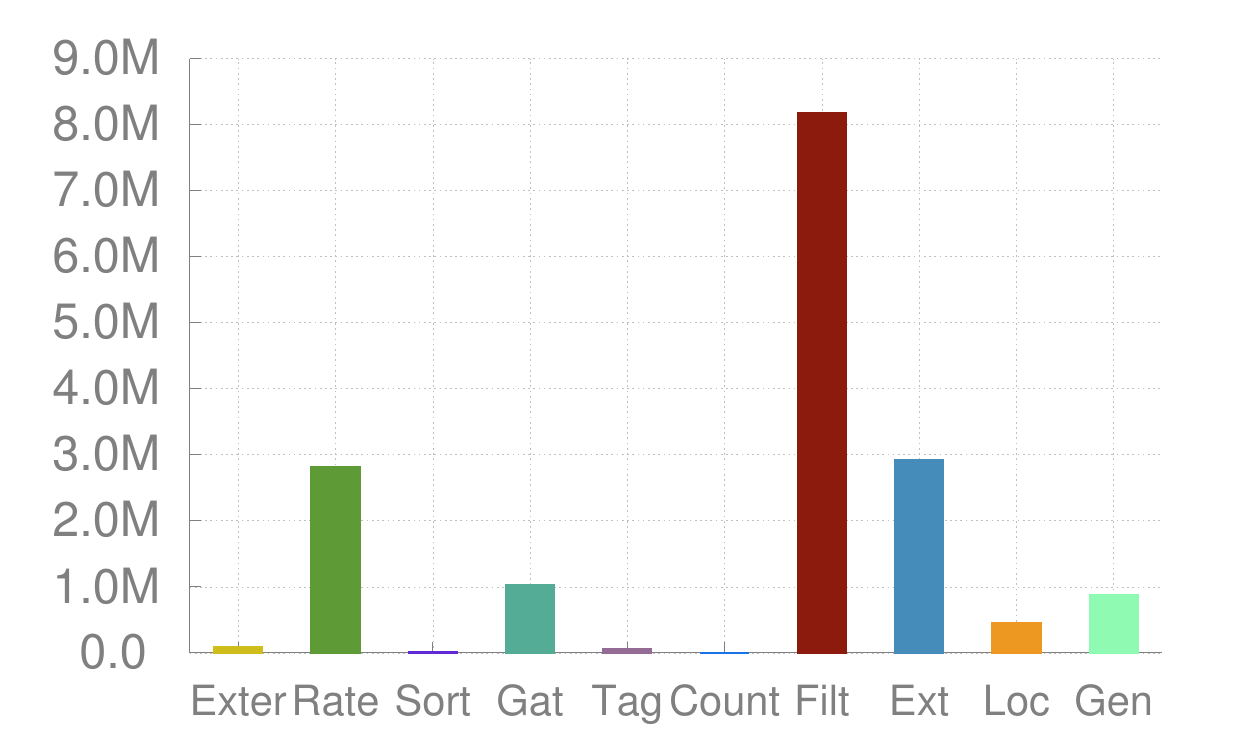}
\vspace{-15pt}
\caption{\label{fig:label_freq_task_type}Popular operators}
\end{subfigure}
\vspace{-5pt}
\caption{Distribution of goals, data types, and operators\label{fig:label_freq_first_row}}
\vspace{-5pt}
\end{figure*}

\smallskip
\noindent
{\bf Label distribution.}
First, we analyze the distribution of labels in different categories across tasks. 
Figure~\ref{fig:label_freq_task_goals} depicts the popular goals. We observe that complex
unstructured data understanding based goals---\textsf{language understanding} and \textsf{transcription} are very common, comprising of over 4 and 3 million tasks, that is around 17\% and 13\% respectively, 
despite not having seen extensive optimization research,
as opposed to traditional, simpler goals like 
\textsf{entity resolution} and \textsf{sentiment analysis} that have been extensively analyzed.
Figure~\ref{fig:label_freq_data_type} shows that \textsf{text} and \textsf{image} are still the main types of data available and analyzed --- 9.6 million (40\%) and 6.3 million (26\%) tasks contained \textsf{text} and \textsf{image} data respectively. \textsf{Audio} and \textsf{video} data are also used, and other richer types of data like \textsf{social media}, \textsf{web pages}, and \textsf{maps} are gaining popularity. 
Figure~\ref{fig:label_freq_task_type} shows the common operators used. 
While the distribution of goals indicates that a significant fraction of tasks have
complex goals,
the underlying operators are still predominantly simple. 
The marketplace is dominated by the fundamental \textsf{filter} and \textsf{rate} operations --- over 8 million (33\%) tasks employ some filtering operator, and nearly 3 million (13\% of) tasks make use of rating operators. 
Among more complex operators, we see that \textsf{gathering}, \textsf{extraction}, \textsf{localization}, and \textsf{generation} are frequently applied, together being used in around 5.3 million, i.e., 22\% of all tasks.

\begin{figure*}[!htbp]
\centering

\begin{subfigure}[b]{0.32\textwidth}
\includegraphics[scale=0.45]{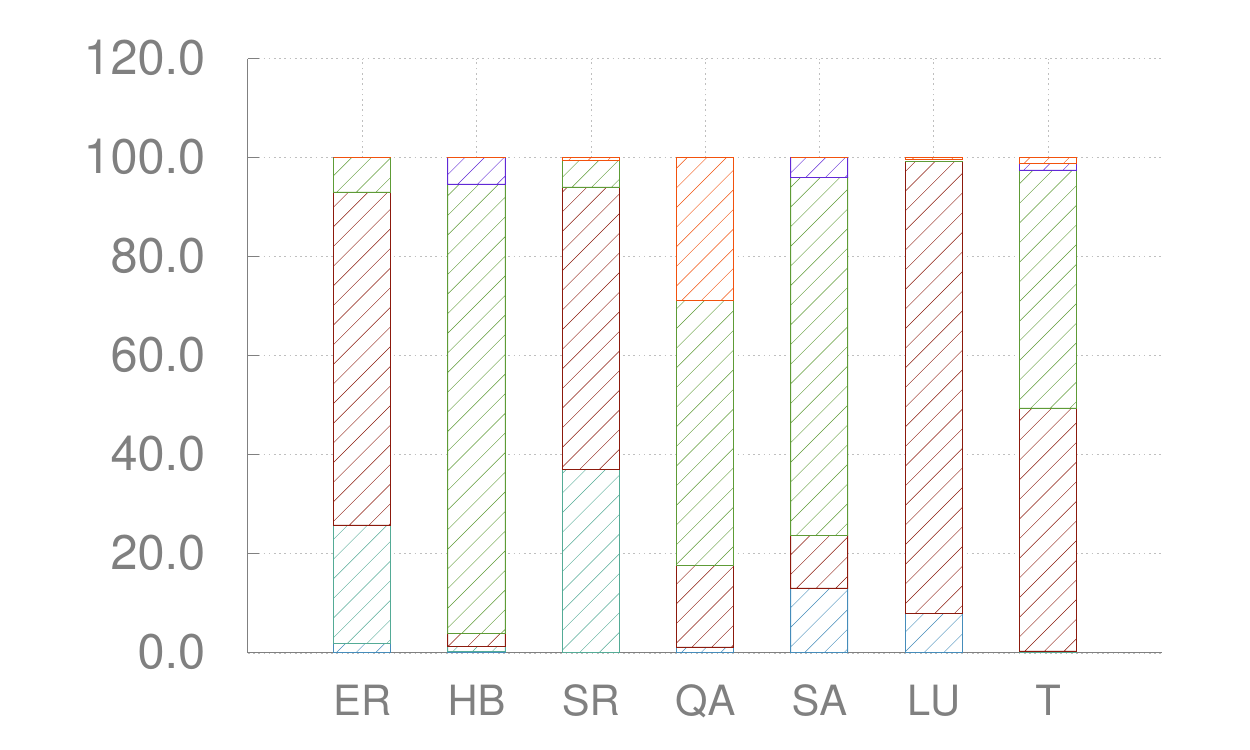}
\vspace{-10pt}
\caption{\label{fig:label_freq_task_goal_data_type}Data used for different task goals}
\end{subfigure}%
\begin{subfigure}[b]{0.32\textwidth}
\includegraphics[scale=0.45]{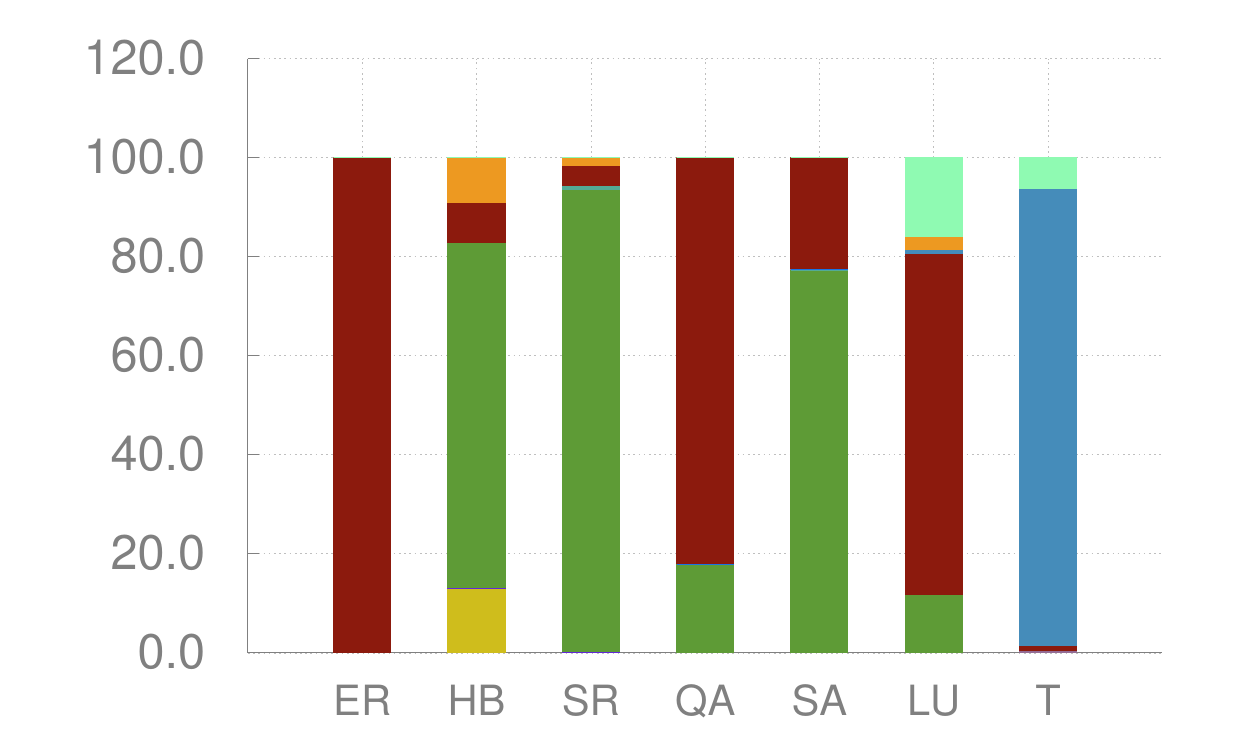}
\vspace{-10pt}
\caption{\label{fig:label_freq_task_goal_task_type}Popular Operators for different task goals}
\end{subfigure}
\begin{subfigure}[b]{0.32\textwidth}
\includegraphics[scale=0.45]{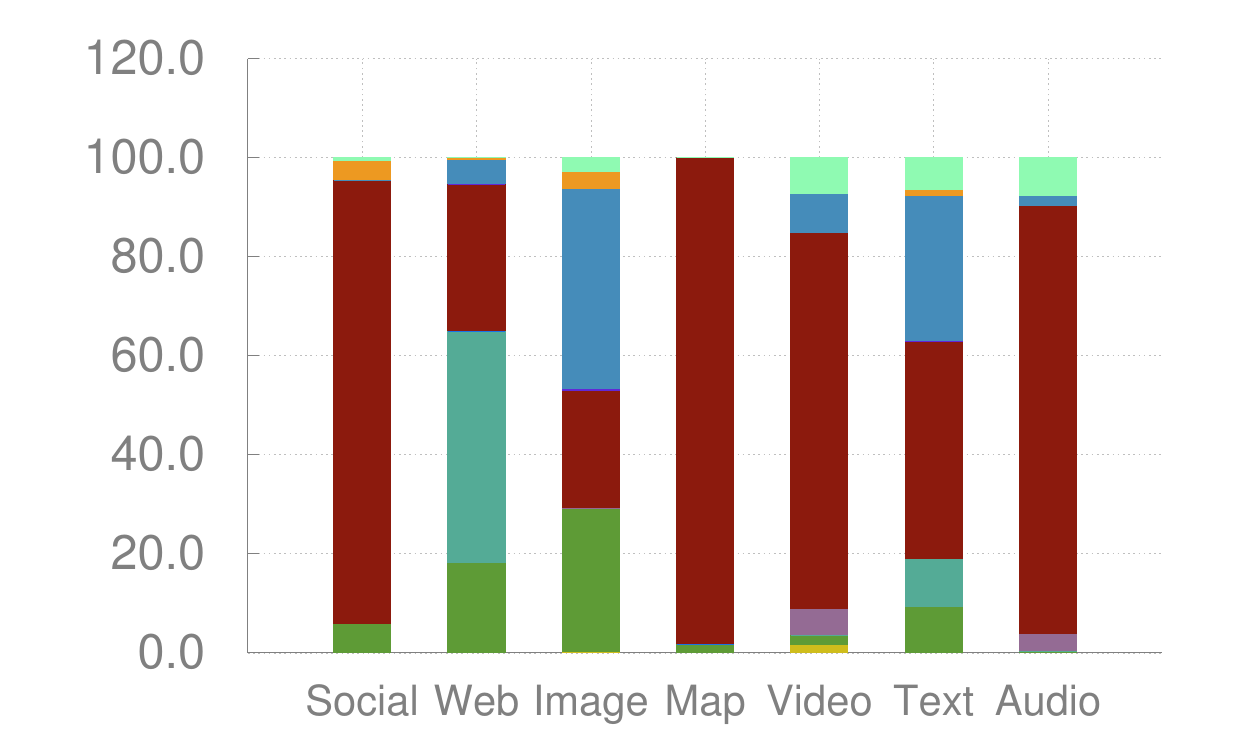}
\caption{\label{fig:label_freq_data_type_task_type}Popular Operators for different data}
\end{subfigure}
\vspace{-5pt}
\caption{Correlations across data and goal, operator and goal, and operator and data\label{fig:label_freq_second_row}}
\vspace{-5pt}
\end{figure*}

\smallskip
\noindent
{\bf Goals, operators and data types that occur frequently together.}
Next, we look at the correlations between the three types of labels for tasks. 
For example, one question we aim to answer 
is what kinds of operators are typically applied 
to different types of data, or used to achieve particular goals?
Looking at such correlations across goals, operators, and data types provides fine-grained 
insights into the structure and design of tasks that is not immediate from our aggregate statistics alone. 
\papertext{Here we present three charts in Figure~\ref{fig:label_freq_second_row} that depict 
the correlation between each pair; the remaining three charts, along with detailed insights,
can be found in the technical report~\cite{techreport}.}\techreport{
The charts depicting the correlation can be found
in Figure~\ref{fig:label_freq_first_row} and \ref{fig:label_freq_second_row}.
}
For instance, Figure~\ref{fig:label_freq_task_goal_task_type} shows the breakdown for each goal by the percentage usage of different operators towards achieving that goal; Figure~\ref{fig:label_freq_task_type} serves as a legend for the stacked bars. \techreport{(Figure~\ref{fig:label_freq_task_type_task_goal}, legend Figure~\ref{fig:label_freq_task_goals}, yields similar insights, but from a slightly different perspective.)}
We observe that \textsf{filter} and \textsf{rate} operators are used in most kinds of tasks, as well as form a significant majority as the constituent building block for most goals. One notable exception is \textsf{transcription} (which, recall, constitutes over 13\% of all tasks by itself, making it a significant exception), where the primary operation employed is \textsf{extraction}.
As another example, Figure~\ref{fig:label_freq_task_goal_data_type} shows that 
\textsf{text} and \textsf{images} are important for all types of task goals, for
certain types, e.g., \textsf{ER, SA, SR}, social media is also quite important.
Lastly, Figure~\ref{fig:label_freq_data_type_task_type} shows that beyond filtering
and rating being important, extraction is used quite prominently on text and image data,
often rivaling that of filtering. 
\techreport{
For \textsf{language understanding} tasks, while \textsf{filter} and \textsf{rate} are the primary operations, \textsf{generate} is also used frequently (16\% of the time). Also, for tasks looking to understand \textsf{human behavior}, 13\% of the tasks involve performing operations at \textsf{external links}  (such as online surveys), and 9\% of the tasks involve \textsf{localization}. As Figure~\ref{fig:label_freq_task_type_data_type} (legend Figure~\ref{fig:label_freq_data_type}) indicates, \textsf{filter} and \textsf{rate} operators have been used to analyze most types of data as well. 

Figure~\ref{fig:label_freq_task_goal_data_type} shows the breakdown for each goal by the percentage of different data types present in tasks having that goal. Figure~\ref{fig:label_freq_data_type} serves as a legend for the stacked bars. (Figure~\ref{fig:label_freq_data_type_task_goal}, legend Figure~\ref{fig:label_freq_task_goals}, yields similar insights, but from a slightly different perspective.) While for most goals, a large fraction of data used in tasks seems to be \textsf{text} and \textsf{image} based, we observe that for \textsf{entity resolution} and \textsf{search relevance}, \textsf{web} data is relevant (serving 24\% and 37\% of \textsf{entity resolution} and \textsf{search relevance} tasks respectively). Also, \textsf{sentiment analysis} and \textsf{language understanding} style of analyses employ \textsf{social} media as a significant fraction of their input data (13\% and 8\% respectively). 
While some efforts are being made towards analyzing other types of data (besides \textsf{text} and \textsf{image}), they are still largely unexplored.
}

\techreport{
\ta{
We observe that the marketplace exhibits a diverse range of tasks spanning across over 7 broad goals, at least 10 distinct operations and 7 data types.
\begin{denselist}
\item \textsf{Text} and \textsf{image} data are by far the most prevalent and utilized across most tasks. \textsf{Web} and \textsf{social media} data are also available and relevant to a small subset of tasks (in particular tasks involving data integration and cleaning for \textsf{web}, and natural language processing for \textsf{social media} data). While \textsf{text} and \textsf{image} data (and to a lesser extent, \textsf{web} data) have been heavily studied using several different operators, there are still many exciting avenues waiting to be explored for the other types of data.
\item \textsf{Filter} and \textsf{rate} are used as basic operators for achieving most goals and analyzing all types of data. It is crucial to understand and optimize the usage of these operators.
\item \textsf{Language understanding}, and \textsf{transcription} seem to be very popular task goals constituting of a large number tasks. Considering the fact that these tasks require complex human operations (\textsf{generation} and \textsf{extraction} as opposed to the simple \textsf{filter} and \textsf{rate} operations), it might be worthwhile to train and maintain a specialized worker pool for such tasks.
\item For the popular goals of \textsf{Language understanding}, and \textsf{transcription}, we expect the heavy percentage of \textsf{text}-based data. It is interesting to observe the high percentage of of \textsf{social media} and \textsf{image} data for these tasks as well.
\end{denselist}
}
}

\techreport{
\begin{figure*}[!htbp]
\centering
\vspace{-10pt}
\begin{subfigure}[b]{0.32\textwidth}
\includegraphics[scale=0.45]{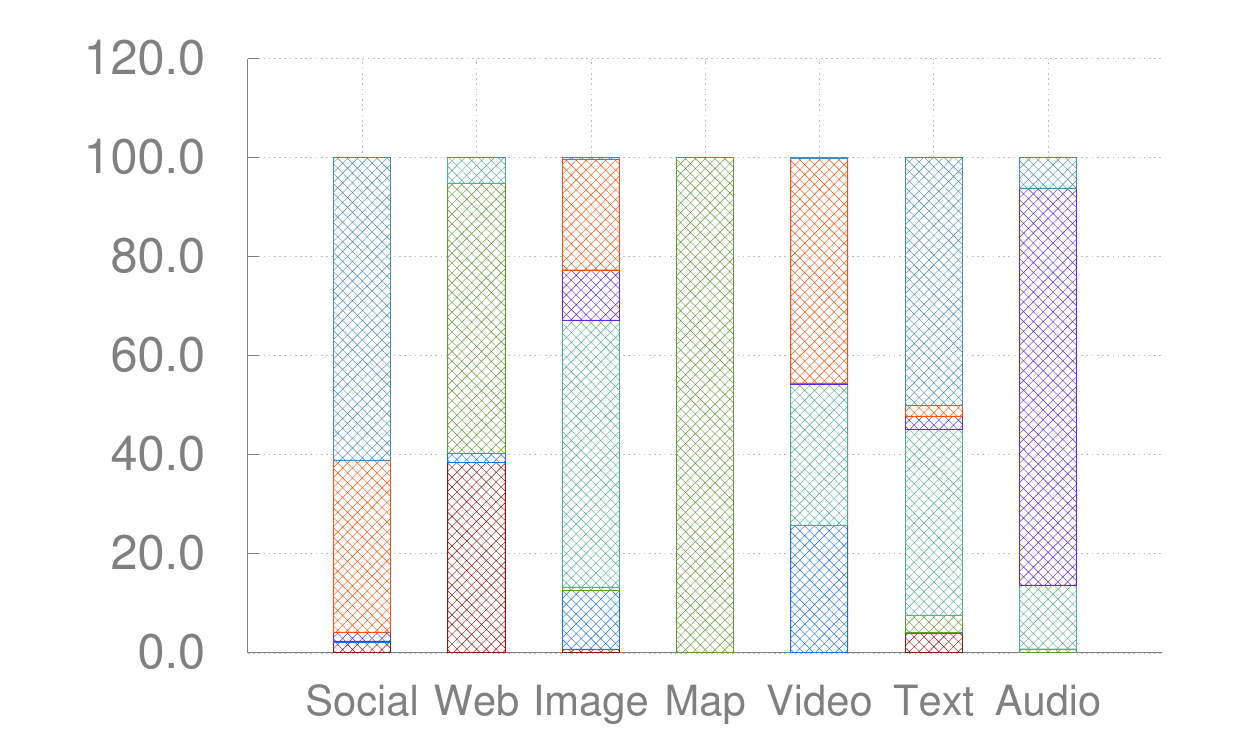}
\vspace{-10pt}
\caption{\label{fig:label_freq_data_type_task_goal}Popular task goals for different data}
\end{subfigure}
\begin{subfigure}[b]{0.32\textwidth}
\includegraphics[scale=0.45]{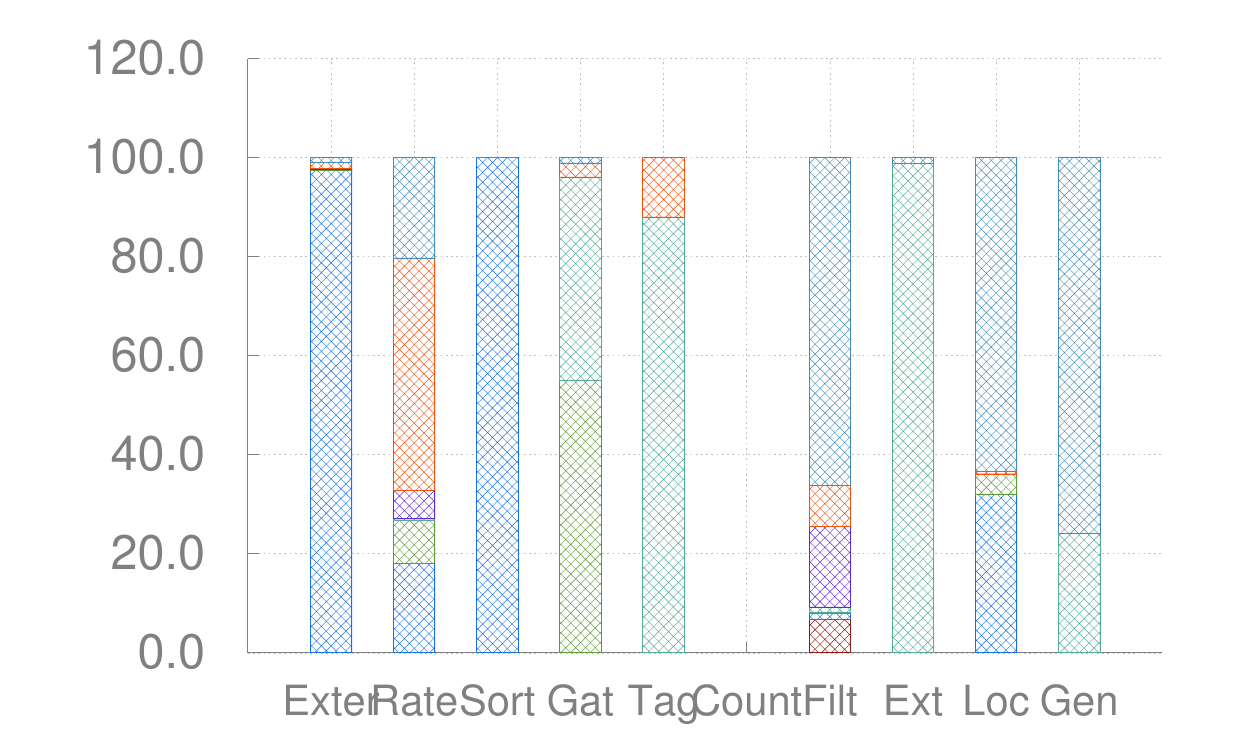}
\vspace{-10pt}
\caption{\label{fig:label_freq_task_type_task_goal}Popular task goals for different operators}
\end{subfigure}%
\begin{subfigure}[b]{0.32\textwidth}
\includegraphics[scale=0.45]{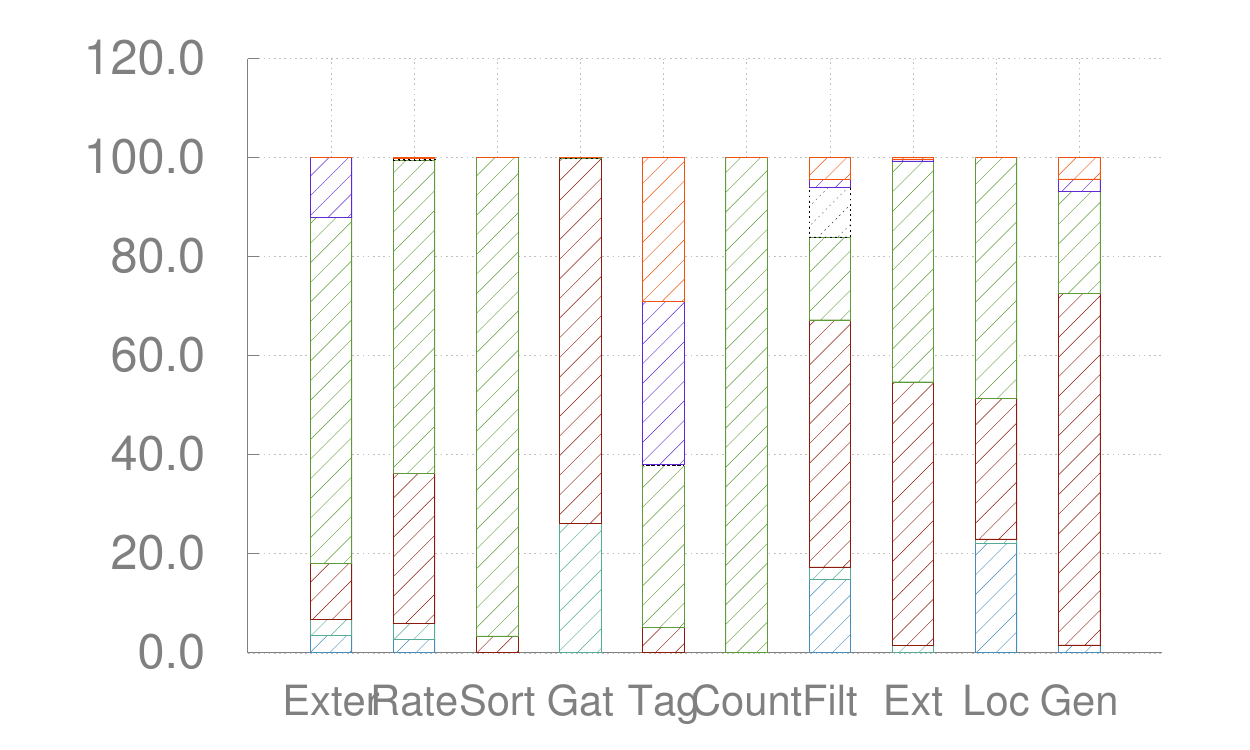}
\vspace{-10pt}
\caption{\label{fig:label_freq_task_type_data_type}Popular data for different operators}
\end{subfigure}
\vspace{-5pt}
\caption{Correlations across goal and data, goal and operator, and data and operator\label{fig:label_freq_third_row}}
\vspace{-5pt}
\end{figure*}
}

\techreport{
\subsection{Trend towards open-ended tasks.}\label{sec:open-ended}
In this section, our aim is to explore the trend in the complexity of crowdsourced tasks over time. That is, we intend to answer the following questions: 
\begin{denselist}
\item Are requesters moving on to more complex, or open-ended goals?
\item Are they looking at more challenging datasets?
\item Are they using more sophisticated tools or utilizing human intelligence more effectively than in the past?
\end{denselist}
We split each of our categories, \textsf{goals}, \textsf{operators}, and \textsf{data} into two classes: {\em simple} and {\em complex}. Among the set of observed goals, we classify \{\textsf{entity resolution}, \textsf{sentiment analysis}, \textsf{quality assurance}\} as simple, and the remaining 7 as complex. For operators, we classify \textsf{filter} and \textsf{rate} as simple and the remaining 8 as complex. For data types, we only consider \textsf{text} as simple, and the remaining 6 types as complex. While this classification is subjective, our high-level observations apply to most reasonable mappings of labels to \{simple, complex\}.

In Figure~\ref{fig:simple_vs_complex}, we compare the trend between number of simple tasks and the number of complex tasks on the marketplace over time. On the x-axis, we plot {\em time in increments of one week}. On the y-axis we plot the {\em cumulative number of clusters of tasks}, that is the number of unique tasks, issued so far -- one line each for simple, versus complex tasks. Note that we deduplicate similar batches and count them as a single point, so these plots represent the interests of all requesters equally.
\begin{figure*}
\centering
    \begin{subfigure}[b]{0.33\textwidth}
    \centering
\includegraphics[width=\textwidth]{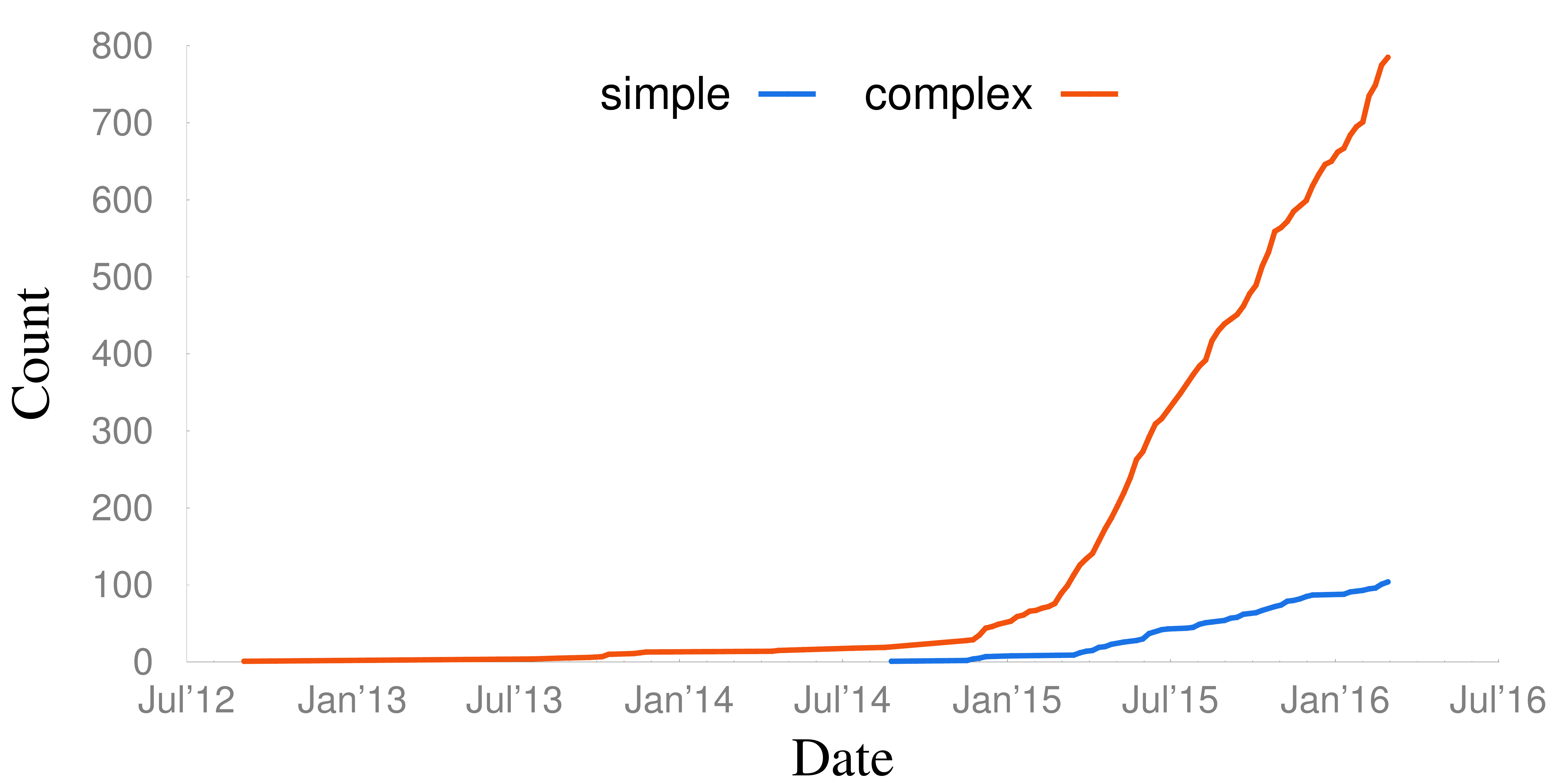}%
    \caption{Goals}
    \label{fig:goals_simple_vs_complex}
    \end{subfigure}%
\hfill
    \begin{subfigure}[b]{0.33\textwidth}
    \centering
\includegraphics[width=\textwidth]{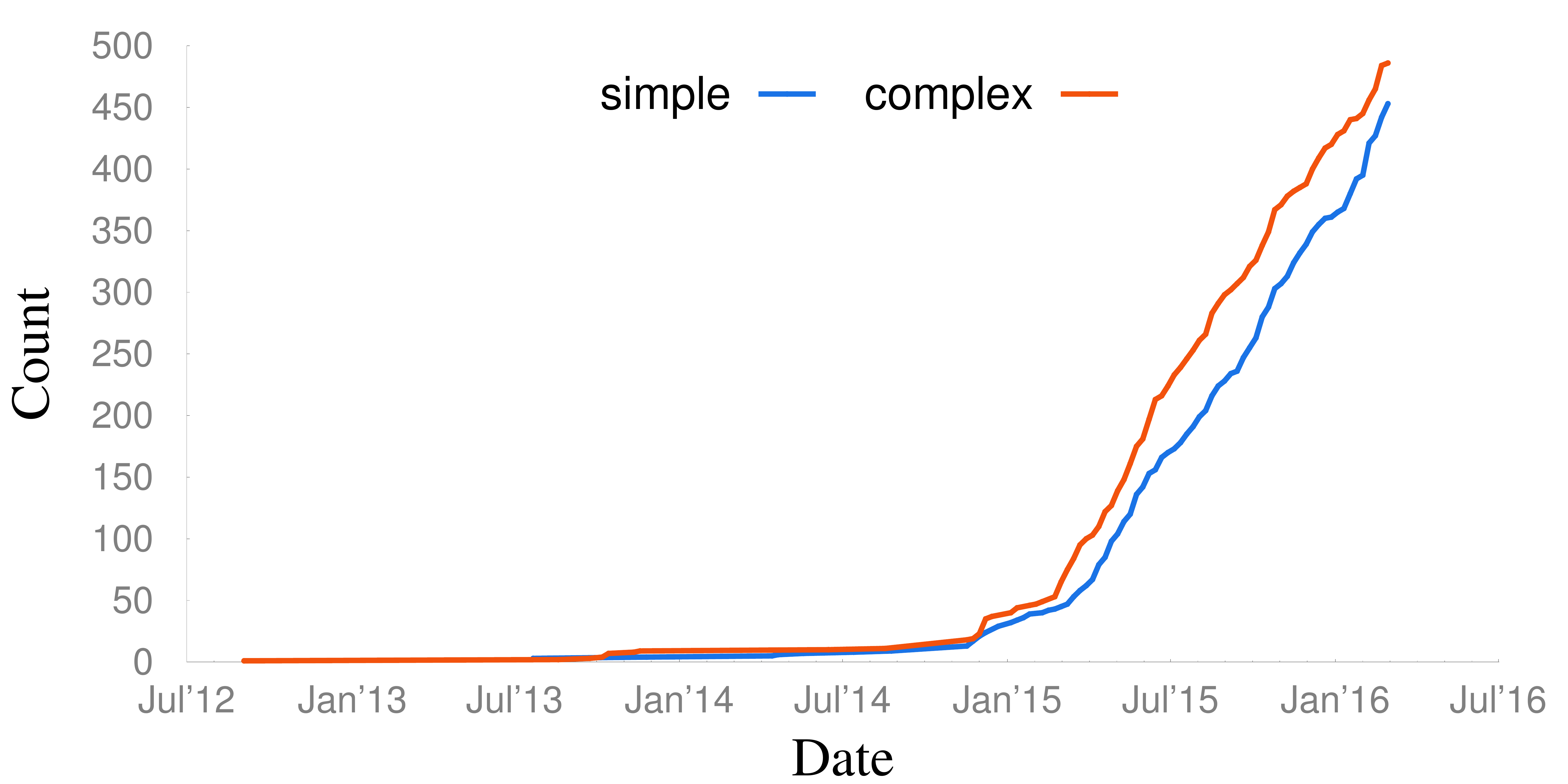}%
    \caption{Operator}
    \label{fig:operator_simple_vs_complex}
    \end{subfigure}%
\hfill
    \begin{subfigure}[b]{0.33\textwidth}
    \centering
\includegraphics[width=\textwidth]{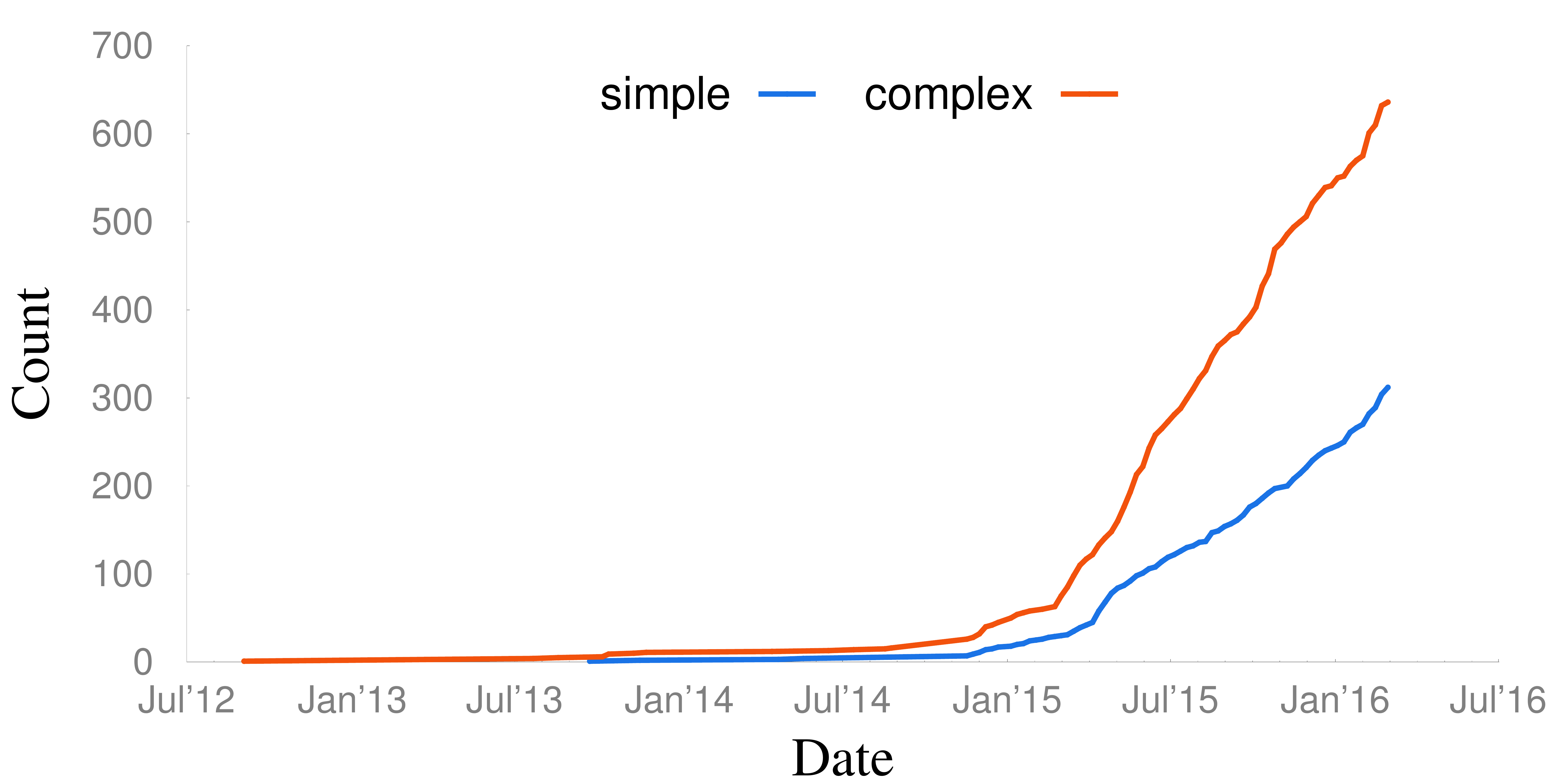}%
    \caption{Data}
    \label{fig:data_simple_vs_complex}
    \end{subfigure}%
\caption{Simple vs complex tasks over time}
\label{fig:simple_vs_complex}
\end{figure*}
From Figures~\ref{fig:goals_simple_vs_complex} and ~\ref{fig:data_simple_vs_complex}, we observe that the number of clusters of tasks involving complex goals and non-textual data types is much larger, and growing faster, than the corresponding numbers of simple clusters. For instance, as of January 2016, there had been around(a) 510 clusters involving non-textual versus about 240 clusters involving text data, and (b) 620 clusters with complex goals, and just 80 clusters with simple goals. By contrast, Figure~\ref{fig:operator_simple_vs_complex} demonstrates that the usage of complex operators is comparable to that of simple operators. Specifically, we observe a total of around 410 clusters using complex operators and 340 clusters involving simple operators issued cumulatively as of January 2016.

\ta{While requesters (and researchers) are more interested in achieving complex goals on complex data (and getting more so with time), the fundamental human-operators of \textsf{filter} and \textsf{rate}, are by themselves still as widely used as all other operators combined. This raises both the need to optimize the existing simple operators as far as possible, as well as the opportunity for the exploration and understanding of more complex operators.}
}
\section{Effective Task Design}\label{sec:task_analyses}
In this section, we address the question of effective task design. Specifically, we (1) characterize and quantify what constitutes an ``effective'' task, (2) make data-driven recommendations on how requesters can design effective tasks, and (3) predict the ``effectiveness'' of tasks based on our hypotheses.
\subsection{Metrics for effective tasks}\label{sec:metrics}
The standard three metrics that are used to measure crowdsourcing 
effectiveness are: error, cost, and latency. 
There are various ways these three metrics could be measured;
we describe our notions below, given what we can calculate.

\smallskip
\noindent {\bf Error: Disagreement Score.}
In our dataset, we have every worker answer provided for 
each question within each task instance, operating on one distinct item,
but not the corresponding ground truth answers. 
We use these answers to quantify how ``confusing'' or 
ambiguous a task is, overall.
The way we quantify this is to consider the worker answers for a given question on a given item.
If the workers disagree on a specific question on a specific item, then the task is likely to be ambiguous---indicating that it is poorly designed, or hard to answer---either way, this information is important
to dictate the task design (e.g., clarify instructions) and the level of redundancy (e.g., more redundancy for confusing questions) that should be adopted by requesters.
Our proxy for error is the {\em average disagreement in the answers for questions on the same item},
across all questions and items in a batch.
We consider all pairs of workers who have operated on the same item, and check if their
answers are the same or different, giving a score of one if they disagree, and zero if they agree;
we then compute the average disagreement score of an item by averaging all these scores;
and lastly, we compute the average disagreement score for a batch by averaging the scores
across items and questions. 
We shall henceforth refer to the ``Disagreement Score'' as \textsf{disagreement}.


There is however, one small wrinkle. 
Some operators, and corresponding worker responses 
may involve textual input.
Two textual responses may be unequal even if they are only
slightly different from each other.
Since textual responses occupy a large fraction of our dataset,
it is not possible to ignore them altogether.
\papertext{(We consider this and other strategies in our technical report~\cite{techreport}.)}
We instead adopt a simple rule: we prune away all
tasks with 
\textsf{disagreement} > 0.5 so as to eliminate tasks with very high variations in worker responses.
This eliminates the subjective textual tasks, while still retaining the textual tasks that
are objective. 

\cut{
It should be noted that there is one caveat here. Some tasks, and correspondingly worker responses may be textual. Two textual responses may be unequal even if they are only slightly different from each other.
Ignoring all tasks with textual fields, or ignoring the textual fields of all tasks are not reasonable strategies, as they account for a very significant fraction of the dataset. Using an edit-distance or partial scoring scheme is also not ideal as many common tasks require precise textual responses, such as retrieving names of businesses and people, or transcribing media.
In our technical report~\cite{techreport} we discuss these previous strategies in greater detail.
For the rest of this paper, we adopt an intermediate approach: we prune away all tasks with \textsf{disagreement} > 0.5 so as to eliminate tasks with very high variations in worker responses.
}

\techreport{
Another way to handle the subjectivity of tasks is to simply ignore text-boxes. This could be done in two ways: (1) only evaluate disagreement for tasks with no text-box fields, and (2) for every task, compute disagreement only on its non textual fields. In our experiments, we tried both these options, but rejected them for reasons we discuss below.

It turns out that a large majority of tasks in our dataset contain at least one text-box field. Eliminating all of them leaves very few tasks, spread out across a large number of features (such as those discussed in the sections to follow) and labels (goals, operators and data types)---the remaining data is too sparse for any statistically significant inferences to be made. 

For the second option of computing disagreement on non-textual fields, we face a problem with the distribution of disagreement values itself. First, for all the tasks that only have textual responses, it is not possible to define a disagreement score; we are unable to compute a disagreement score in this manner for over 60\% of all batches. Second, ignoring text fields misrepresents the true distribution of disagreements for the remaining datasets. It is possible that we represent tasks with high disagreement as having low disagreement simply because they have a small number of non-textual fields. 

A third approach would be an edit-distance or partial scoring scheme; however, this approach is not ideal since in practice crowdsourcing requesters require high exact agreement, not partial agreement, so that the answers can be easily aggregated via conventional majority vote type schemes. Furthermore, many tasks with textual responses are objective. Some common examples that we see include transcription, captcha, image labeling, and retrieving URLs. For such textual but objective tasks, it is not clear if an edit-distance based agreement scheme is the right approach.
}

\smallskip
\noindent
{\bf Cost: Median Task Time.}
A typical measure for how much effort a worker has put into a task instance
is the amount of time taken to complete it. 
Since we do not have information about the actual payments made to workers, 
we use the median amount of time taken ({\em in seconds}) 
by workers to complete tasks in a batch as a proxy for the cost of the batch. 
This can be calculated from the data that is available, given that we have the start 
and end times for each task in a batch.
We shall subsequently denote the ``Median Task Time'' by \textsf{task-time}.

\smallskip
\noindent
{\bf Latency: Median Pickup Time.}
To characterize latency, we use pickup time, i.e., how quickly tasks are picked up by workers, on average. 
Pickup time for a batch is computed as follows:
$\textsf{pickup-time}$ $=\text{median}(<\text{start time of task}_i - \text{batch start time}>)$ ({\em in seconds}). Here, we use the start time of the earliest batch, i.e. $\text{start time of task}_1$, 
as a proxy for the batch start time.
We justify this choice for the latency metric quantitatively in 
\papertext{our technical report~\cite{techreport}}\techreport{below.
}.
\techreport{
Our reasons are twofold. First, we observe that the \textsf{pickup-time} of tasks is typically orders of magnitude higher than that of \textsf{task-time}, which might otherwise seem like a reasonable proxy for latency. This means that the actual turnaround time for a task is dominated by when workers start its instances, rather than how long they take to complete them once started. Figures~\ref{fig:batch-latency} and~\ref{fig:task-latency} support this claim. For each of the figures, we compare the \textsf{pickup-time} against the \textsf{task-time}, both on the y-axis with varying \textsf{end-to-end-time} along the x-axis. Figure~\ref{fig:batch-latency} shows this distribution at a batch level, with the median values for \textsf{pickup-time} and \textsf{task-time} being plotted against each batch's \textsf{end-to-end-time}. Figure~\ref{fig:task-latency} shows this distribution at a task instance level, with each task's individual \textsf{pickup-time} and \textsf{task-time} being plotted against its \textsf{end-to-end-time}, which in this case is simply $(\textsf{pickup-time} + \textsf{task-time})$ (to reduce the number of points in the plot, we only plot the median of \textsf{pickup-time} and \textsf{task-time} corresponding to a vertical splice, that is, we plot one median point for all instances having a common \textsf{end-to-end-time}. We observe that in both plots, the \textsf{pickup-time} is orders of magnitude higher than the \textsf{task-time}. 
\begin{figure}[h]
\centering
\begin{subfigure}[b]{0.48\linewidth}
        \centering
\includegraphics[width=\textwidth]{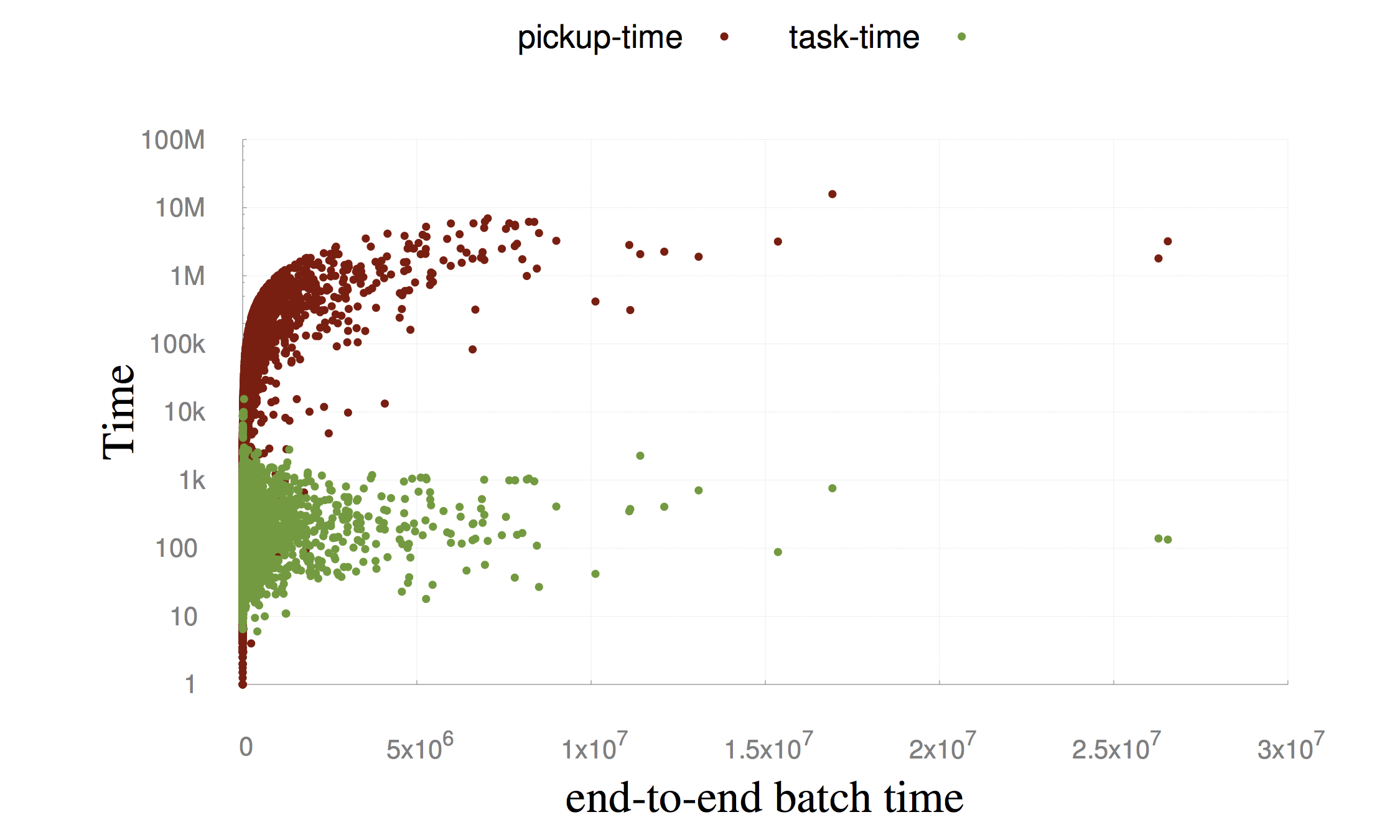}
\caption{Batch-level}\label{fig:batch-latency}
\end{subfigure}%
\hfill
\begin{subfigure}[b]{0.48\linewidth}
        \centering
\includegraphics[width=\textwidth]{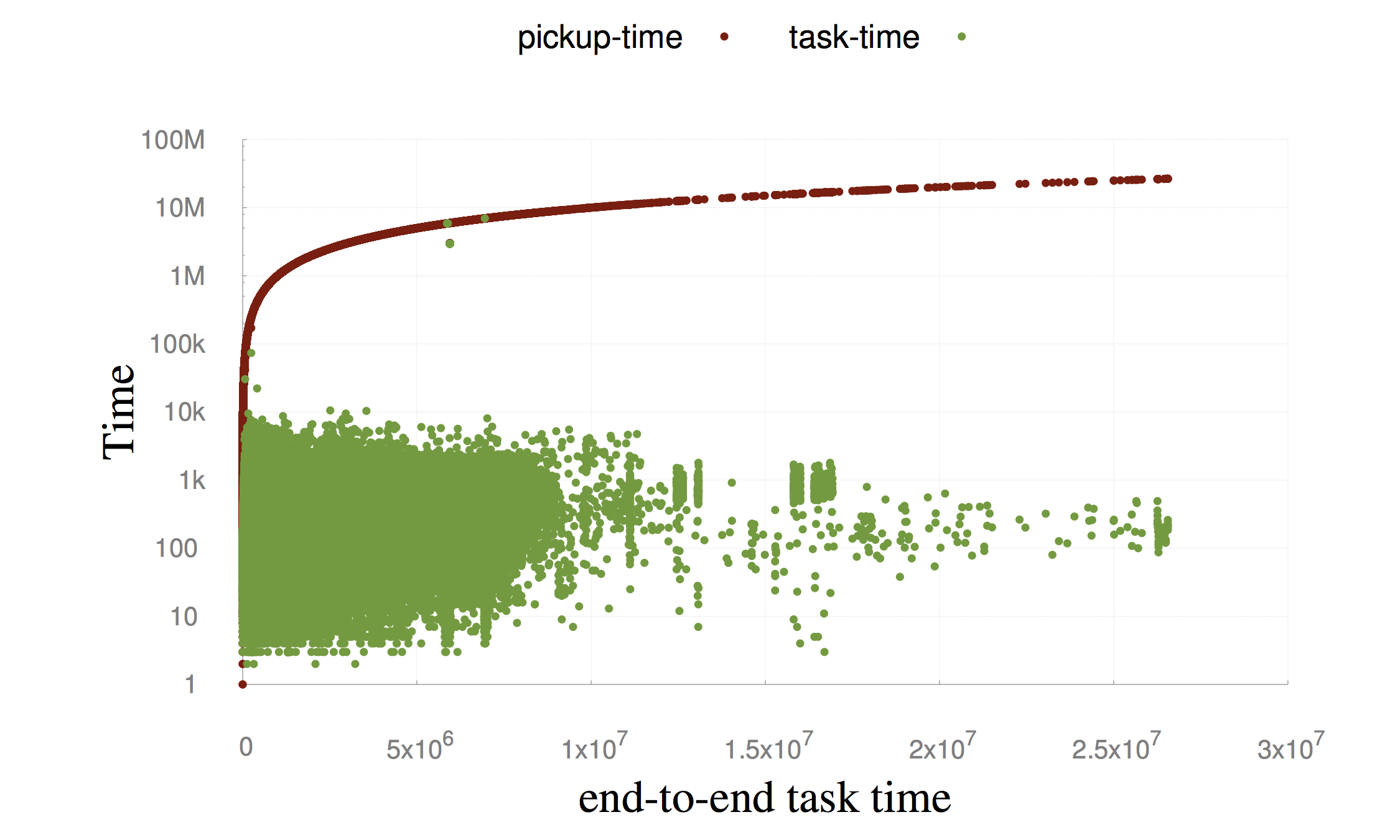}
\caption{Task-level}\label{fig:task-latency}
\end{subfigure}%
\caption{Latency}\label{fig:latency}
\end{figure}
Secondly, most measures of time that we can obtain from our available data strongly depend on features, such as the size and difficulty of a task. Since \textsf{pickup-time} only looks at the time taken for workers to start a task and not how long they spend on it (which, as we have seen, is anyway an insignificant fraction of time), it is relatively independent of such features. This helps separate out the influence of features that requesters often cannot control, and that we cannot quantify, from our latency metric, making our subsequent quantitative analyses more statistically meaningful.
} 
In short, we observe that in general the \textsf{pickup-time} for batches is orders of magnitude higher than the \textsf{task-time}, indicating that 
the latency or total turnaround time of a task is in fact dictated
by the rate at which workers accept and start the task instances.
We denote the ``Median Pickup Time'' by \textsf{pickup-time}.

\subsection{Correlation Analysis Methodology}
In the next set of subsections, 
we examine some influential features or parameters that a requester can tune, 
to help improve a task's error (\textsf{disagreement}), cost (\textsf{task-time}) and latency (\textsf{pickup\--time}).
For instance, features of a task include the length of the task, 
or the number of examples within it.
For each feature, we look at the correlation between the feature and each of the three metrics. 
We perform a series of (correlation-investigating) experiments, each of which corresponds to one \{feature, metric\} pair. 
All our experiments follow the following structure:
\begin{denselist}
\item {\bf Cluster:} 
We first cluster batches based on the task in order to not have
the ``heavy-hitter'' tasks that appear frequently in multiple batches 
across the dataset to dominate and bias our findings.
Since our analysis will also involve {\em matching}, or clustering tasks further based on labels, we restrict our focus to the set of around 3,200 labeled clusters corresponding to 83\% of all batches and 89\% of all task instances.
Subsequently, for each cluster, we take the median of metric values across batches,
as well as the median of the feature being investigated.
\item {\bf Binning:} We separate the clusters into two bins based on their feature value --- all clusters with feature value lower than the global median feature value go into Bin-1 (say), 
while the ones with feature value higher than the median go into Bin-2. 
(Clusters with feature value exactly equal to the median are all put into either Bin-1 or Bin-2 while keeping the bins as balanced as possible.)  
For each metric, we then examine its value distribution in the two bins --- in particular, we look for significant differences between the average, median, or distribution of metric values in the two bins. 
A significant difference indicates a correlation between the feature we have binned on, and the metric being looked at. 
\techreport{We then hypothesize about the underlying reason(s) behind the correlation.}
\item {\bf Statistical significance:} We perform a {\em t-test} to check 
whether the metric value distribution in our two feature-value-separated 
bins is statistically significant. 
We use a threshold {\em p-value} of 0.01 to determine significance, 
that is, we only reject the null hypothesis (that bins have similar metric values) 
if the p-value is less than 1\%.
\item {\bf Visualization:} For each feature-metric pair, we plot a cumulative distribution (CDF) plot, with the metric value plotted along the $x$-axis. Each of the two bins corresponds to one line in the plot. 
For $x=m$, the corresponding $y$ value on each of the lines represents the probability that a batch will have metric value better than $m$. 
Thus, a higher value is preferable; and we compare the two bins (or lines) in this plot.
\end{denselist}
Below, in Sections~\ref{sec:feature-num-words}-~\ref{sec:feature-image}, we look at the results for some of the significant correlations we found. 
\papertext{In our technical report~\cite{techreport}, we further support our claims from these sections by providing examples of real tasks issued on the marketplace that are dissimilar in individual features, but similar in other respects, and comparing their qualities.}

\begin{figure*}
\centering
 \begin{subfigure}[b]{0.25\textwidth}
\centering
\includegraphics[width=\textwidth]{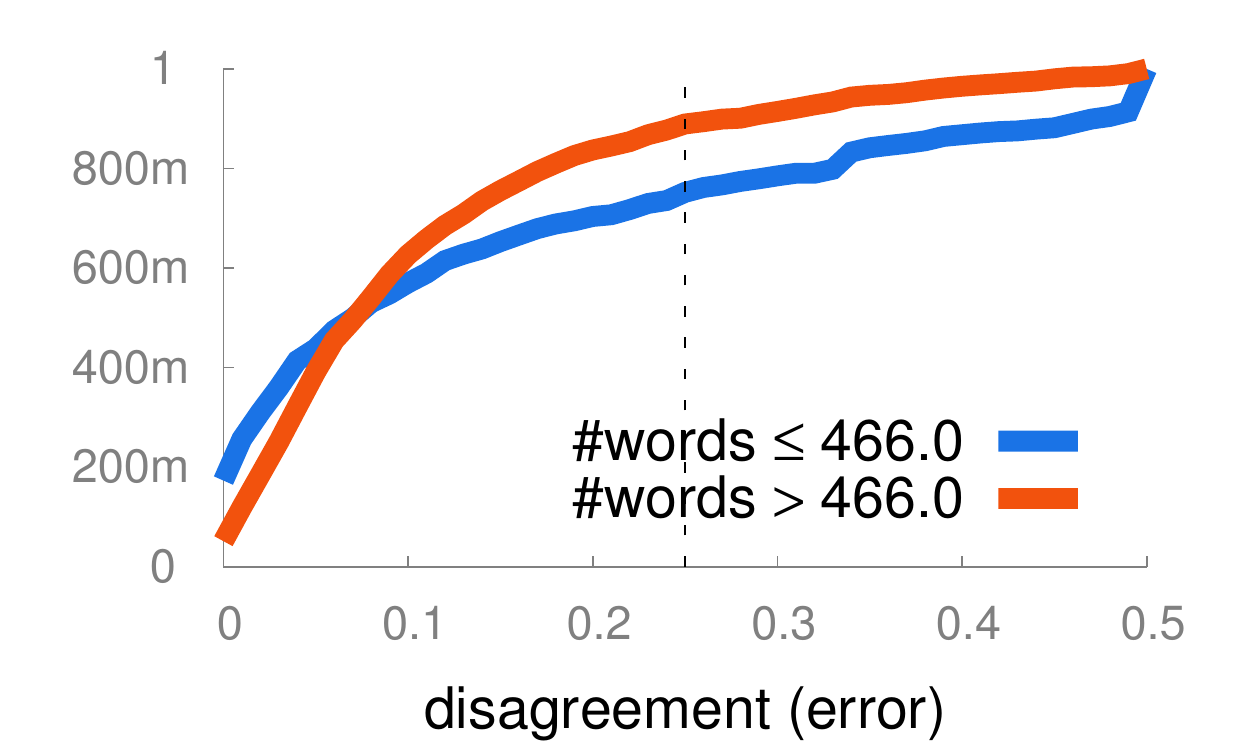}%

\caption{CDF: \#\textsf{words} vs \textsf{disagreement}}
\label{fig:words_vs_disagreement}
    \end{subfigure}   
 \begin{subfigure}[b]{0.5\textwidth}
\centering
\includegraphics[width=0.5\textwidth]{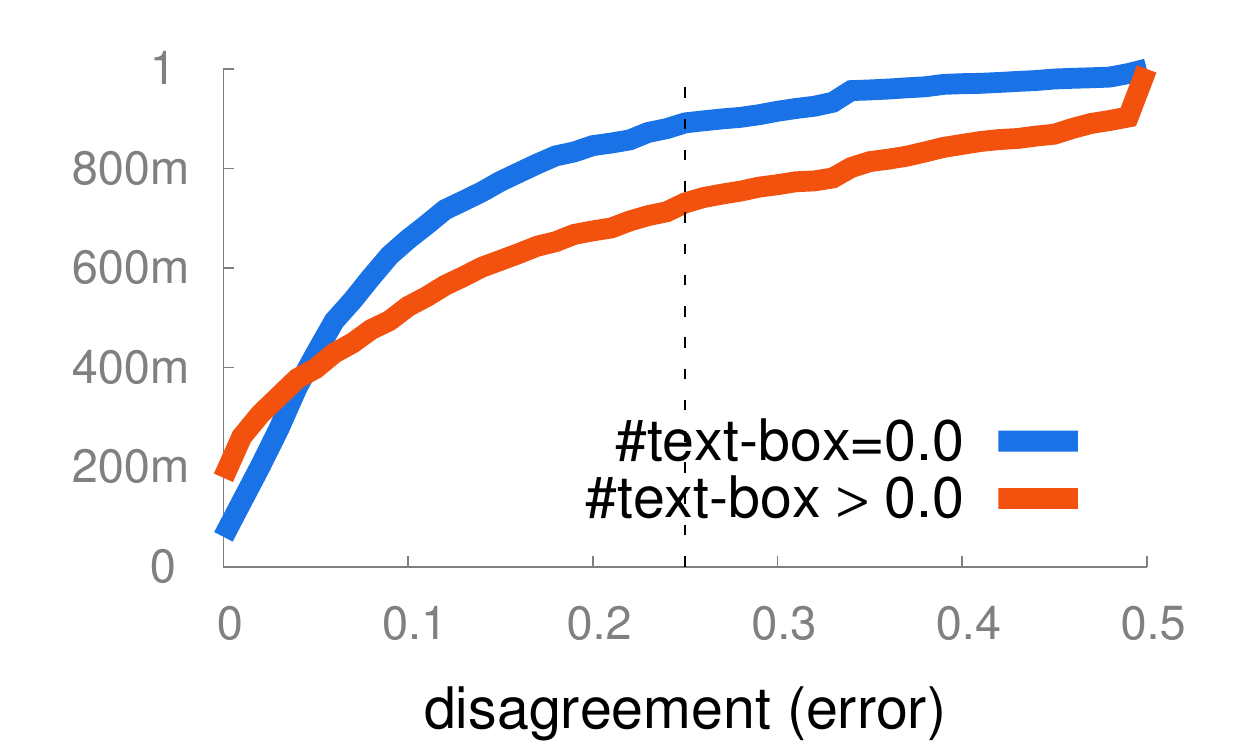}%
\includegraphics[width=0.5\textwidth]{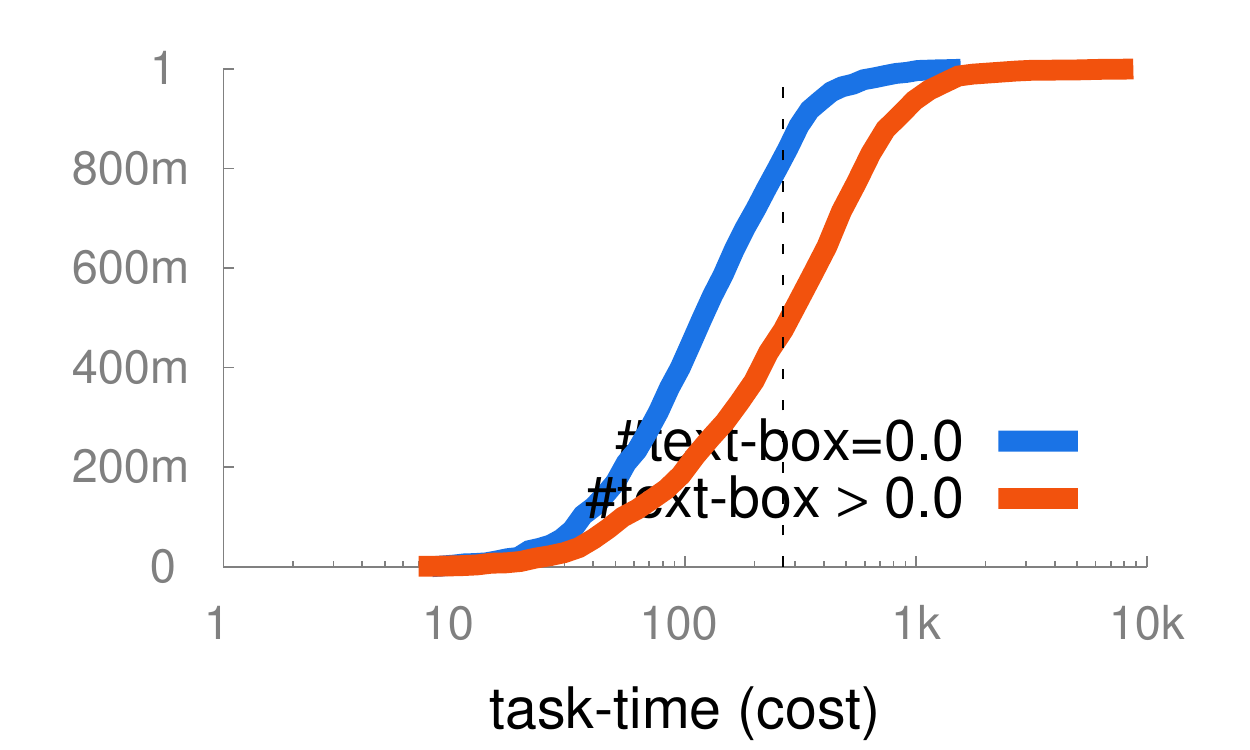}%
\caption{CDF: \#\textsf{text-box} vs (1) \textsf{disagreement}, (2) \textsf{task-time}}
\label{fig:boxes_vs_disagreement_task}
    \end{subfigure}
    \vspace{5pt}
    \begin{subfigure}[b]{0.75\textwidth}
\centering
\includegraphics[width=0.33\textwidth]{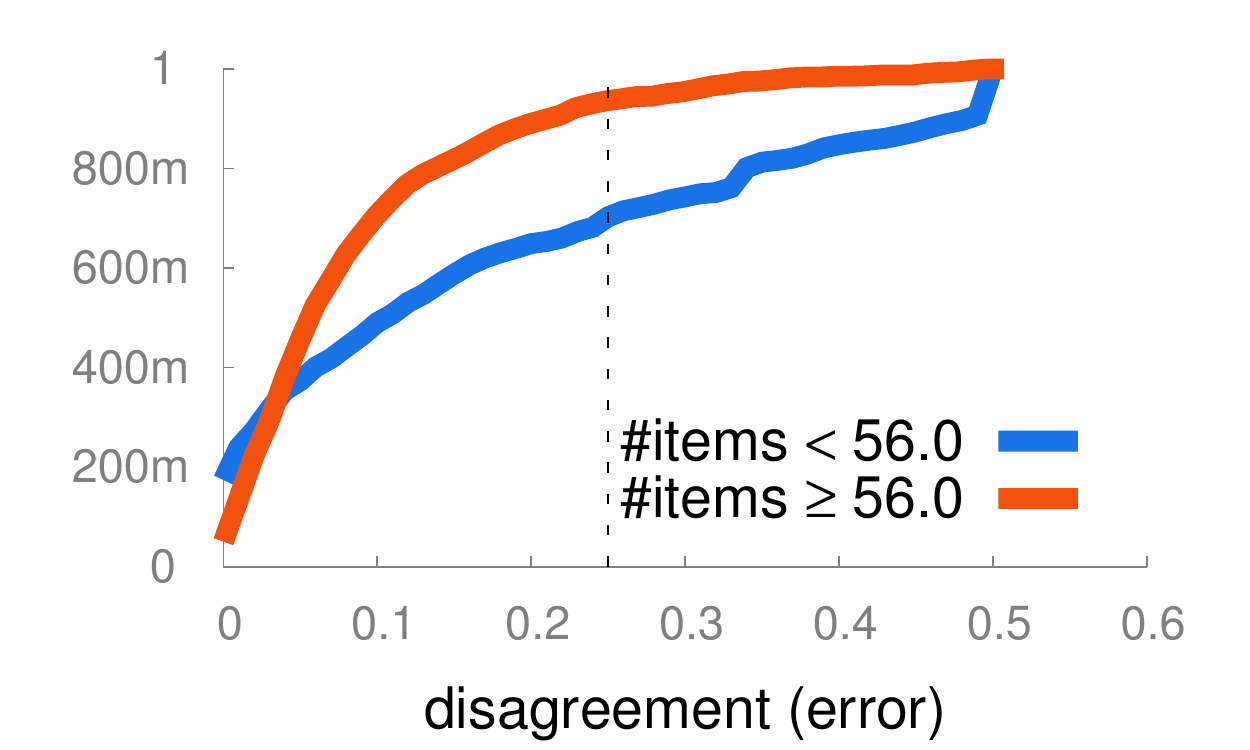}%
\includegraphics[width=0.33\textwidth]{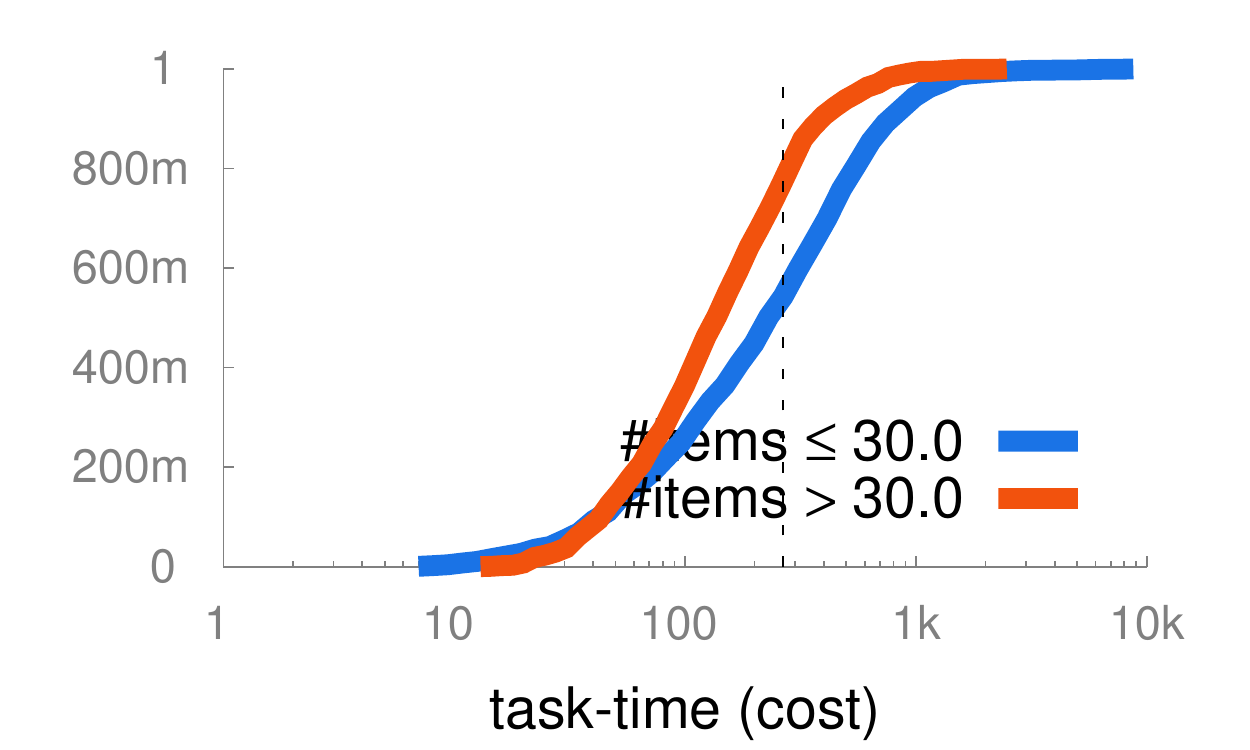}%
\includegraphics[width=0.33\textwidth]{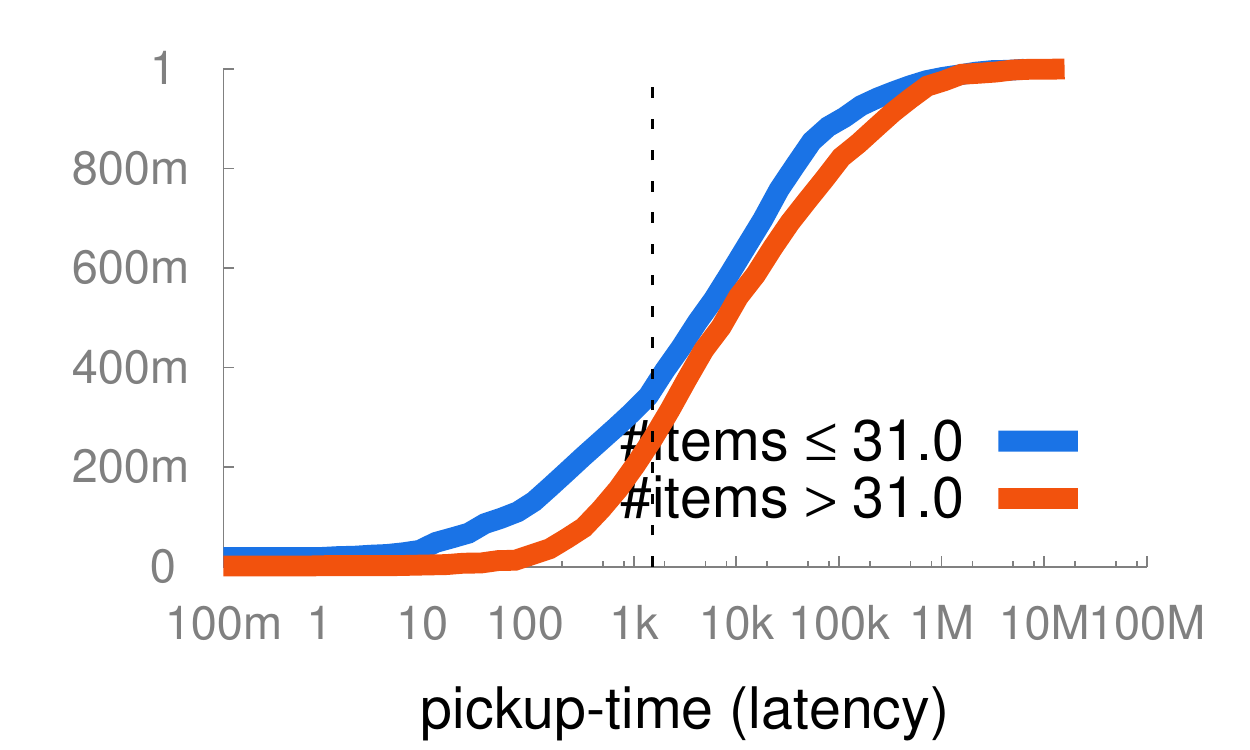}%
          \caption{CDF: \#\textsf{items} vs (1) \textsf{disagreement}, (2) \textsf{task-time}, (3) \textsf{pickup-time}}
\label{fig:items_vs_disagreement_pickup_task}
    \end{subfigure}%

    \vspace{1pt}

    \begin{subfigure}[b]{0.5\textwidth}
\centering
\includegraphics[width=0.5\textwidth]{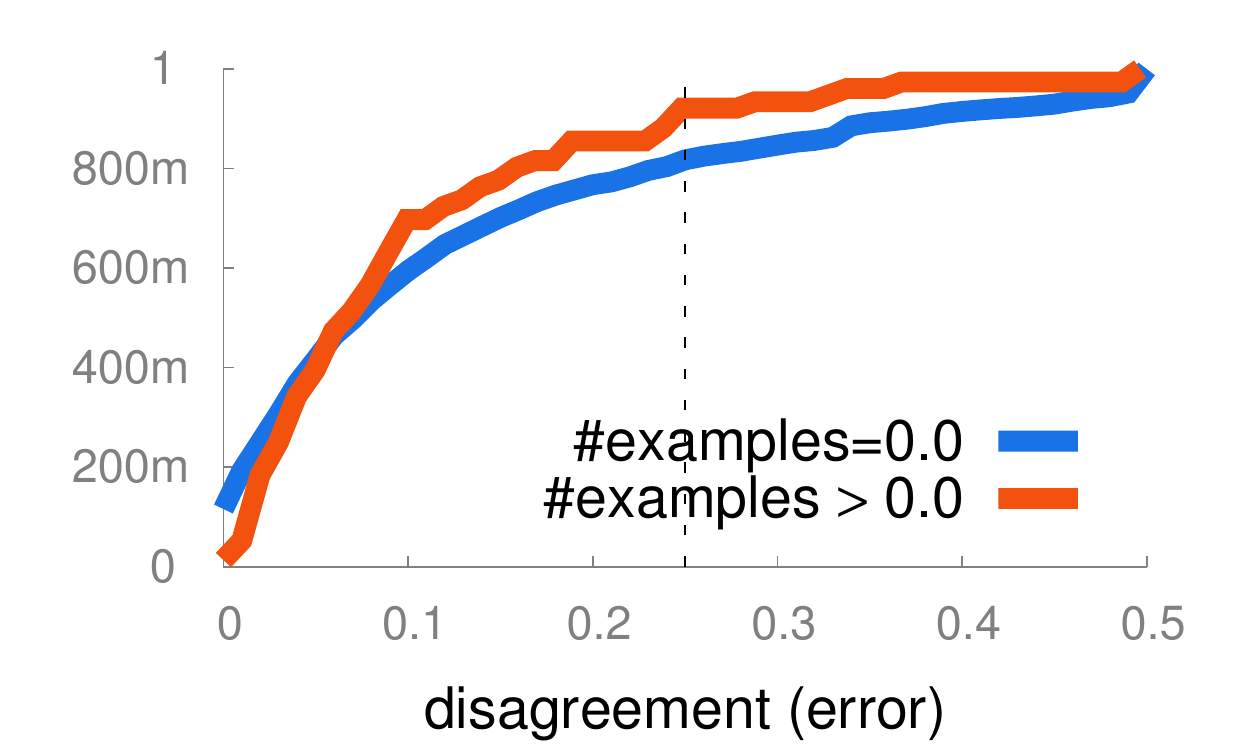}%
\includegraphics[width=0.5\textwidth]{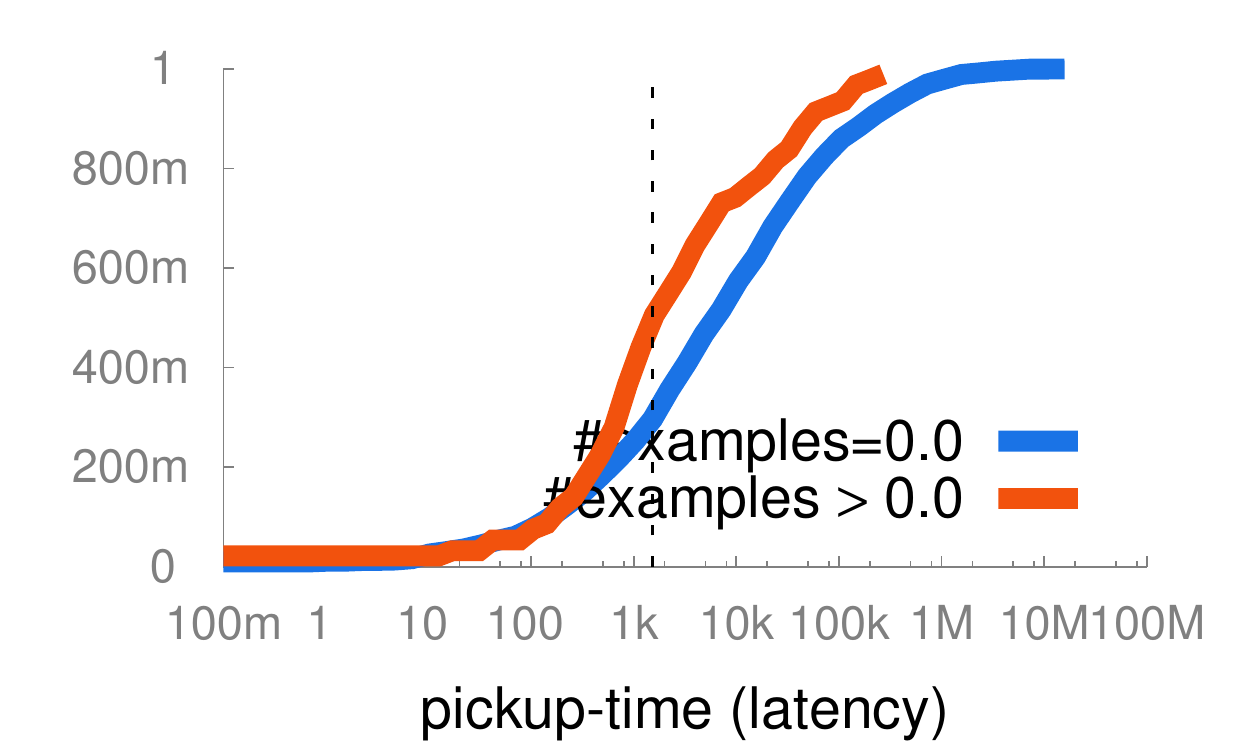}%
\caption{CDF: \#\textsf{examples} vs (1) \textsf{disagreement}, (2) \textsf{pickup-time}}
\label{fig:examples_vs_disagreement_pickup}
    \end{subfigure}%
\hfill
    \begin{subfigure}[b]{0.5\textwidth}
\centering
\includegraphics[width=0.5\textwidth]{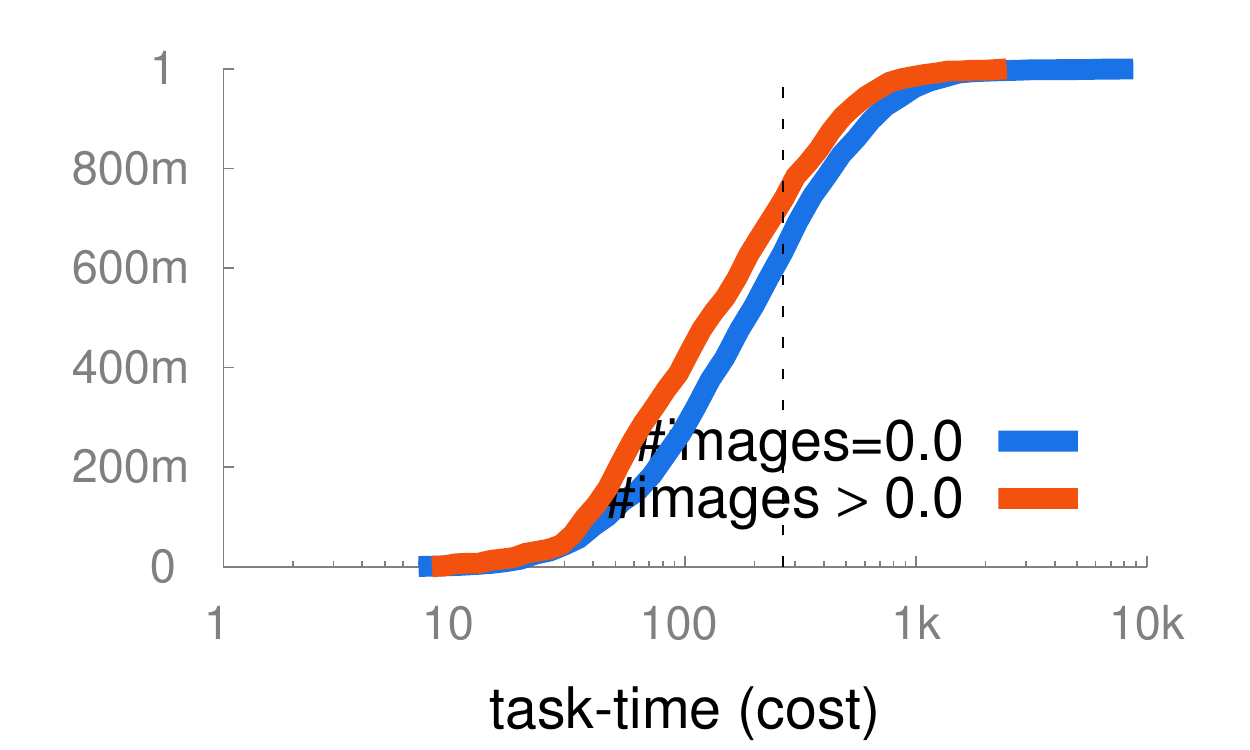}%
\includegraphics[width=0.5\textwidth]{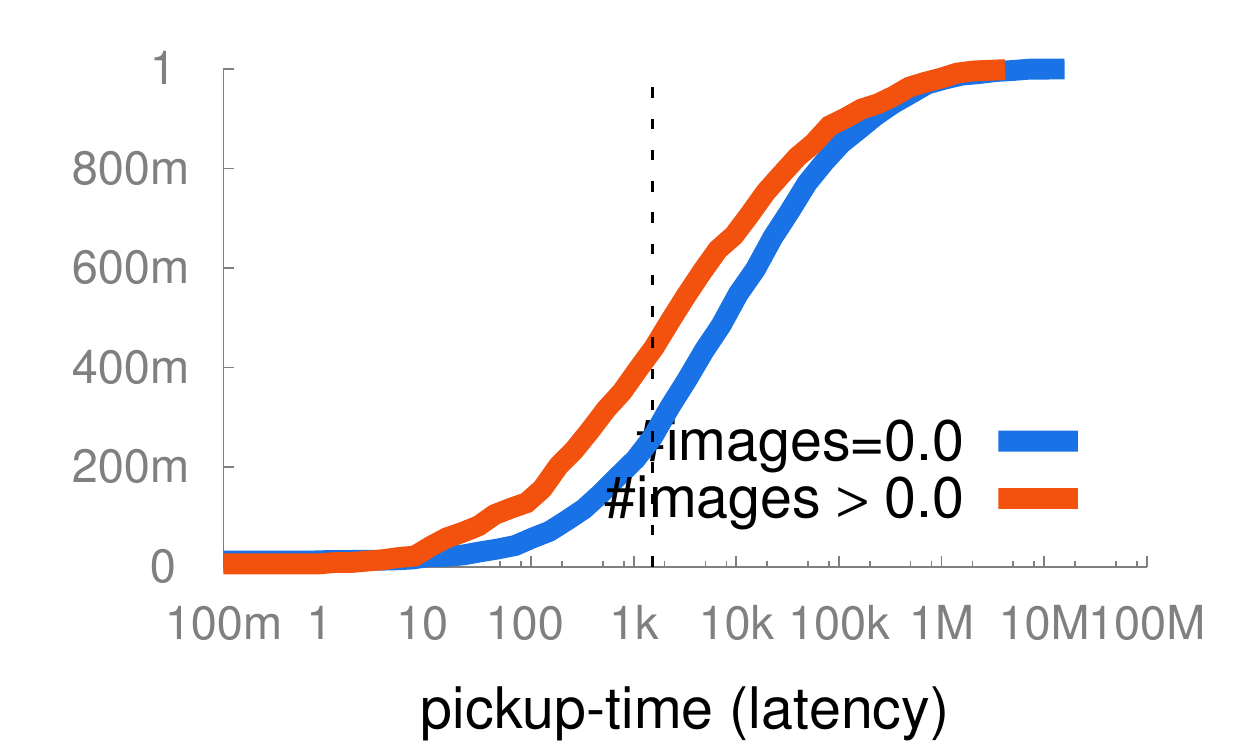}%
\caption{CDF: \#\textsf{images} vs (a) \textsf{task-time}, (b) \textsf{pickup-time}}
\label{fig:images_vs_task_pickup}
    \end{subfigure}
\vspace{-10pt}
\caption{Task Design Parameters and Metrics}
\vspace{-5pt}
\end{figure*}

\subsection{Number of HTML words}\label{sec:feature-num-words}
We examine how the length of task---defined as
the number of words in the HTML page, and denoted as \#\textsf{words}---impacts the effectiveness of the task. 
We show the effect of length of task on our metrics in Figure~\ref{fig:words_vs_disagreement}.
We observe that the line for clusters with higher \#\textsf{words} in their HTML interface dominates, or is above the line for the clusters with fewer \#\textsf{words}. 
\techreport{
We see that the median value of \textsf{disagreement} for tasks with $\textsf{\#words}\leq 466$ is 0.147, while that for tasks with $\textsf{\#words}>466$ is 0.108.}
This may be because longer tasks tend to be more descriptive, and the detailed instructions help reduce ambiguity in tasks, train workers better, 
and thereby reduce mutual disagreement in answers.
We also note that the length of the task does not significantly affect either the \textsf{pickup-time} or \textsf{task-time} metrics. Thus, workers are neither discouraged nor slowed down by longer textual descriptions of tasks.

While increasing the number of words in the HTML source
of tasks helps reduce disagreement in general, this benefit 
may be more pronounced for particular types of tasks. 
Intuitively, we expect detailed instructions to help more for harder tasks, 
and have less impact on easier tasks. To test this hypothesis, 
we separate tasks into buckets by their labels 
(recall \textsf{goal}, \textsf{operator} and \textsf{data}),
 and test the effect of our feature, \textsf{\#words}. 
From Figure~\ref{fig:num_words_gather}, we see that for (relatively hard) \textsf{gather} tasks, \textsf{\#words} has a pronounced effect 
on \textsf{disagreement} with higher \textsf{\#words} 
leading to significantly lower \textsf{disagreement}. 
On the other hand, Figure~\ref{fig:num_words_rating} seems 
to indicate that for (relatively simple) \textsf{rating} tasks, 
\textsf{\#words} has no significant impact on \textsf{disagreement}.

\techreport{
\sampar{Example.}
To demonstrate the effect of having more detailed description, or higher number of words in a task's HTML interface on \textsf{disagreement}, we compare two actual tasks which are both from the domain of \textsf{Language Understanding}, but differ in their descriptiveness and \textsf{\#num-words}. We look at two different tasks that require workers to find urls or email IDs of businesses through basic web searches. Both have extremely similar interfaces, and ask similar questions. Neither employs examples (which we shall see later has a significant impact on disagreement). The main difference between the two tasks is that the first (having 970 instances), has median number of words = 233, while the second (having 1254 instances) has a median of 6072 words in its HTML interface. Figure~\ref{fig:num_words-f844264} depicts the first task and Figure~\ref{fig:num-words-f750831} depicts the second. The first task uses these extra words to give detailed instructions (shown in Figure~\ref{fig:num-words-f750831-1}) on how to go about the task. In contrast, the second task has almost no description at all. It requires workers to enter the ``synonymy'' of correct sentences, and to correct incorrect sentences, without giving any examples or input for what these tasks entail.
While the first task has a median \textsf{disagreement} of 0.26, the second shows a median disagreement of 0.08. This demonstrates the power of examples in reducing task ambiguity.

\begin{figure}[h]
\centering
\includegraphics[width=0.5\textwidth]{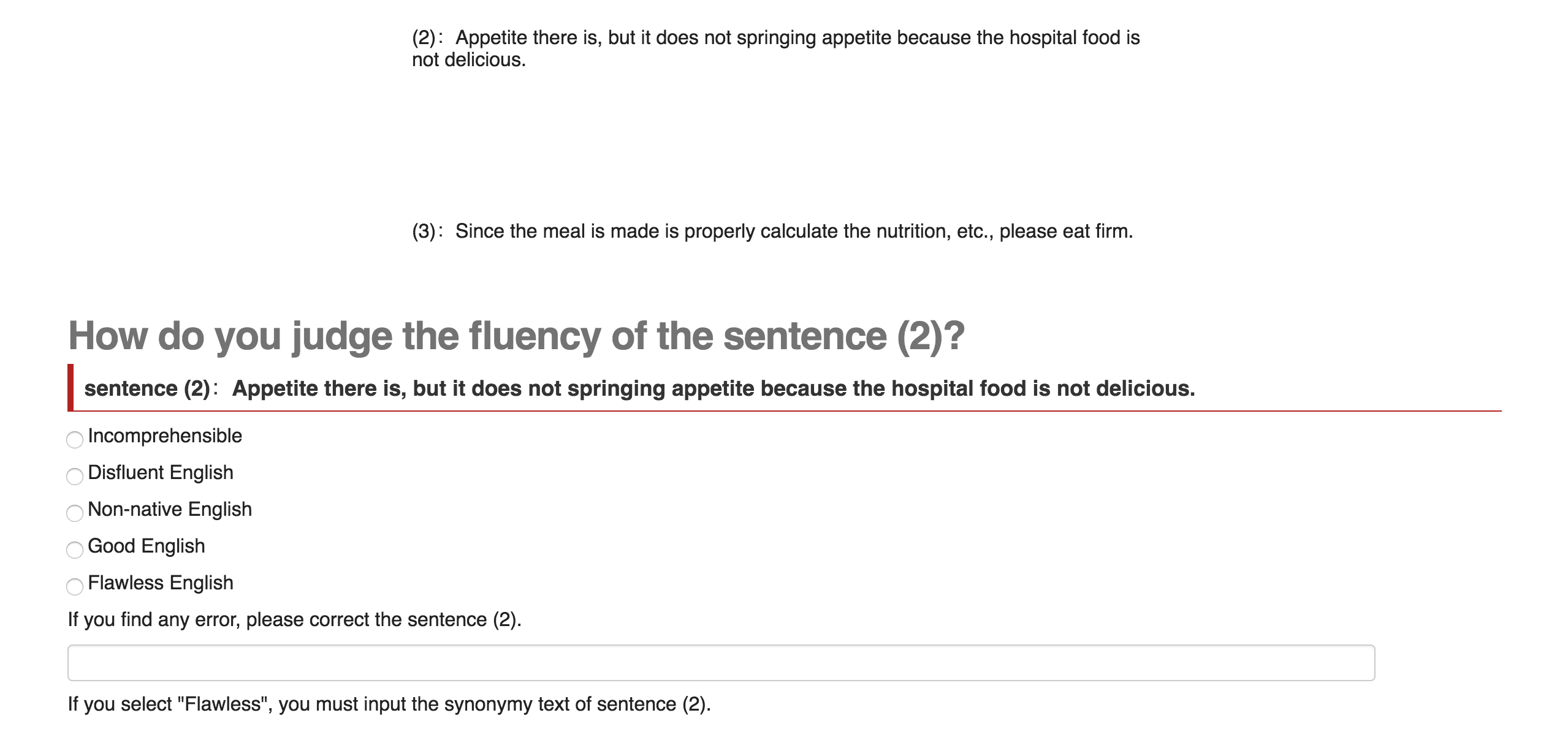}
\caption{\label{fig:num_words-f844264}Task with low \textsf{\#words}}
\end{figure}%
\begin{figure}[h]
\centering
\begin{subfigure}[b]{0.5\textwidth}
\includegraphics[width=\textwidth]{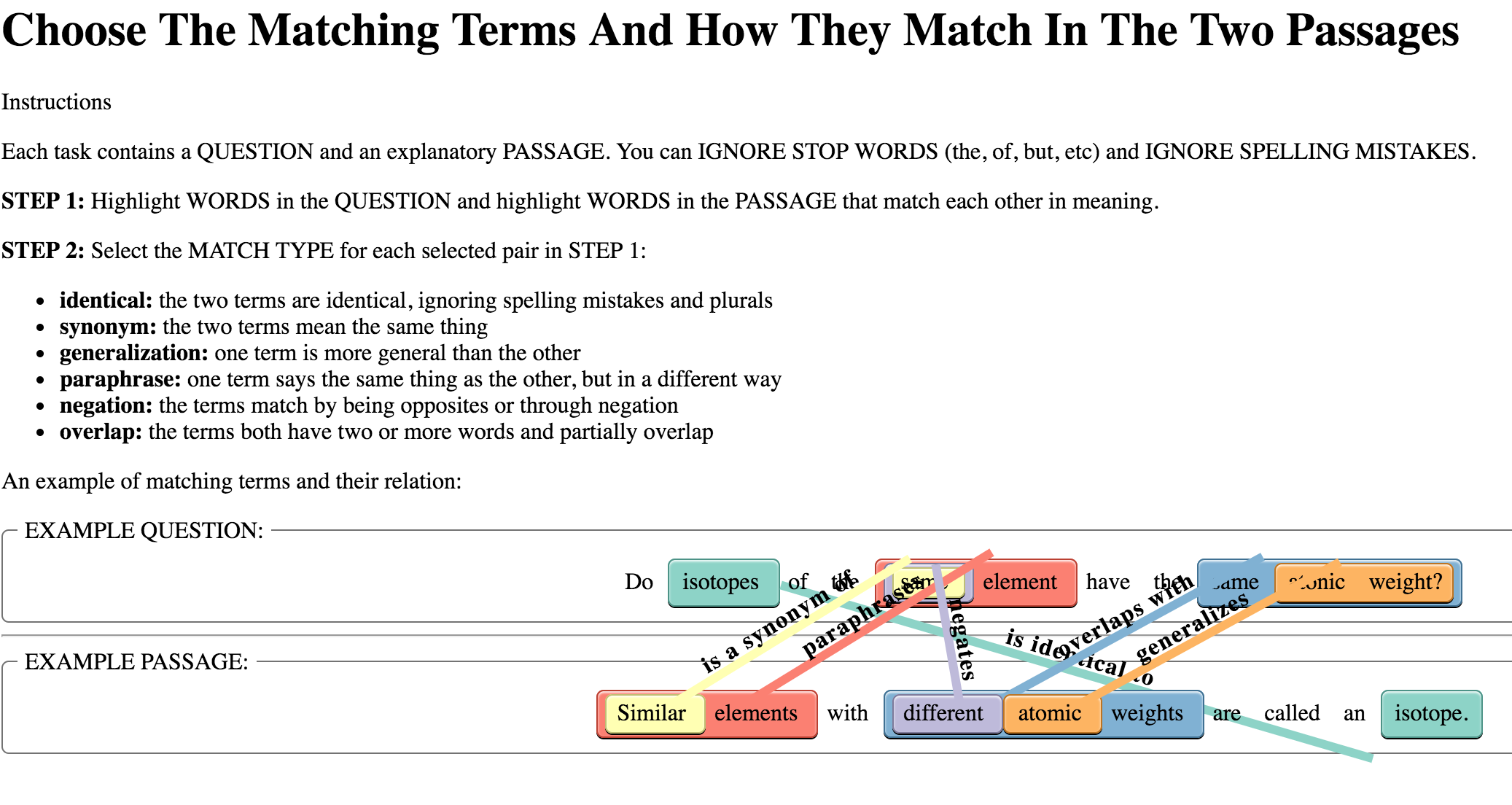}
\caption{\label{fig:num-words-f750831-1}Description}
\end{subfigure}
\begin{subfigure}[b]{0.5\textwidth}
\includegraphics[width=\textwidth]{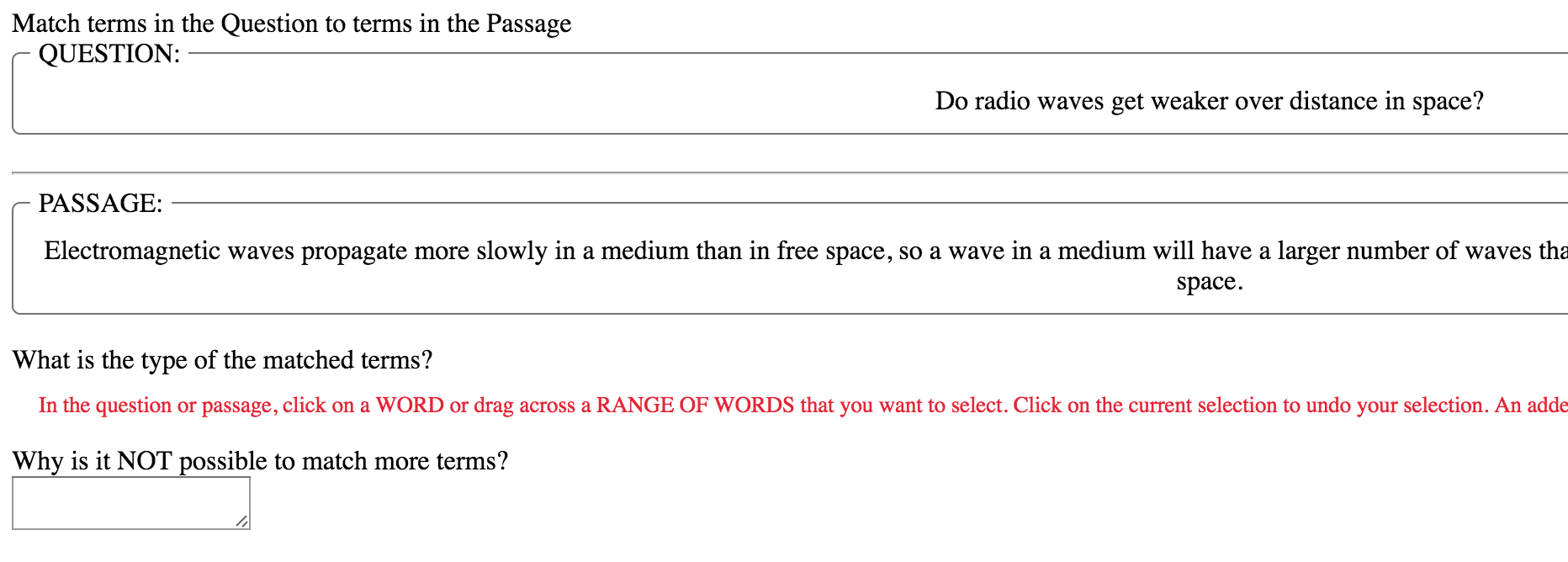}
\caption{\label{fig:num-words-f750831-2}Task}
\end{subfigure}
\caption{\label{fig:num-words-f750831}Task with high \textsf{\#words}}
\end{figure}%
}

\techreport{\ta{Tasks with higher \#\textsf{words} in their HTML sources are typically the ones with more detailed instructions or examples. We see that this has the effect of decreasing \textsf{disagreement} amongst workers, particularly for complex tasks.}}
\subsection{Presence of input text-boxes}\label{sec:num-text-boxes}
Next, we explore the effect of including text boxes as input fields. 
We denote the number of text boxes present in the HTML interface as \#\textsf{text-box}, 
and show its effect on \textsf{disagreement} in Figure~\ref{fig:boxes_vs_disagreement_task}. Specifically, we compare the set of tasks having non-zero text-boxes, i.e. $\#\textsf{text-box}>0$, against tasks with no text-boxes, i.e. $\#\textsf{text-box}=0$.
Not surprisingly, Figure~\ref{fig:boxes_vs_disagreement_task} shows that there tends to be higher \textsf{disagreement} between workers for tasks with text-boxes. 
\techreport{We see that the median value of \textsf{disagreement} for tasks with $\textsf{\#text-box}=0$ is 0.102, while that for tasks with $\textsf{\#text-box}>0$ is 0.160.}
This could be due to the fact that \textsf{disagreement} is agnostic to 
the input operator type and looks for an exact match of worker answers,
while also possibly being affected by the fact that textual tasks may be
more subjective (we have however, filtered out all tasks with very high disagreement).
We also observe that workers tend to take longer to complete such tasks. 
\techreport{We see that the median value of \textsf{task-time} for tasks with $\textsf{\#text-box}=0$ is 119s, while that for tasks with $\textsf{\#text-box}>0$ is 286s.}
Again, this is not surprising, as we expect it to typically take longer to fill out text than to choose from a list of options.

As in Section~\ref{sec:feature-num-words}, we match tasks based on their labels 
and dig deeper to check if the insights obtained from our correlation analyses 
on the complete dataset hold true on individual classes of tasks as well. 
From Figure~\ref{fig:textbox-sa}, we see that for \textsf{sentiment analysis} tasks, the presence of \textsf{text-boxes} significantly increases the \textsf{task-time}. 
Checkboxes or multiple-choice style interfaces are likely to yield much lower \textsf{task-time}s 
than ones based on \textsf{text-boxes}.

\techreport{
\sampar{Example.}
As a concrete example, we consider two different tasks aimed towards the goal of \textsf{Sentiment Analysis}. Both have extremely simple interfaces, and ask simple questions. The primary difference between the two is that the first, depicted in~\ref{fig:tb-example-f836296} contains text-boxes while the second, depicted in~\ref{fig:tb-example-f775761} doesn't. 

\begin{figure}[h]
\centering
\includegraphics[width=0.5\textwidth]{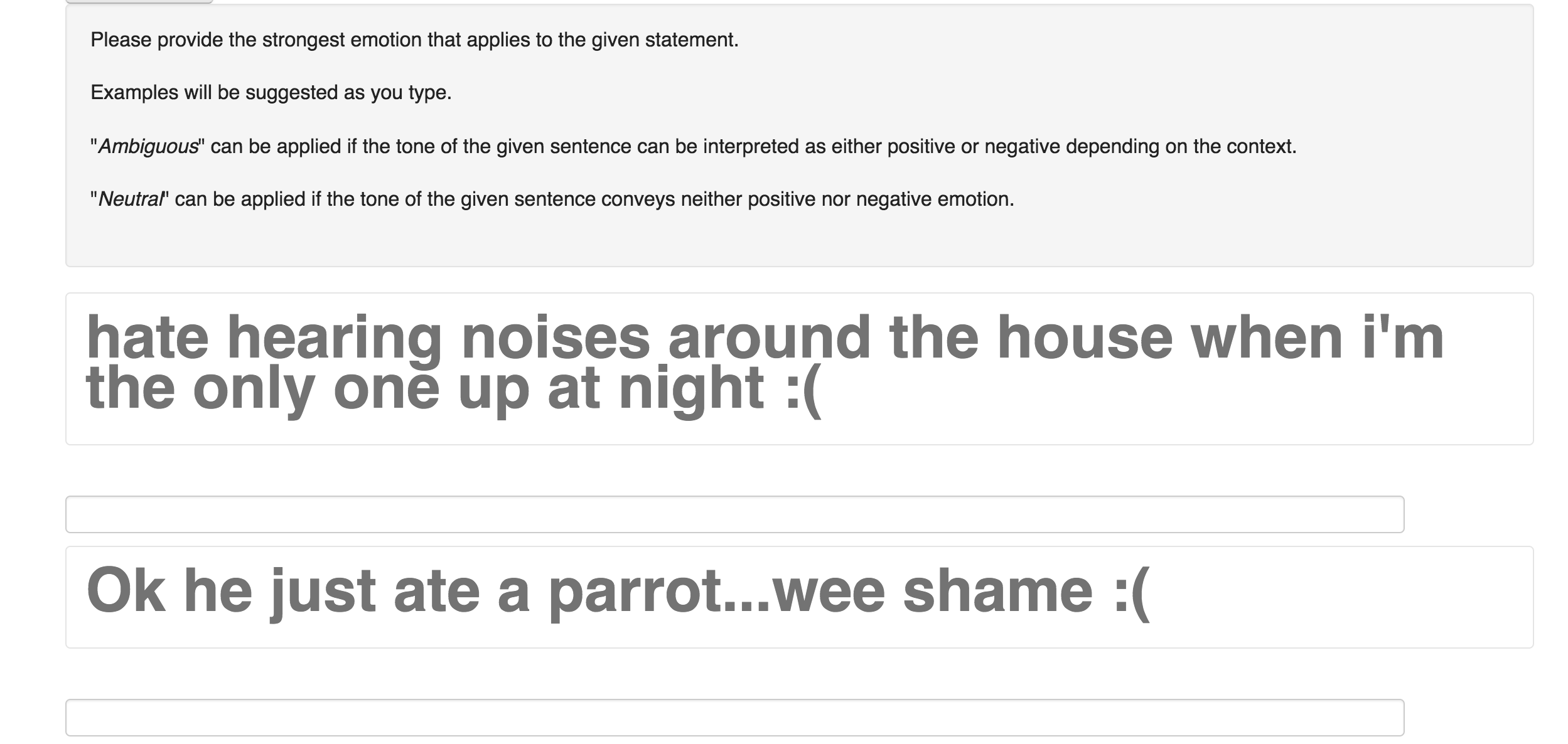}
\caption{\label{fig:tb-example-f836296}Task with text-boxes}
\end{figure}%
\begin{figure}[h]
\centering
\includegraphics[width=0.5\textwidth]{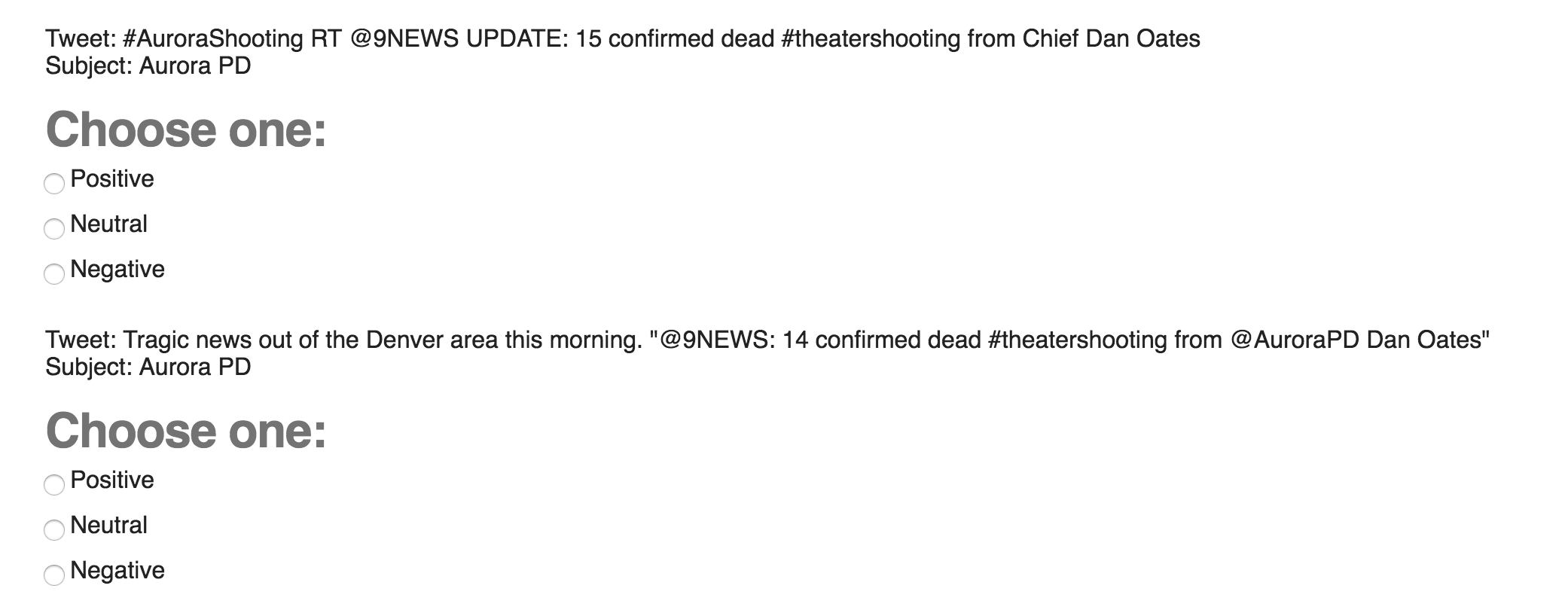}
\caption{\label{fig:tb-example-f775761}Task without text-boxes}
\end{figure}%

Both tasks represent a significant number of instances (around 2680 and 8455 respectively). While the first has a median \textsf{task-time} of 141 seconds, the second displays a median \textsf{task-time} of only 37 seconds.If the goal was to truly infer the sentiment of the pieces of text (and not, say, try to understand what different words workers use to describe the same thing), requesters could just have easily provided a list of sentiments to choose from, and thereby reduced the \textsf{task-time}.
}

\techreport{\ta{
Tasks with a higher number of text-based questions or input require more worker effort (higher \textsf{task-time}) and show higher \textsf{disagreement} between workers. Thus, it pays to simplify questions down to a set of alternatives rather than leaving it open-ended, if possible.
}}
\subsection{Number of items\label{sec:feature-items}}
Another parameter of interest is how many items are operated on in a batch across
many instances and questions. Anecdotally, the number of items in a batch is known to attract workers, since they can read instructions once and work for longer without having to switch context. 
We use \#\textsf{items} to denote this feature. 
We observe that when the \#\textsf{items} is increased, both the \textsf{task-time} as well as the \textsf{disagreement} metrics improve. That is, tasks get done faster, and workers show lower disagreement when tasks have a higher \#\textsf{items} (see Figure~\ref{fig:items_vs_disagreement_pickup_task}). 
\techreport{
We see that the median value of \textsf{disagreement} for tasks with $\textsf{\#items}\geq 56$ is 0.086, while that for tasks with $\textsf{\#items}<56$ is 0.169.
}
One potential reason for this is that tasks with a high \#\textsf{items} 
attract better and more serious workers. 
Another explanation is that workers get better with experience (both faster and more accurate). 
Increasing \#\textsf{items}, however, has the effect of increasing the \textsf{pickup\--time} of a task for similar reasons---this is probably due to the fact that even though 
there may be a higher number of items and therefore task instances, 
the number of available workers (and therefore the parallelism) is still fixed, 
and therefore the same worker
may end up working on different instances in sequence, leading to higher pickup times for the task instances later on in the worker's sequence. 
\techreport{
We see that the median value of \textsf{task-time} for tasks with $\textsf{\#items}>30$ is 136s, while that for tasks with $\textsf{\#items}\leq 30$ is 230s.
}

Further, we believe that having larger \textsf{\#items} would help more for harder tasks, and have less impact on easier tasks. This is supported by our observations from Figure~\ref{fig:items_gather}. We see that \textsf{\#items} has a pronounced effect on \textsf{disagreement} for (relatively hard) \textsf{gather} tasks with higher \textsf{\#items} leading to significantly lower \textsf{disagreement}. Figure~\ref{fig:items_rating} on the other hand, seems to indicate that for (relatively simple) \textsf{rating} tasks, \textsf{\#items} has insignificant impact on \textsf{disagreement}.

\techreport{
\sampar{Example.}
We look at two different tasks that require workers to find urls or email IDs of businesses through basic web searches. Both have extremely similar interfaces, and ask similar questions. Neither employs examples (which as we have mentioned, and shall see later has a significant impact on disagreement). The main difference between the two tasks is that the first (having 540 instances), has median number of items = 1, while the second (having 115425 instances) has a median of 1171 items. Figure~\ref{fig:item-example-f760308} depicts the first task and Figure~\ref{fig:item-example-f790847} depicts the second. (Our snapshot of the task does not depict the high number of items, but does demonstrate the similarity of the two tasks in other respects.) While the first task has a median \textsf{disagreement} of 0.25, the second shows a median disagreement of only 0.04! This demonstrates the power of examples in reducing task ambiguity.

\begin{figure}[h]
\centering
\includegraphics[width=0.5\textwidth]{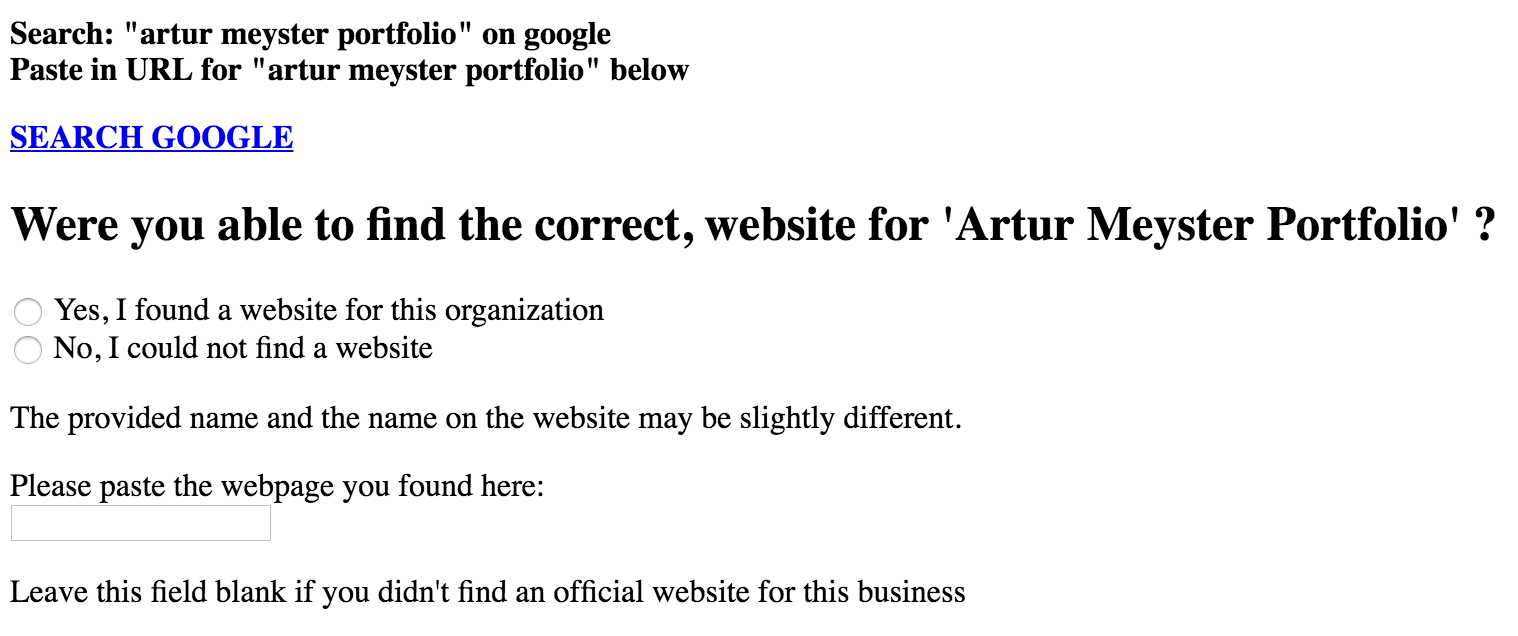}
\caption{\label{fig:item-example-f760308}Task with high \textsf{\#items}}
\end{figure}%
\begin{figure}[h]
\centering
\includegraphics[width=0.5\textwidth]{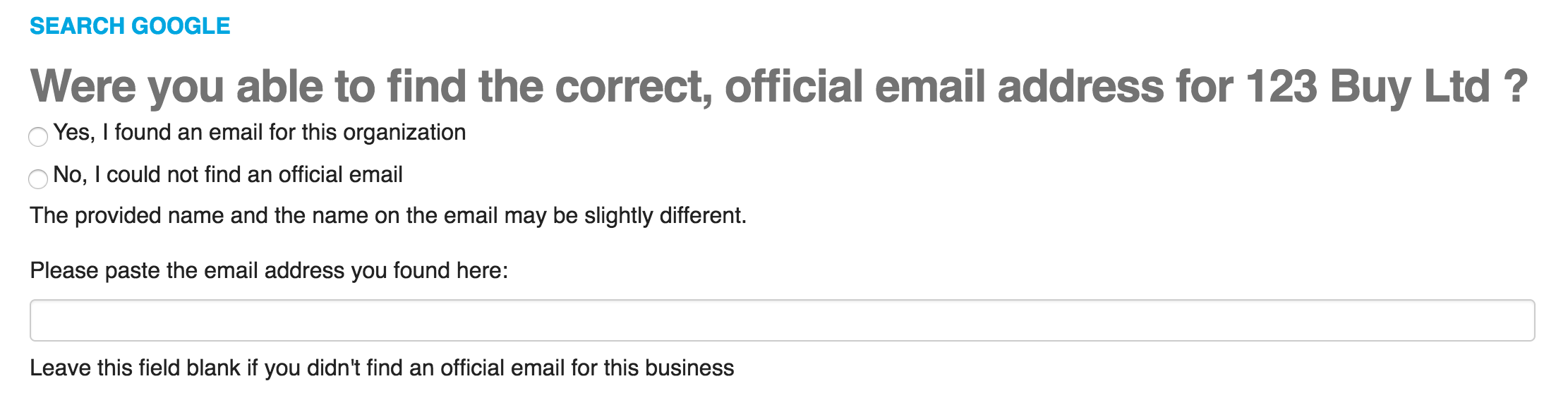}
\caption{\label{fig:item-example-f790847}Task with low \textsf{\#items}}
\end{figure}%
}

\techreport{\ta{Increasing the \#\textsf{items} or instances, improves the quality (reduces \textsf{disagreement} between workers) and reduces the cost (\textsf{task-time}) of a task, but does not help reduce the latency, due to the limited parallelism available in the marketplace.}}

\subsection{Using examples}\label{sec:feature-examples}
It is well-known that examples can have a huge influence on the effectiveness of a task, by training workers on how to answer questions. 
To study how many examples are used in a task, we count the number of times the word ``example'' comes wrapped in a tag of its own in the HTML, indicating that the example is prominently displayed\techreport{. This excludes small or easy-to-miss examples hidden in tasks' textual descriptions, and only counts examples that workers are likely to look at} --- we denote this parameter by \#\textsf{examples}. 
Figure~\ref{fig:examples_vs_disagreement_pickup} demonstrates that examples have the effect of improving worker agreement. 
\techreport{
We see that the median value of \textsf{disagreement} for tasks with $\textsf{\#examples}=0$ is 0.128, while that for tasks with $\textsf{\#examples}>0$ is 0.101.
}
We also observe that examples have the effect of reducing pickup times. 
\techreport{
We see that the median value of \textsf{pickup-time} for tasks with $\textsf{\#examples}=0$ is 6303s, while that for tasks with $\textsf{\#examples}>0$ is 1353s.
} 
It is possible that workers are more inclined to pick up ones 
that seem more ``well-defined'' or clear, thereby choosing the ones 
with examples preferentially over others.
\techreport{\papertext{Despite these obvious benefits, we observe that only 
around 100 task clusters employ explicit examples, as compared to the around 2500 clusters that don't.}}
We observe no significant correlation between the \#\textsf{examples} and the \textsf{task-time} --- this may be because, the time taken to read and understand examples trades off against the improved speed of performing tasks ``post-training''.
Finally, we match tasks based on their labels and dig deeper into individual categories of tasks. From Figure~\ref{fig:example_LU}, we see that \textsf{examples} have a significant effect on \textsf{disagreement} for the most popular task goal, \textsf{Language Understanding}.

\techreport{
\sampar{Example.}
To demonstrate the power of examples, we turn to a similar setting as that in Section~\ref{sec:feature-items}. 
We look at two different tasks that both require workers to find urls of businesses or people through basic web searches (one of which we have seen earlier). Both have extremely simple interfaces, and ask simple questions. A crucial difference between the two tasks is that the first (spanning a significant 1743 instances) provides a detailed example, while the second (having 1006 instances) does not. Figure~\ref{fig:example-example-f788649-1} depicts the example provided by requesters in the first task, and Figure~\ref{fig:example-example-f788649-2} shows the actual task itself. Figure~\ref{fig:example-example-f782907} depicts the second ``example-less'' task. While the first task has a median \textsf{disagreement} of only 0.16, the second shows a median disagreement of 0.45! This demonstrates the power of examples in reducing task ambiguity.

\begin{figure}[h]
\centering
\begin{subfigure}[b]{0.5\textwidth}
\includegraphics[width=\textwidth]{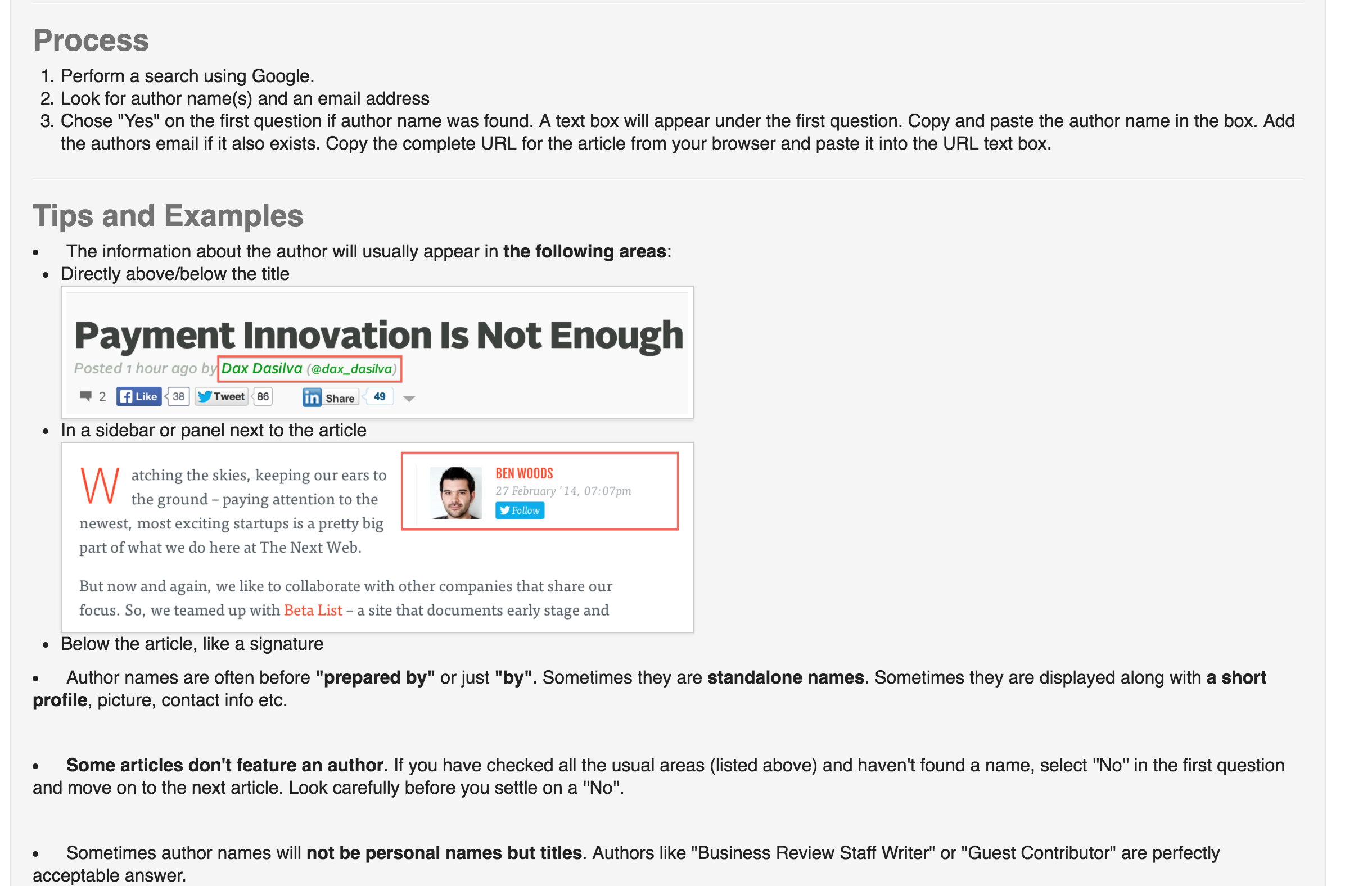}
\caption{\label{fig:example-example-f788649-1}Example.}
\end{subfigure}
\begin{subfigure}[b]{0.5\textwidth}
\includegraphics[width=\textwidth]{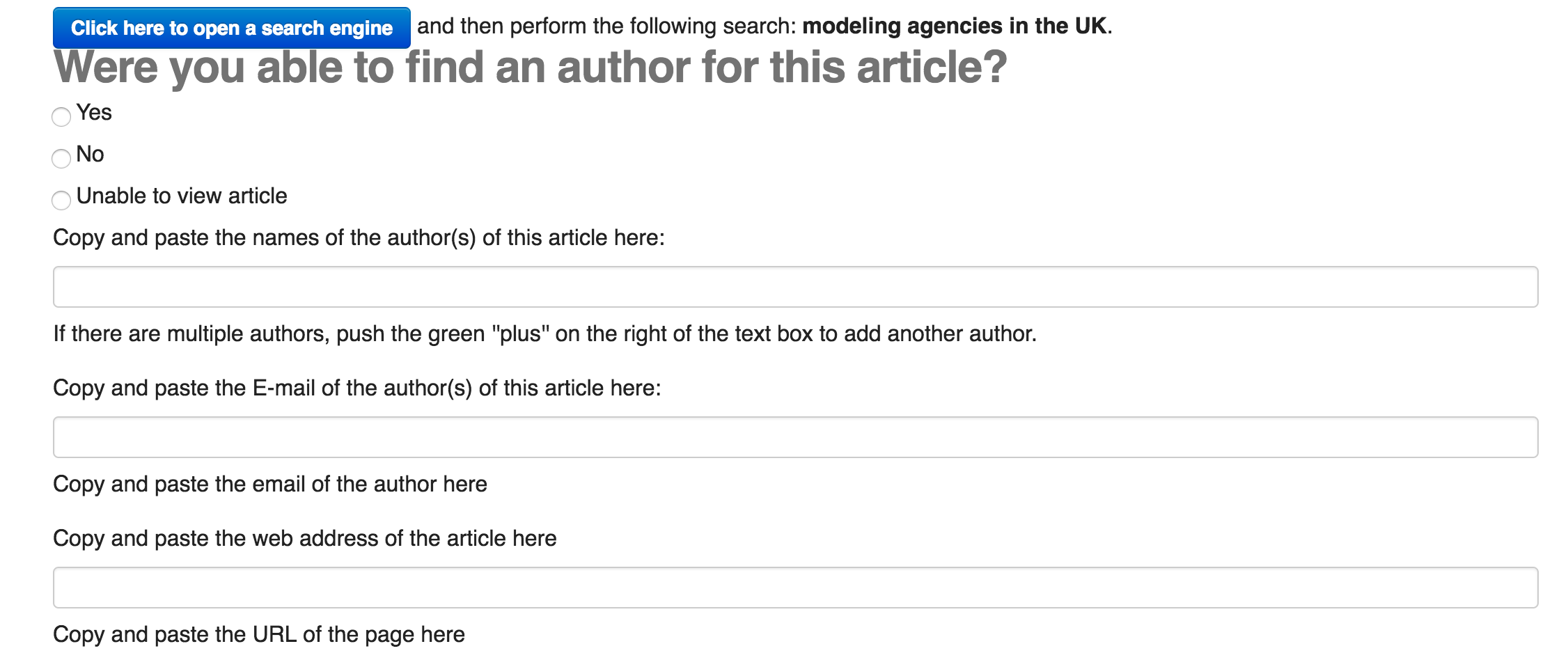}
\caption{\label{fig:example-example-f788649-2}Task.}
\end{subfigure}
\caption{\label{fig:}Task with example.}
\end{figure}%
\begin{figure}[h]
\centering
\includegraphics[width=0.5\textwidth]{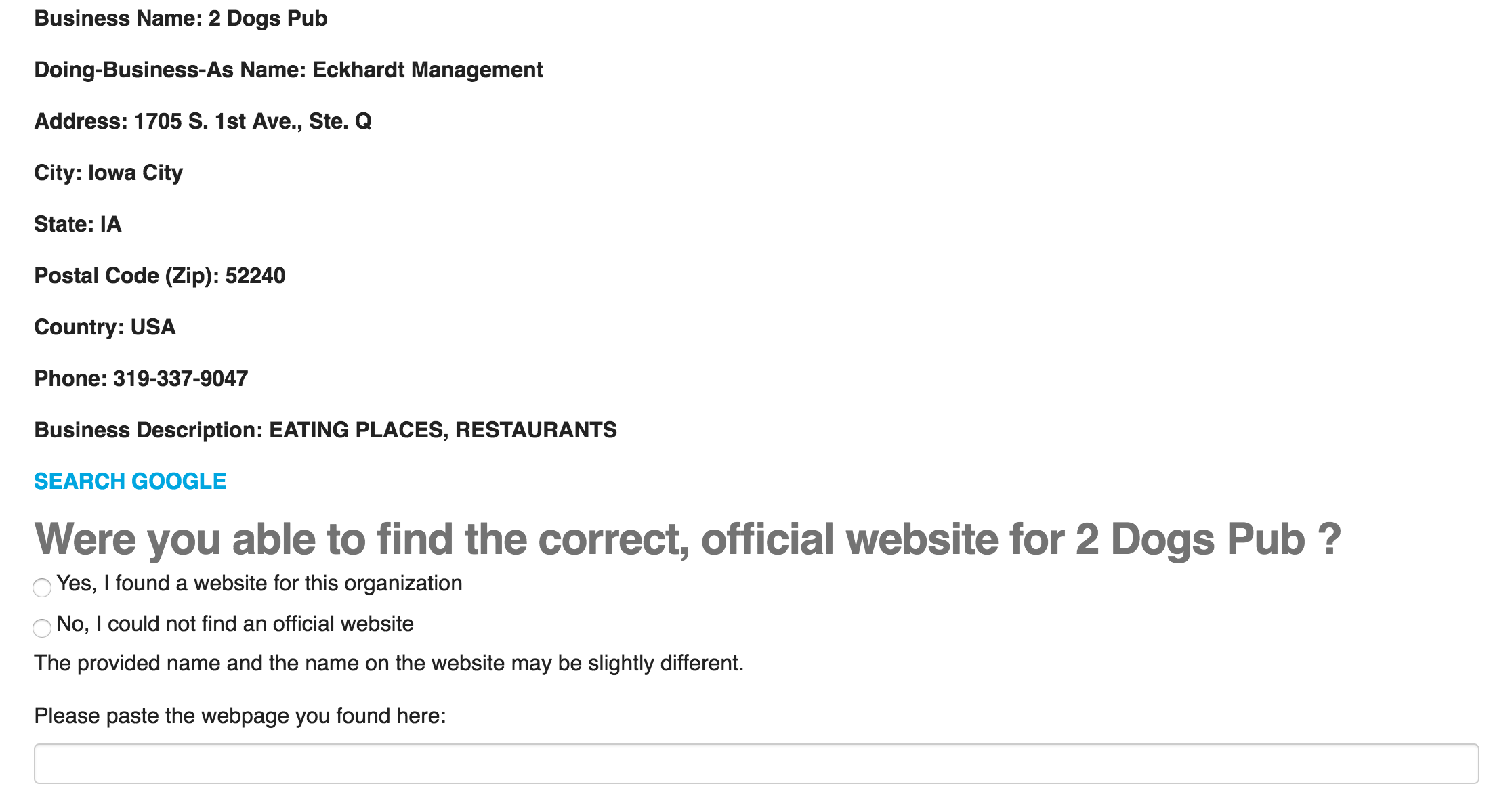}
\caption{\label{fig:example-example-f782907}Task without examples.}
\end{figure}%

Both tasks represent a significant number of batches (14 and 6 respectively) as well as instances (around 8000, 40000). While the first has a median \textsf{pickup-time} of only 233 seconds, the second displays a median \textsf{pickup-time} of about 20,000 seconds! This supports our hypothesis that tasks with images get picked up faster.
}

\techreport{\ta{Examples are very important; not only do they help reduce \textsf{disagreement} (or task ambiguity, resulting in more confident answers), but they also reduce \textsf{pickup-time} (latency) significantly; tasks with examples attract workers much more quickly than tasks without examples. Despite this we observe that only around 200 task clusters employ explicit examples, as compared to the around 3500 clusters that don't.}}
\subsection{Adding images}\label{sec:feature-image}
\begin{figure*}[!t]
\centering

\begin{subfigure}[b]{0.25\textwidth}
\centering
\includegraphics[width=\textwidth]{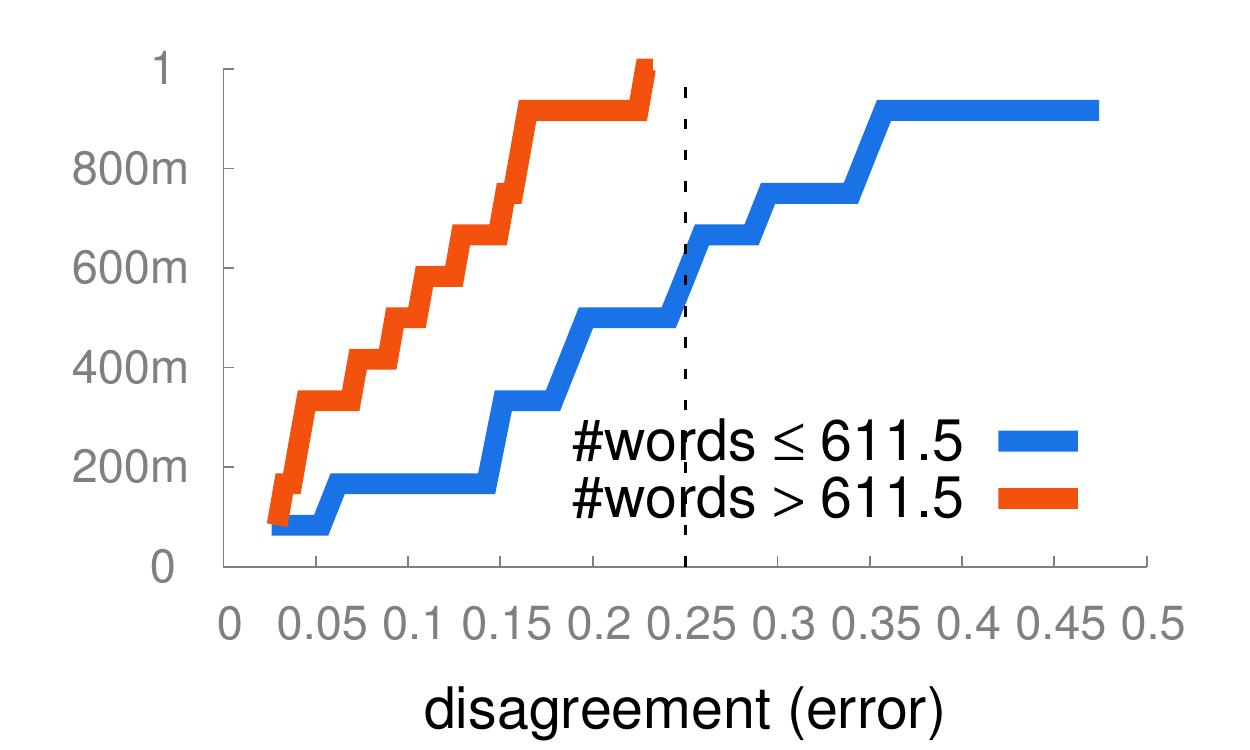}%
\caption{\label{fig:num_words_gather}\textsf{\#words}-\textsf{disagreement}: \textsf{Gather}}
\end{subfigure}%
\hfill
\begin{subfigure}[b]{0.25\textwidth}
\centering
\includegraphics[width=\textwidth]{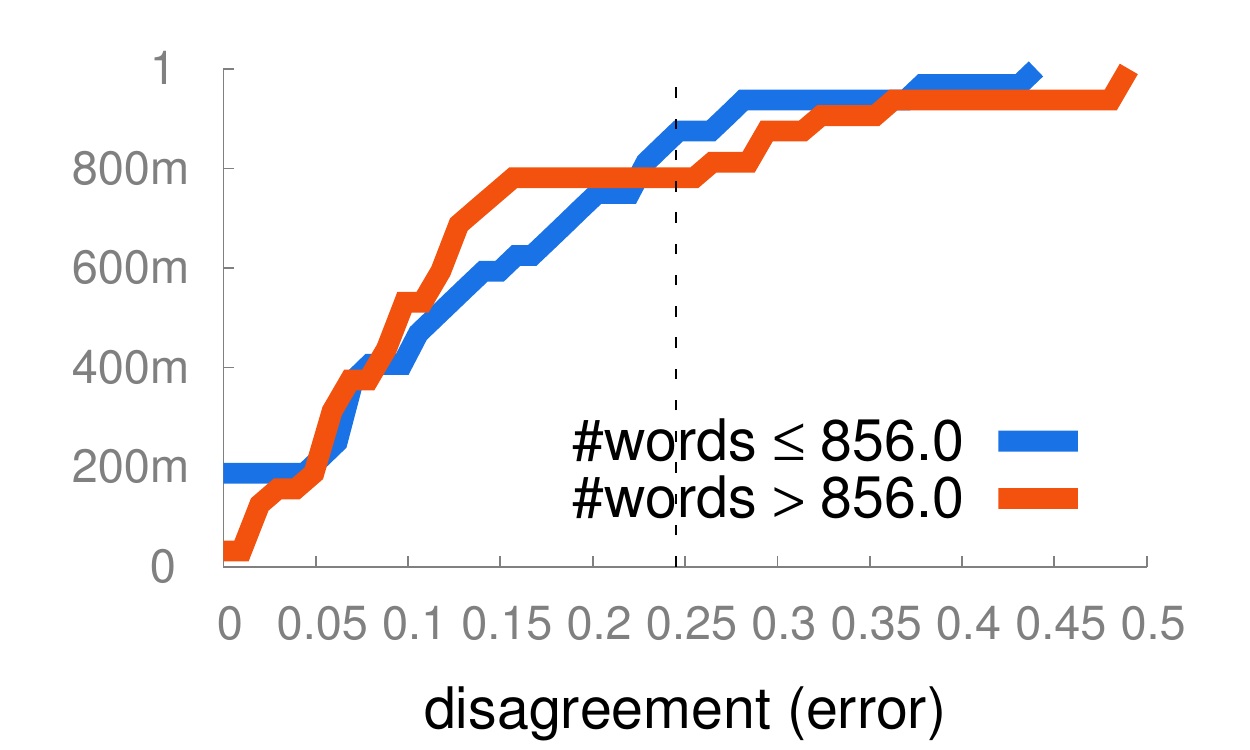}%
\caption{\label{fig:num_words_rating}\textsf{\#words}-\textsf{disagreement}: \textsf{Rating}}
\end{subfigure}%
\hfill
\begin{subfigure}[b]{0.25\textwidth}
\centering
\includegraphics[width=\textwidth]{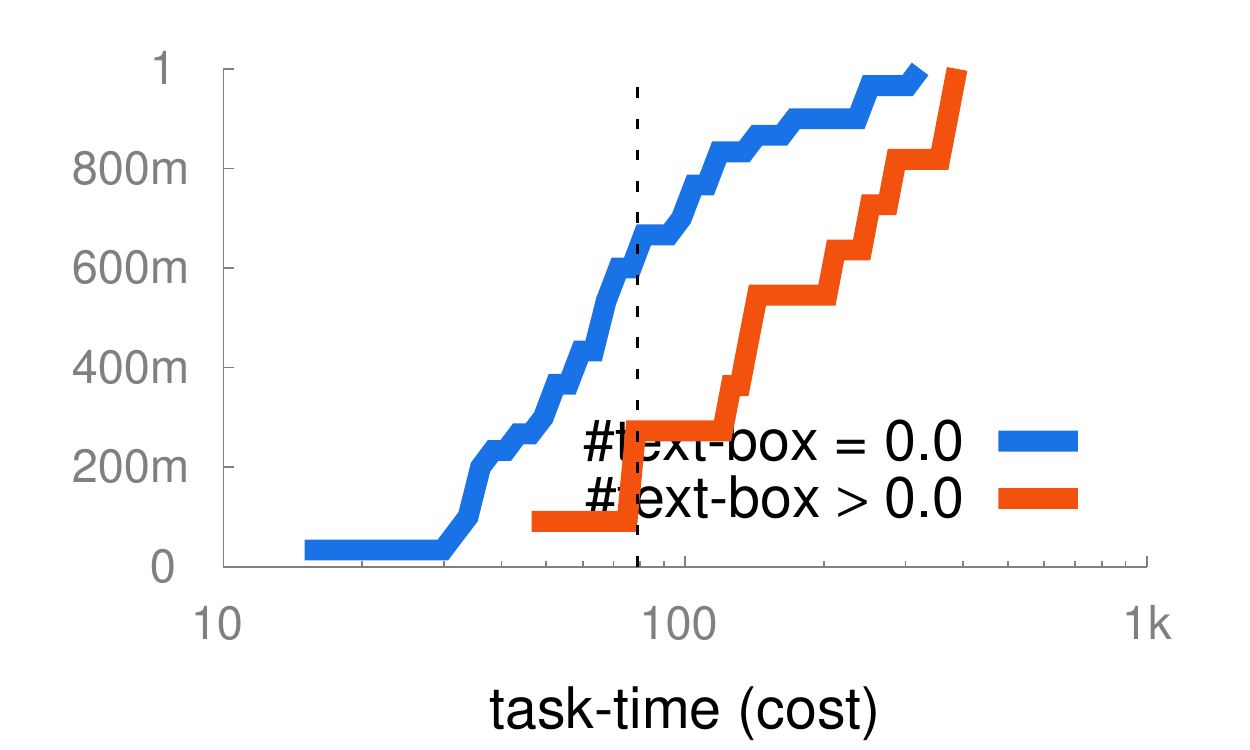}%
\caption{\label{fig:textbox-sa}\textsf{\#text-boxes}-\textsf{task-time}: \textsf{SA}}
\end{subfigure}%
\hfill
\begin{subfigure}[b]{0.25\textwidth}
\centering
\includegraphics[width=\textwidth]{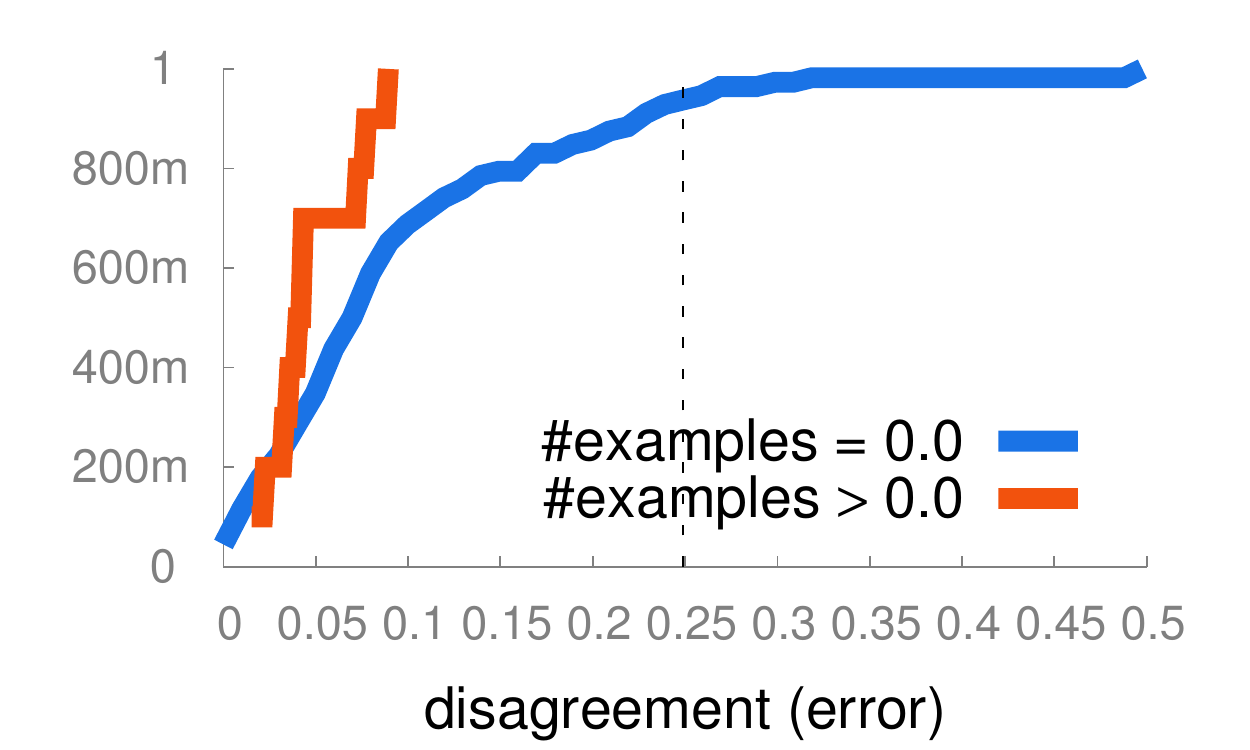}%
\caption{\label{fig:example_LU}\textsf{\#examples}-\textsf{disagreement}: \textsf{LU}}
\end{subfigure}%


\begin{subfigure}[b]{0.25\textwidth}
\centering
\includegraphics[width=\textwidth]{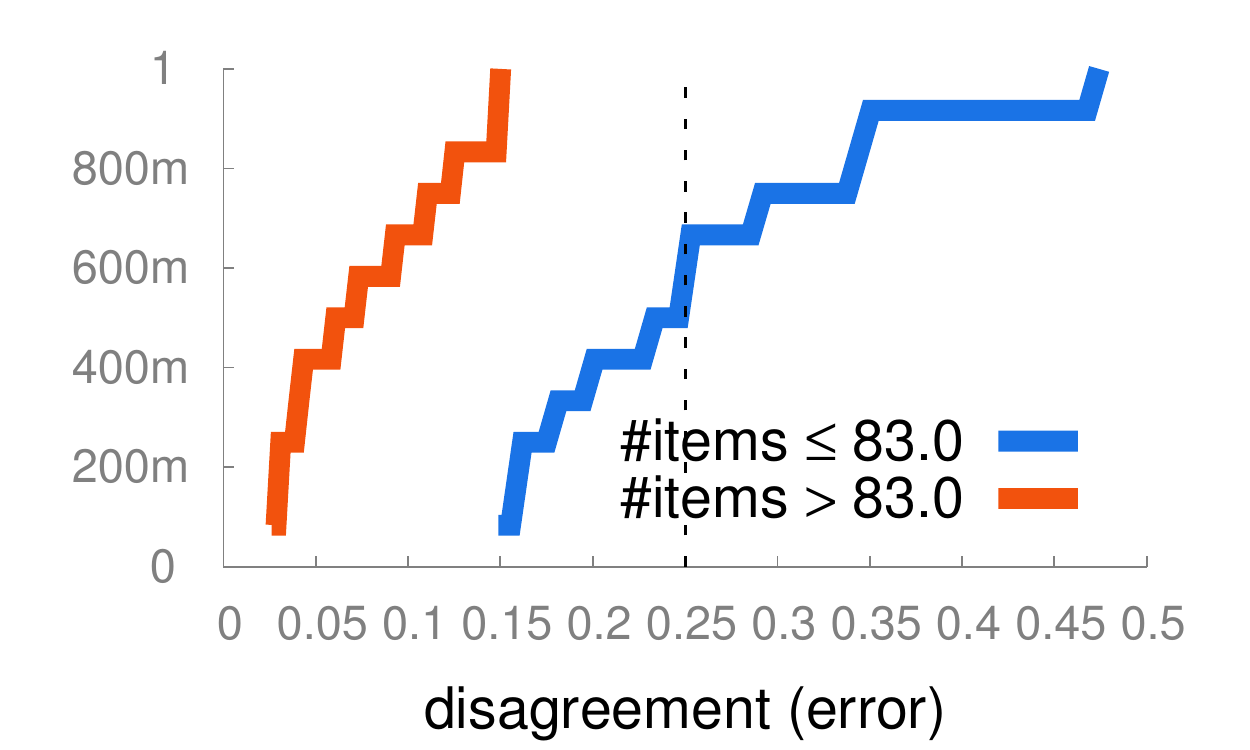}%
\caption{\label{fig:items_gather}\textsf{\# items}-\textsf{disagreement}: \textsf{Gather}}
\end{subfigure}%
\hfill
\begin{subfigure}[b]{0.25\textwidth}
\centering
\includegraphics[width=\textwidth]{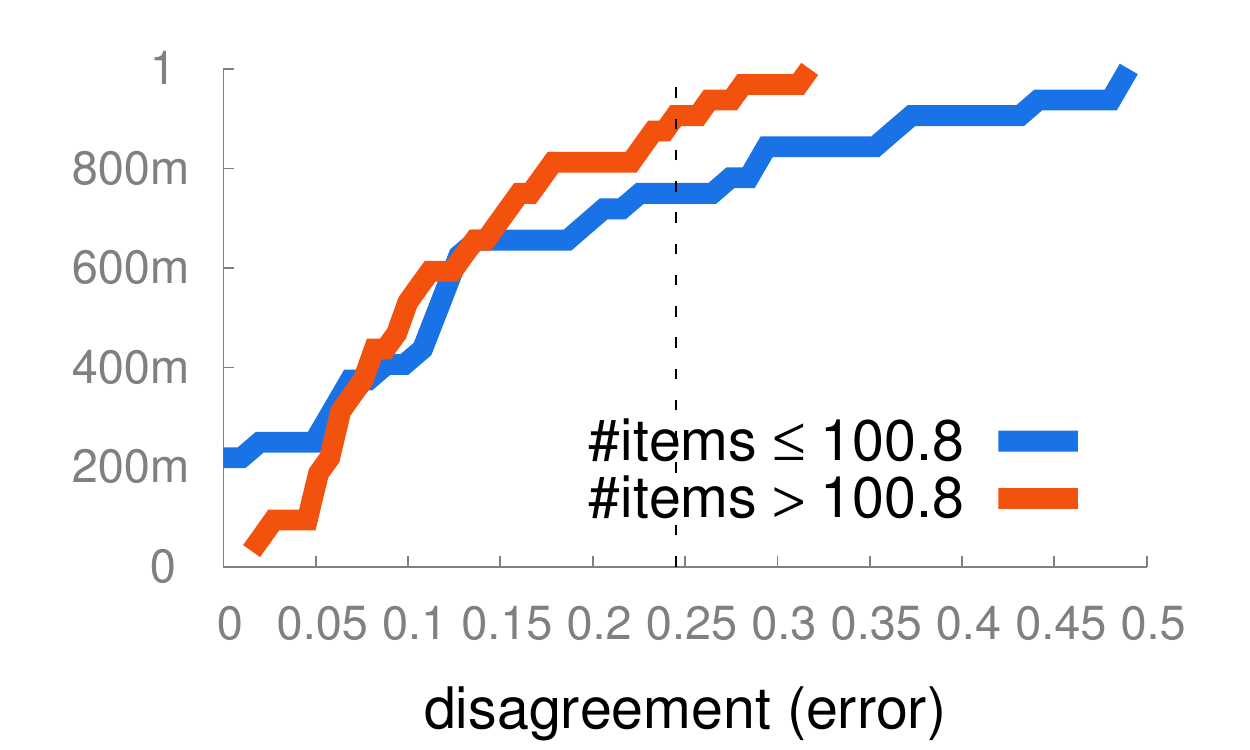}%
\caption{\label{fig:items_rating}\textsf{\# items}-\textsf{disagreement}: \textsf{Rating}}
\end{subfigure}%
\hfill
\begin{subfigure}[b]{0.25\textwidth}
\centering
\includegraphics[width=\textwidth]{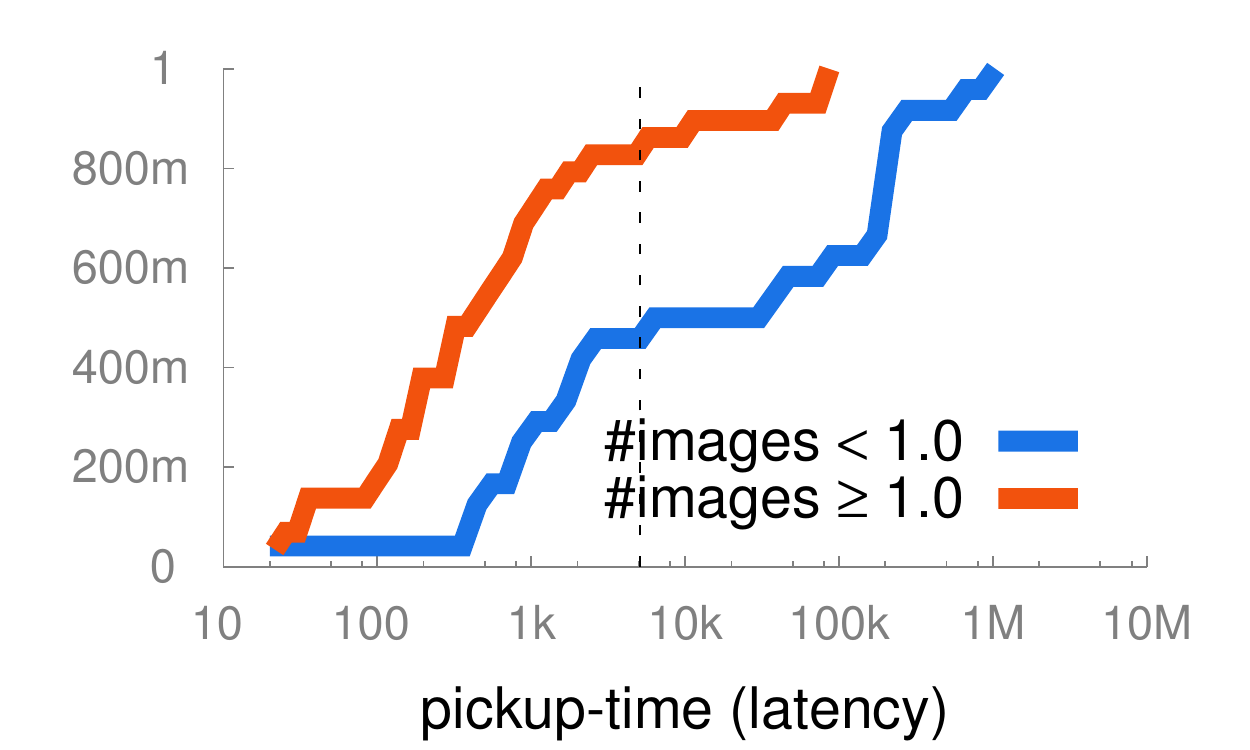}%
\caption{\label{fig:images_extract}\textsf{\#images}-\textsf{pickup-time}: \textsf{Extract}}
\end{subfigure}%
\hfill
\begin{subfigure}[b]{0.25\textwidth}
\centering
\includegraphics[width=\textwidth]{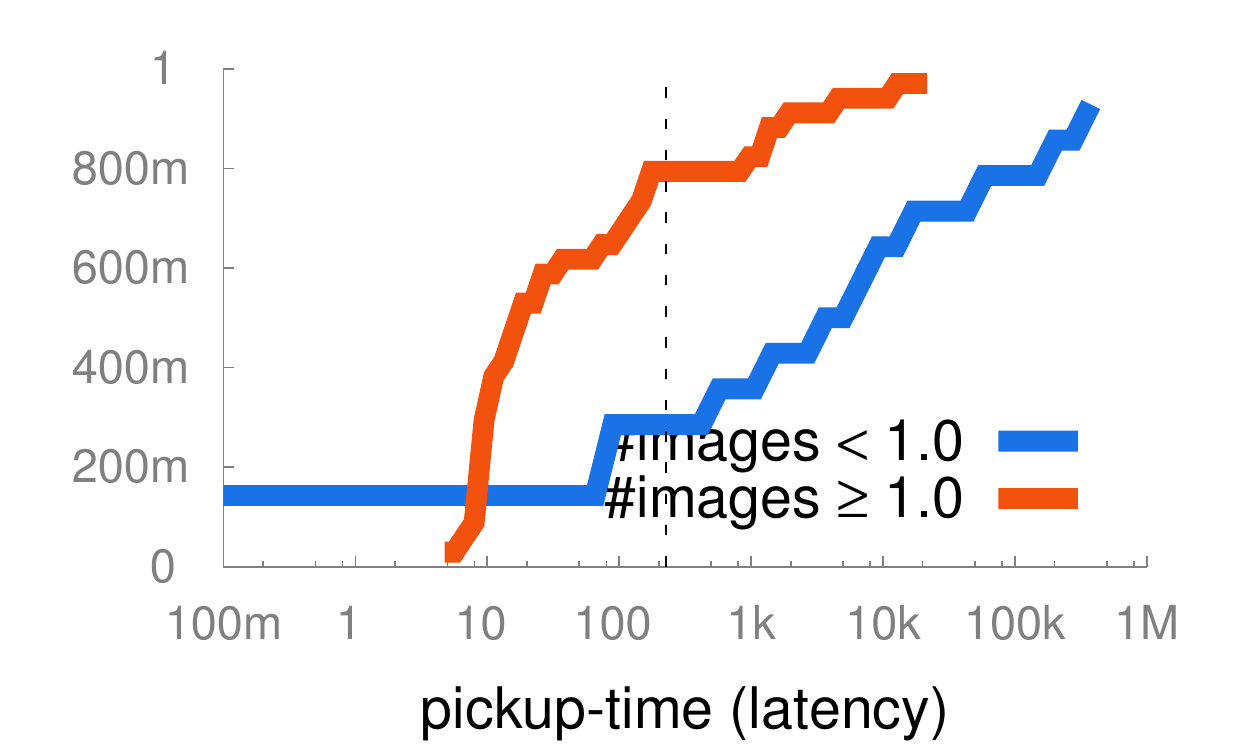}%
\caption{\label{fig:images_quality_control}\textsf{\#images}-\textsf{pickup-time}: \textsf{Quality}}
\end{subfigure}%

\caption{\label{fig:cdf_drill}Features-Metrics CDF: Drill down by match on labels}
\end{figure*}

We speculate that images can play a role in capturing worker interest, and improving the overall worker experience. To evaluate this aspect, we first count the number of image tags present in the HTML source---we denote this feature as \#\textsf{images}. We find that around 700 clusters contain at least one image, while around 2200 contain none. Figure~\ref{fig:images_vs_task_pickup} shows that tasks with $\#\textsf{images}>0$ are picked up faster than those with $\#\textsf{images}=0$. 
\techreport{
We see that the median value of \textsf{pickup-time} for tasks with $\textsf{\#images}=0$ is 7838s, while that for tasks with $\textsf{\#images}>0$ is 2431s.
} 
We believe that this is due to a similar reason as with \#\textsf{examples} --- workers are attracted to more interesting and well-designed tasks, and images go a long way to help with that.
We also drill-down our dataset on task categories to check if the above insight holds true even for specific categories. We plot our observation for tasks with (i) \textsf{operator} \textsf{Extract} in Figure~\ref{fig:images_extract}, or (ii) \textsf{goal} \textsf{Data Quality Control} in Figure~\ref{fig:images_quality_control}. These categories have a significant number of tasks with and without images and the figures show that our hypothesis that tasks are picked up faster due to the presence of images holds true even when we focus on particular operators or goals.

\techreport{
\sampar{Example.}
As a concrete example, we consider two different tasks aimed towards the goal of \textsf{Language Understanding}. Both have extremely simple interfaces, and ask simple questions. The first is related to the relevance of text to a given image and the readability of the text. The second is related to discovering events in text. Both use single choice radio buttons as their choice for worker response. The primary difference between the two is that the first, depicted in~\ref{fig:image-example-f653143} contains images while the second, depicted in~\ref{fig:image-example-f649491} doesn't. 
(Note that while we have access to the HTML source files, we do not have access to the images referred to in them, {\em unless they are embedded} --- therefore in this particular interface we can only see that there is an image present, but cannot see the actual image.)

\begin{figure}[h]
\centering
\includegraphics[width=0.5\textwidth]{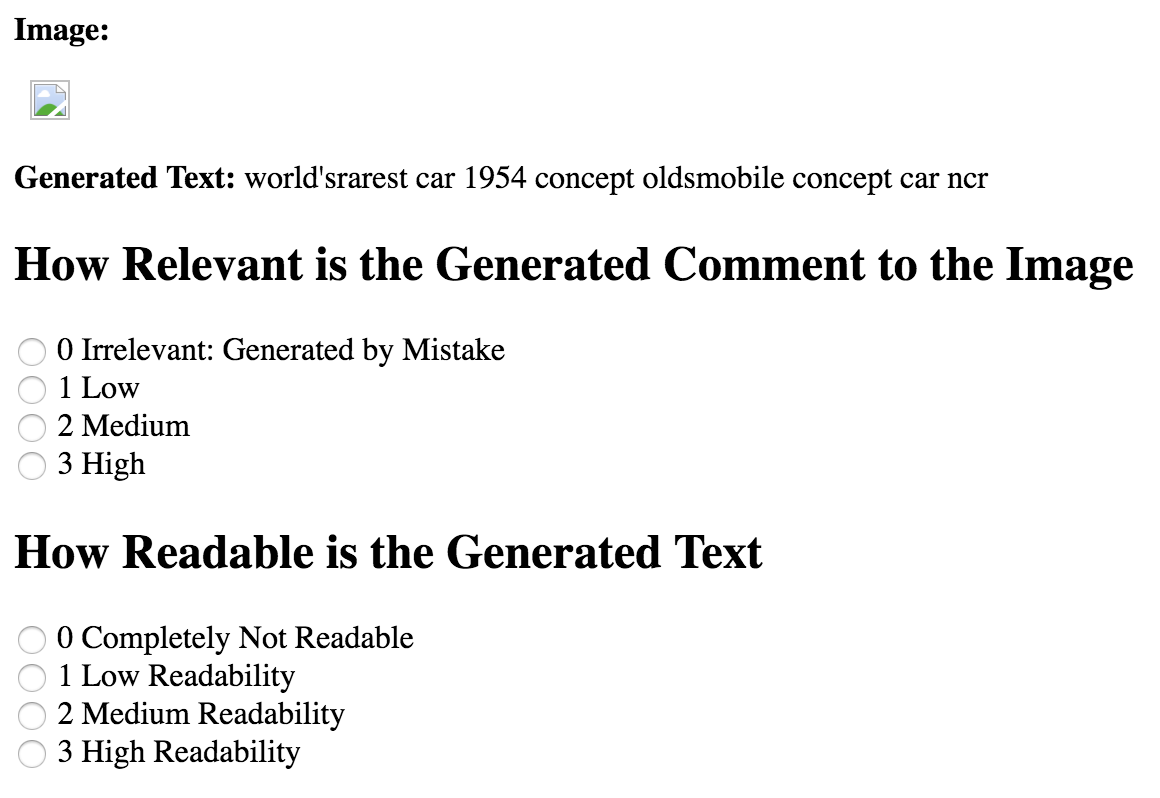}
\caption{\label{fig:image-example-f653143}Task with images}
\end{figure}%
\begin{figure}[h]
\centering
\includegraphics[width=0.5\textwidth]{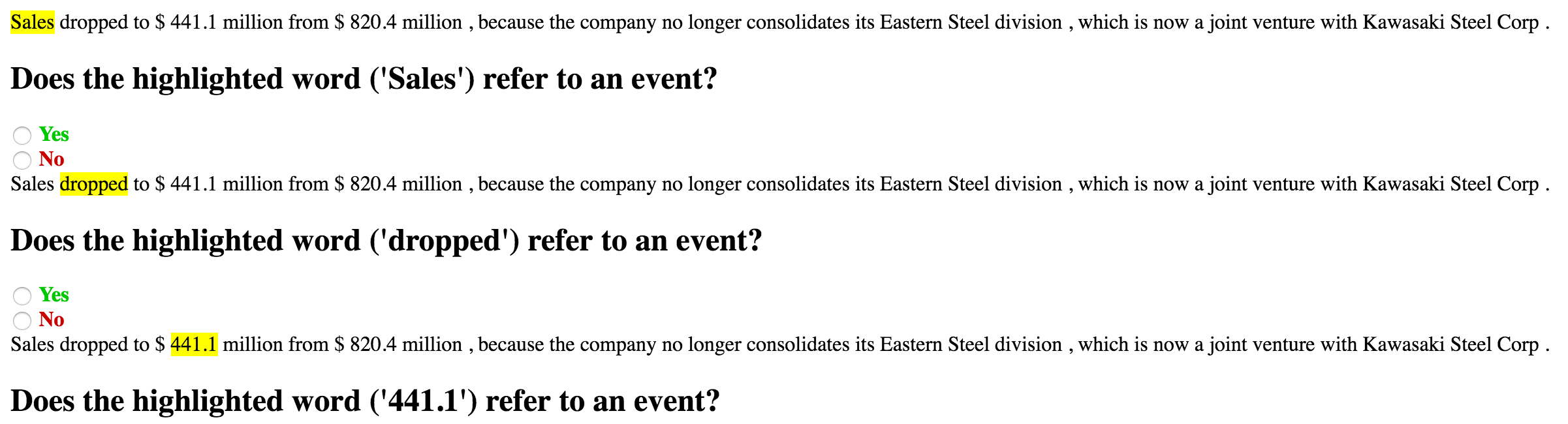}
\caption{\label{fig:image-example-f649491}Task without images}
\end{figure}%

Both tasks represent a significant number of batches (14 and 6 respectively) as well as instances (around 8000, 40000). While the first has a median \textsf{pickup-time} of only 233 seconds, the second displays a median \textsf{pickup-time} of about 20,000 seconds! This supports our hypothesis that tasks with images get picked up faster.
}

We also observe that tasks with images tend to get completed faster. 
\techreport{
We see that the median value of \textsf{task-time} for tasks with $\textsf{\#images}=0$ is 184s, while that for tasks with $\textsf{\#images}>0$ is 129s.
}
One possible explanation for this is that for tasks with \#\textsf{images} $>0$, workers are more energetic or ``enthusiastic'' in completing the task,
and visual understanding often takes less time than textual understanding. We observe no significant correlation between the \#\textsf{images} and the \textsf{disagreement} of tasks, indicating that these tasks are not inherently easier.

\techreport{\ta{Tasks with images attract workers much more quickly than ones (lower \textsf{pickup-time}) without, making them a very powerful tool in reducing latency. Further, workers tend to perform tasks with images faster than those without (lower \textsf{task-time}).}}

\subsection{Summary from a metric point of view\label{sec:task-summary}}In addition to the features we have seen so far, we also looked for correlations between other features and the target metrics. For instance, we examined whether batches were issued on weekdays or weekends, what time of day they were issued at, and how many input fields they had. We observed no significant correlations between these features and any of our metrics. (Recall that for a correlation to be considered statistically significant, we perform a t-test and only those observations with a sufficiently small p-value are considered. For the correlations that we summarize for each of our metrics, the p-values are all significantly below our threshold of 0.01.) We present the quantitative observations corresponding to the noticeable correlations in Tables~\ref{tab:disagreement},~\ref{tab:task} and~\ref{tab:pickup}, and discuss the underlying insights below. 

\smallskip
\noindent
{\bf Disagreement Score.}
Table~\ref{tab:disagreement} summarizes the effect of features that show correlation with the \textsf{disagreement} of tasks. 
Based on our observations, we draw the following conclusions:
Providing detailed instructions for workers can be crucial. If we have multiple items or questions, we should issue them together in one batch (as opposed to scattered across batches) in order to benefit from more experienced workers and workers who get better with experience. Interfaces should use multiple-choice questions to phrase tasks rather than text-based ones wherever possible. Examples are also crucial in reducing errors.

\smallskip
\noindent
{\bf Median Task Time.}
Table~\ref{tab:task} summarizes the effect of features that show correlation with the \textsf{task-time} of tasks. Based on our observed correlations, we note that similar to \textsf{disagreement}, it is beneficial to issue items all at once to benefit from workers with experience. Interfaces should use multiple-choice questions to phrase tasks rather than text-based ones wherever possible, as they also affect the typical task time, and correspondingly, worker effort. Adding images not only makes tasks look more pleasing, but also improves worker experience and latency.


\smallskip
\noindent
{\bf Median Pickup Time.}
Table~\ref{tab:pickup} summarizes the effect of features that show correlation with the \textsf{pickup-time} of tasks. 
Including examples and images is observed to help increase pick-up rate (reduce latency), probably because workers are attracted to more interesting and well-structured tasks.
At the same time, issuing more task instances in parallel will lead to increases in the pickup time due to limited parallelism in the marketplace.


\begin{table}
\scriptsize
   \centering
   \begin{tabular}{|c|c|c|c|c|c|c|} 
\hline
        & \multicolumn{4}{c|}{Cluster Bins} & \multicolumn{2}{c|}{} \\ 
        Feature & \multicolumn{4}{c|}{(split at median(feature-value))} & \multicolumn{2}{c|}{\textsf{disagreement}} \\ 
        \cline{2-7}
        & Bin-1 & \# clusters & Bin-2 & \# clusters &Bin-1 &Bin-2\\ 
        \hline
        \#\textsf{words} & $\leq 466$ & $1150$ & $>466$ & $1149$ & \cellcolor{red!25}0.147 & \cellcolor{green!25}0.108 \\ 
        \#\textsf{items} & $<56$ & $1148$ & $\geq 56$ & $1151$ & \cellcolor{red!25}0.169 & \cellcolor{green!25}0.086 \\ 
        \#\textsf{text-boxes} & $=0$ & $1283$ & $>0$ & $1014$ & \cellcolor{green!25}0.102 & \cellcolor{red!25}0.160 \\ 
        \#\textsf{examples} & $=0$ & $2221$ & $>0$ & $76$ & \cellcolor{red!25}0.128 & \cellcolor{green!25}0.101 \\ 
\hline
      \end{tabular}
      \caption{Disagreement Score: summary}
   \label{tab:disagreement}
\end{table}

\begin{table}
\scriptsize
   \centering
   \begin{tabular}{|c|c|c|c|c|c|c|} 
\hline
        & \multicolumn{4}{c|}{Cluster Bins} & \multicolumn{2}{c|}{} \\ 
        Feature & \multicolumn{4}{c|}{(split at median(feature-value))} & \multicolumn{2}{c|}{\textsf{task-time}} \\ 
        \cline{2-7}
        & Bin-1 & \# clusters & Bin-2 & \# clusters &Bin-1 &Bin-2\\ 
        \hline
        \#\textsf{items} & $\leq 30$ & $1511$ & $> 30$ & $1469$ & \cellcolor{red!25}230s & \cellcolor{green!25}136s \\ 
        \#\textsf{text-boxes} & $=0$ & $1565$ & $>0$ & $1412$ & \cellcolor{green!25}119.0s & \cellcolor{red!25}285.7s \\ 
        \#\textsf{images} & $=0$ & $2268$ & $>0$ & $709$ & \cellcolor{red!25}183.6s & \cellcolor{green!25}129.0s \\ 
\hline
      \end{tabular}
      \caption{Median Task Time: summary}
   \label{tab:task}
\end{table}

\begin{table}
\scriptsize
   \centering
   \begin{tabular}{|c|c|c|c|c|c|c|} 
\hline
        & \multicolumn{4}{c|}{Cluster Bins} & \multicolumn{2}{c|}{} \\ 
        Feature & \multicolumn{4}{c|}{(split at median(feature-value))} & \multicolumn{2}{c|}{\textsf{pickup-time}} \\ 
        \cline{2-7}
        & Bin-1 & \# clusters & Bin-2 & \# clusters &Bin-1 &Bin-2 \\ 
        \hline
        \#\textsf{items} & $\leq 31$ & $1471$ & $> 31$ & $1470$ & \cellcolor{green!25}4521s & \cellcolor{red!25}8132s \\ 
        \#\textsf{examples} & $=0$ & $2845$ & $>0$ & $93$ & \cellcolor{red!25}6303s & \cellcolor{green!25}1353s \\ 
        \#\textsf{images} & $=0$ & $2230$ & $>0$ & $708$ & \cellcolor{red!25}7838s & \cellcolor{green!25}2431s \\ 
\hline
      \end{tabular}
      \caption{Median Pickup Time: summary}
   \label{tab:pickup}
\end{table}

\subsection{Predictive Setting\label{sec:prediction}}

We further concretize our findings from the previous section
by exploring the use of the features for prediction.
We demonstrate that using just these features
allows for an accurate approximate estimation
of various metrics.
Due to the high variability in the range of values of our metrics, 
it is not possible to predict the exact value of a metric for any given task. 
Instead, we bucketize the range of values into 10 buckets, 
and try to predict which bucket any given task will fall into. 
For example, instead of trying to predict \textsf{disagreement} 
for a given task, we predict whether the disagreement 
would fall into the buckets $[0, 0.1), [0.1, 0.2), \ldots, [0.9, 1.0]$. 
There are many different ways in which we could bucketize 
the range of values---each bucketization also corresponds to distributing
 tasks into buckets. In the following, we shall use the 
term bucketization to refer to the bucketization 
of the metric's range of values, as well as tasks interchangeably. 
In our experiments, we consider the two most natural ones: 
(1) bucketization by range, where we evenly divide the range of metric values 
into buckets of uniform width, and (2) bucketization by percentiles, 
where we divide the range of metric values into buckets such that 
all buckets contain roughly equal number of tasks.
For each of these two cases, we divide all three of our metrics into 10 buckets.
We run a simple decision tree classifier with the following feature sets: (1) features for \textsf{disagreement}: \{\textsf{\#items}, \textsf{has-example}, \textsf{\#words}, \textsf{\#text-boxes}\}, (2) features for \textsf{task-time}: \{\textsf{\#items}, \textsf{has-image}, \textsf{\#text-boxes}\}, (3) features for \textsf{pickup-time}: \{\textsf{\#items}, \textsf{has-example}, \textsf{has-image}\}.

\techreport{
\sampar{Bucket distributions.}
For each of the metrics, we now discuss the distribution of clusters across the 10 buckets for both bucketization strategies. For the case of bucketization by range, we have:
\begin{denselist}
\item \textsf{(pickup-time)} Upper bounds of metric value in buckets (in seconds): [$1.6\times 10^6$, $3.2\times 10^6$, $4.7\times 10^6$, $6.3\times 10^6$, $7.9\times 10^6$, $9.5\times 10^6$, $1\times 10^6$, $1.3\times 10^7$, $1.4\times 10^7$, $1.6\times 10^7$], and number of clusters in respective buckets: [2906, 17, 8, 5, 1, 0, 0, 0, 0, 1]
\item \textsf{(task-time)} Upper bounds of metric value in buckets (in seconds): [882, 1756, 2631, 3506, 4380, 5255, 6130, 7004, 7879, 8754], and number of clusters in respective buckets: [2842, 120, 8, 3, 1, 1, 1, 0, 0, 1]
\item \textsf{(disagreement)} Upper bounds of metric value in buckets: [0.1, 0.2, 0.3, 0.4, 0.5, 0.6, 0.7, 0.8, 0.9, 1.0], and number of clusters in respective buckets: [1360, 390, 181, 155, 143, 145, 82, 74, 43, 150]
\end{denselist}
For the case of bucketization by range, we have:
\begin{denselist}
\item \textsf{(pickup-time)} Upper bounds of metric value in buckets (in seconds): [157, 579, 1486, 2955, 5946, 12202, 24796, 53876, 179358, $1.6\times 10^7$], and number of clusters in respective buckets: [294, 294, 294, 293, 294, 294, 293, 294, 294, 294]
\item \textsf{(task-time)} Upper bounds of metric value in buckets (in seconds): [43, 67, 96, 127, 171, 227, 297, 407, 608, 8754], and number of clusters in respective buckets: [300, 297, 298, 296, 301, 297, 296, 297, 297, 298]
\item \textsf{(disagreement)} Upper bounds of metric value in buckets: [0.002, 0.019, 0.038, 0.064, 0.1, 0.158, 0.281, 0.497, 0.693, 1.0], and number of clusters in respective buckets: [273, 272, 272, 272, 273, 272, 272, 272, 272, 273]
\end{denselist}
}

We perform a 5-fold cross-validation to test the accuracy of our models.

\sampar{Bucketization by range.}
We observe that we are able to predict the {\em exact} bucket 
for tasks of \textsf{disagreement} with accuracy 39\%, of \textsf{task-time} 
with 95\%, and of \textsf{pickup-time} with 98\%. \techreport{This is not so surprising given the high skew in their distributions, but knowing the expected range of time for a task to be completed is still useful for requesters.}
Note that here accuracy is averaged across the 5 test cases in our cross-validation. 
For \textsf{disagreement}, we obtain an high accuracy of 62\% if we allow 
an error tolerance of 1 bucket---that is, using just these features alone, we are
able to predict within a tolerance of 1 bucket the disagreement for majority of
the tasks on average.
Given the extremely high dimensional nature of this prediction problem, 
with a very large number of hidden variables that we have not considered, 
even the 39\% accuracy seen for \textsf{disagreement} is very high. 
To verify that the accuracies for \textsf{task-time} and \textsf{pickup-time} 
are not heavily biased by a skew in the distribution of tasks across buckets
 for these metrics, we also perform a similar cross-validation test for the 
percentile-based bucketization. We observe that even in this harder case, 
our model is able to make predictions with reasonable accuracy\papertext{---we 
discuss the percentile-bucketization setting in our technical report~\cite{techreport}}.

\techreport{

\sampar{Bucketization by percentiles.} For the percentile-bucketization, where clusters are divided equally across buckets, the classification problem is much harder because the buckets are of very different and uneven sizes. 
We observe an accuracy of about 16\% for \textsf{task-time}, 15\% for \textsf{pickup-time}, and 20\% for \textsf{disagreement}. The drop in accuracy for \textsf{disagreement} is less pronounced than that for the remaining two metrics since it has a lower skew in its value distribution.
Allowing for a tolerance of one bucket, we see an accuracy of 40\% for \textsf{task-time}, 39\% for \textsf{pickup-time}, and 44\% for \textsf{disagreement}, which is extremely high given the high dimensional nature and small feature set of our classification.
}

\section{Worker Analyses}\label{sec:worker}
In this section, we adopt a worker-centric view of the marketplace and evaluate the worker demographics and behavior patterns. Specifically, we look at (1) distribution of workers across different sources and regions, (2) lifetimes and attention spans of workers.
\subsection{Where do the workers come from?}\label{sec:worker-demo}
\sampar{Labor Sources.}
As described earlier, the marketplace we focus on, unlike Mechanical Turk, gets crowd workers from multiple sources: specifically, the marketplace has {\em 139 different sources} for crowd labor, altogether supplying around 69,000 workers across the period of our collected data. 
These sources---all distinct from each other---are listed in \papertext{our technical report~\cite{techreport}}\techreport{Table~\ref{table:channel_list}}. 
These sources all link to, and allow workers to sign-up with the marketplace. The marketplace directly compensates workers through one of many mechanisms: money, gift cards, or bitcoins. 
Some of these sources (e.g., {\sf imerit\_india}, {\sf yute\_~jamaica}, {\sf taskhunter}) are specific to certain locations in the world, while others provide workers tailored to specific domains of tasks (e.g., {\sf ojooo} provides workers for advertising and marketing campaigns). 
In addition, the marketplace also has its own dedicated worker pool (called {\sf internal}), performing $484k$ tasks, that is about $2\%$ of all tasks in our collected sample. 

\techreport{
\ta{There are a large (over 100) number of active sources of crowd labor, with varying payment schemes; many of these sources are specialized either in terms of demographics, or in terms of the task types.}
}

\techreport{
\begin{table*}[h]
\scriptsize
\centering
\resizebox{\textwidth}{!}{
\begin{tabular}{|cccccccc|}
\hline
neodev  &   clixsense  &   prodege  &   elite  &   instagc  &   tremorgames  &   internal  &   bitcoinget  \\
amt  &   superrewards  &   eup\_slw  &   gifthunterclub  &   taskhunter  &   prizerebel  &   hiving  &   fusioncash  \\  
points2shop  &   clicksfx  &   getpaid  &   cotter  &   coinworker  &   vivatic  &   piyanstantrewards  &   inboxpounds  \\  
imerit\_india  &   personaly  &   stuffpoint  &   errtopc  &   taskspay  &   zoombucks  &   crowdgur  &   gifthulk  \\   
tasks4dollars  &   dollarsignup  &   indivillagetest  &   cbf  &   mycashtasks  &   sendearnings  &   treasuretrooper  &   pokerowned  \\   
diamondtask  &   pforads  &   quickrewards  &   uniquerewards  &   extralunchmoney  &   cashcrate  &   wannads  &   gptbanks  \\   
listia  &   gradible  &   dailyrewardsca  &   clickfair  &   superpayme  &   memolink  &   rewardok  &   snowcirrustechbpo  \\   
pedtoclick  &   rewardingways  &   callmemoney  &   pocketmoneygpt  &   goldtasks  &   dollarrewardz  &   surveymad  &   sharecashgpt  \\  
irazoo  &   zapbux  &   ptcsolution  &   ptc123  &   content\_runner  &   jetbux  &   qpr  &   cointasker  \\   
point\_dollars  &   meprizescf  &   keeprewarding  &   gptking  &   dollarsgpt  &   prizeplank  &   yute\_jamaica  &   onestopgpt  \\   
gptway  &   trial\_pay  &   task\_ph  &   golddiggergpt  &   prizezombie  &   daproimafrica  &   aceinnovations  &   getpaidto  \\   
globalactioncash  &   piyoogle  &   supersonicads  &   poin\_web  &   rewardsspot  &   giftgpt  &   giftcardgpt  &   northclicks  \\   
fastcashgpt  &   dealbarbiepays  &   dailysurveypanel  &   points4rewards  &   gptpal  &   rewards1  &   new\_rules  &   surewardsgpt  \\   
zorbor  &   steamgameswap  &   buxense  &   surveywage  &   offernation  &   probux  &   freeride  &   ojooo  \\   
luckytaskz  &   medievaleurope  &   proudclick  &   steampowers  &   paiddailysurveys  &   wrkshop  &   simplegpt  &   realworld  \\   
surveytokens  &   bemybux  &   onestop  &   plusdollars  &   gptbucks  &   fepcrowdflower  &   embee  &   makethatdollar  \\   
ayuwage  &   luckykoin  &   pointst  &   sedgroup  &   easycashclicks  &   candy\_ph  &   piggybankgpt  &   peoplesgpt  \\   
matomy  &   earnthemost  &   fsprizes &  &  &  &  & \\
\hline
 \end{tabular}
}
\caption{\label{table:channel_list}Sources for crowd workers}
\end{table*}
}

\begin{figure}[!ht]
\centering
\begin{subfigure}[b]{0.5\linewidth}
\centering
\includegraphics[width=1.11\linewidth] {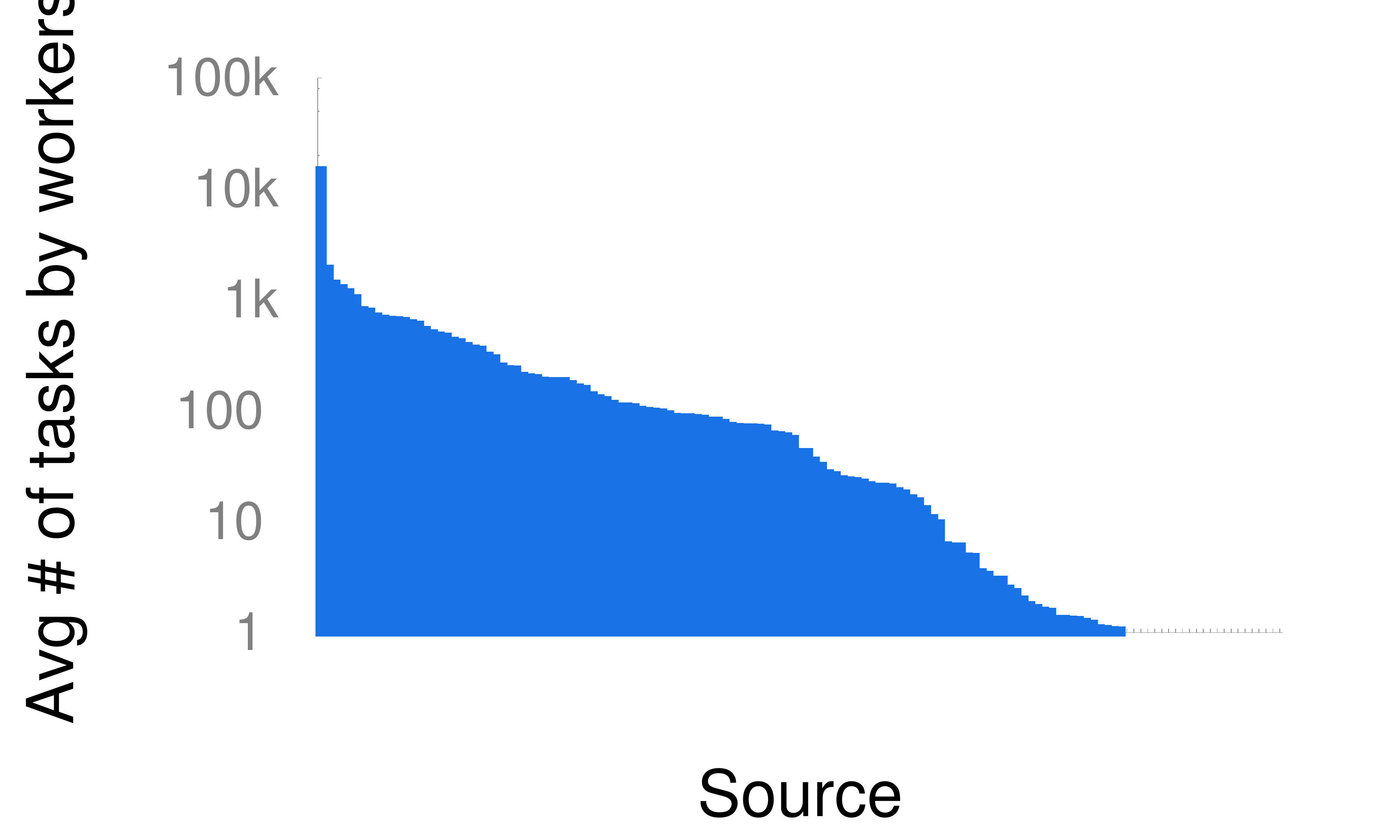}
\vspace{-20pt}
\caption{\label{fig:channels_avg_tasks} Average number of tasks by workers from different sources}
\end{subfigure}
\begin{subfigure}[b]{\linewidth}
\centering
\includegraphics[width=0.8\linewidth] {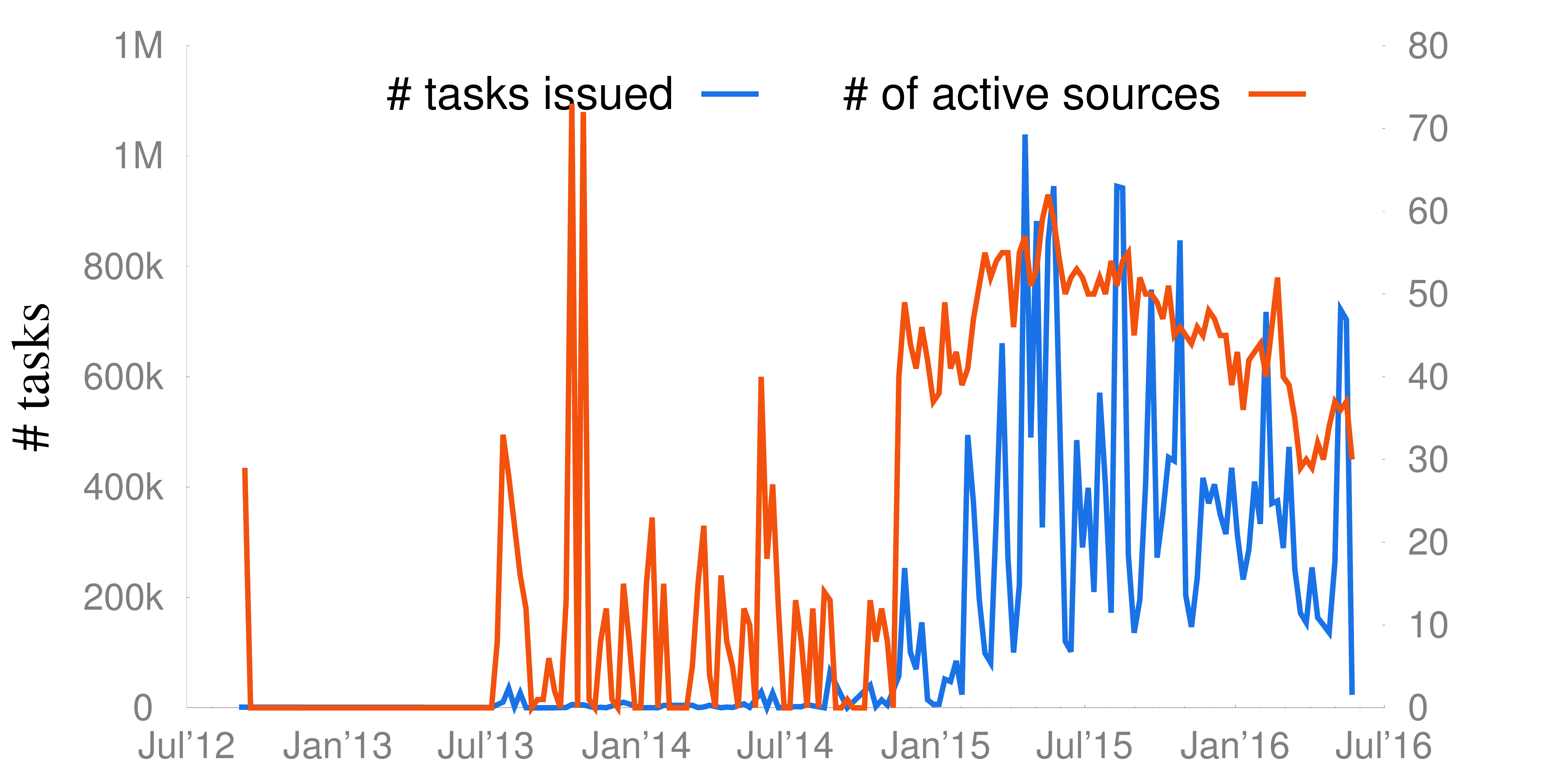}
\caption{\label{fig:channels_weekly} Number of sources used over the evaluation period}
\end{subfigure}
\caption{Tasks performed by workers across sources across weeks}
\end{figure}

We also plot the average number of tasks performed by workers on different sources in Figure~\ref{fig:channels_avg_tasks}. 
Each vertical splice on the x-axis represents a labor source, and the height of the splice indicates the number of tasks performed by a worker from that source on average. 
We see a significant variation in the worker loads across source (the y-axis is log-scale). 
For some sources, workers typically perform more than 10,000 tasks each whereas on the other end of the spectrum, 40\% of the sources have workers performing $\leq 20$ tasks each. 
The variation in number of tasks per worker suggests the presence of two types of sources --- sources having (a) a dedicated workforce performing a large number of tasks per worker, and (b) an on-demand workforce, performing few tasks per worker.
Thus, the availability of these two types of sources is an essential load balancing strategy --- the dedicated workforce is supplemented by on-demand workforce in periods of high task load. 
To study this further, Figure~\ref{fig:channels_weekly} shows the number of sources active every week overlaid on the number of tasks issued. 
This plot seems to indicate that after January 2015, while the marketplace has a relatively fixed number of active sources, the number of tasks issued varied quite a bit. 
\techreport{Thus, these active sources are able to absorb the variation in the number of tasks issued.  }
\papertext{Thus, by using a combination of sources, the marketplace is able to absorb the varying task load.  }

\techreport{\ta{Different sources have different characteristics: some involve an engaged workforce, while in others, participation is more one-off. 
By using a combination of sources, the marketplace is able to absorb a varying task load. } }

\begin{figure*}[!t]
\centering

\begin{subfigure}[b]{0.33\textwidth}
\centering
\includegraphics[width=\textwidth]{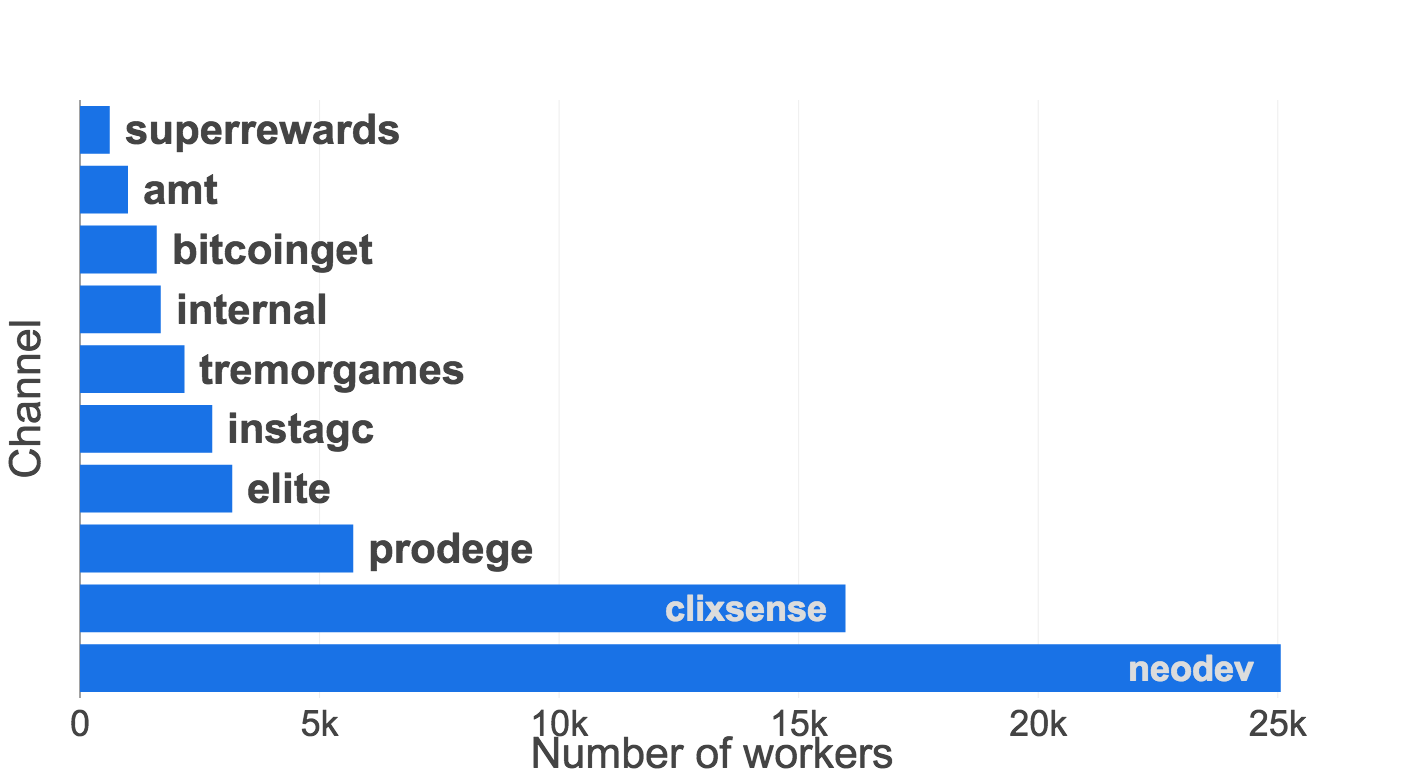}
\caption{\label{fig:channels_workers}Number of workers from top sources}
\end{subfigure}
\hfill
\begin{subfigure}[b]{0.33\textwidth}
\centering
\includegraphics[width=\textwidth]{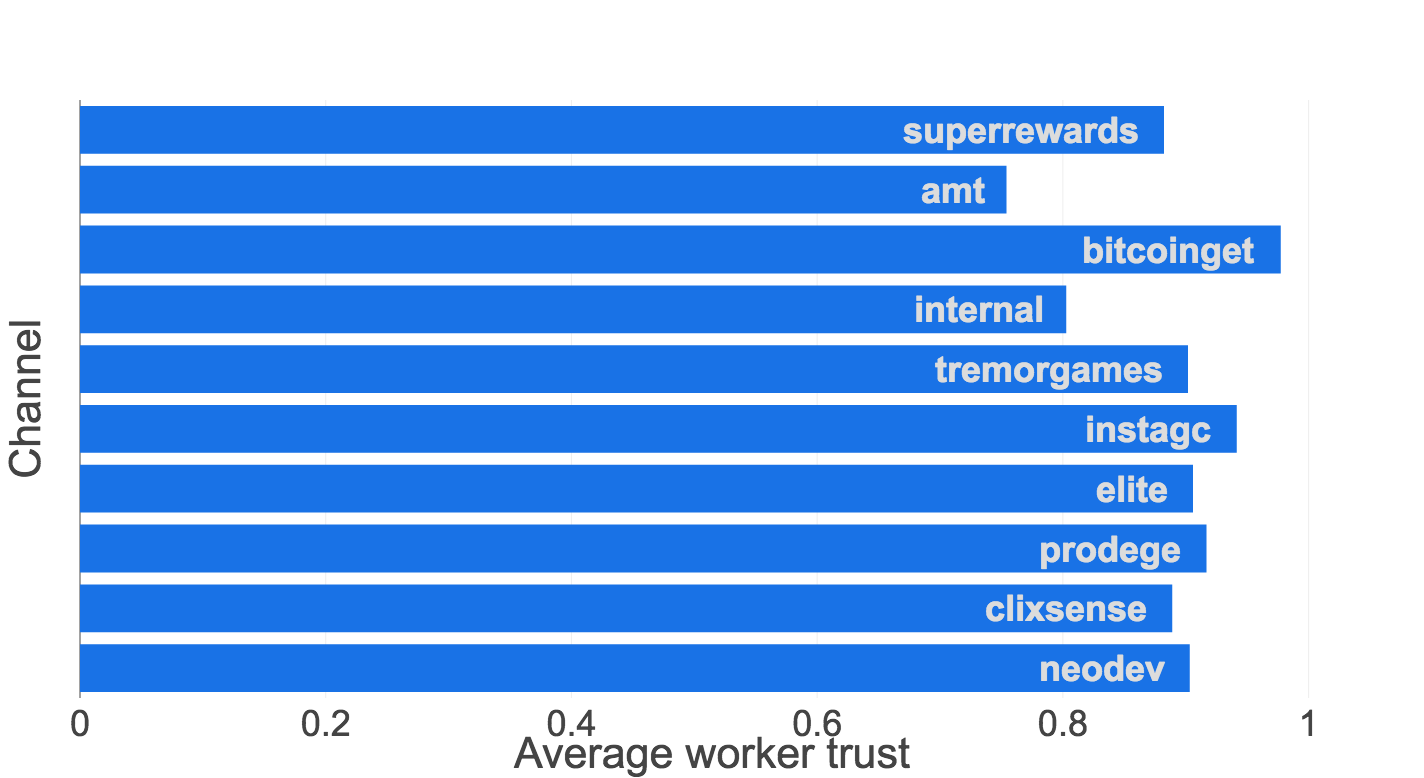}
\caption{\label{fig:major_channels_trust}Mean trust of top sources}
\end{subfigure}
\hfill
\begin{subfigure}[b]{0.33\textwidth}
\centering
\includegraphics[width=\textwidth]{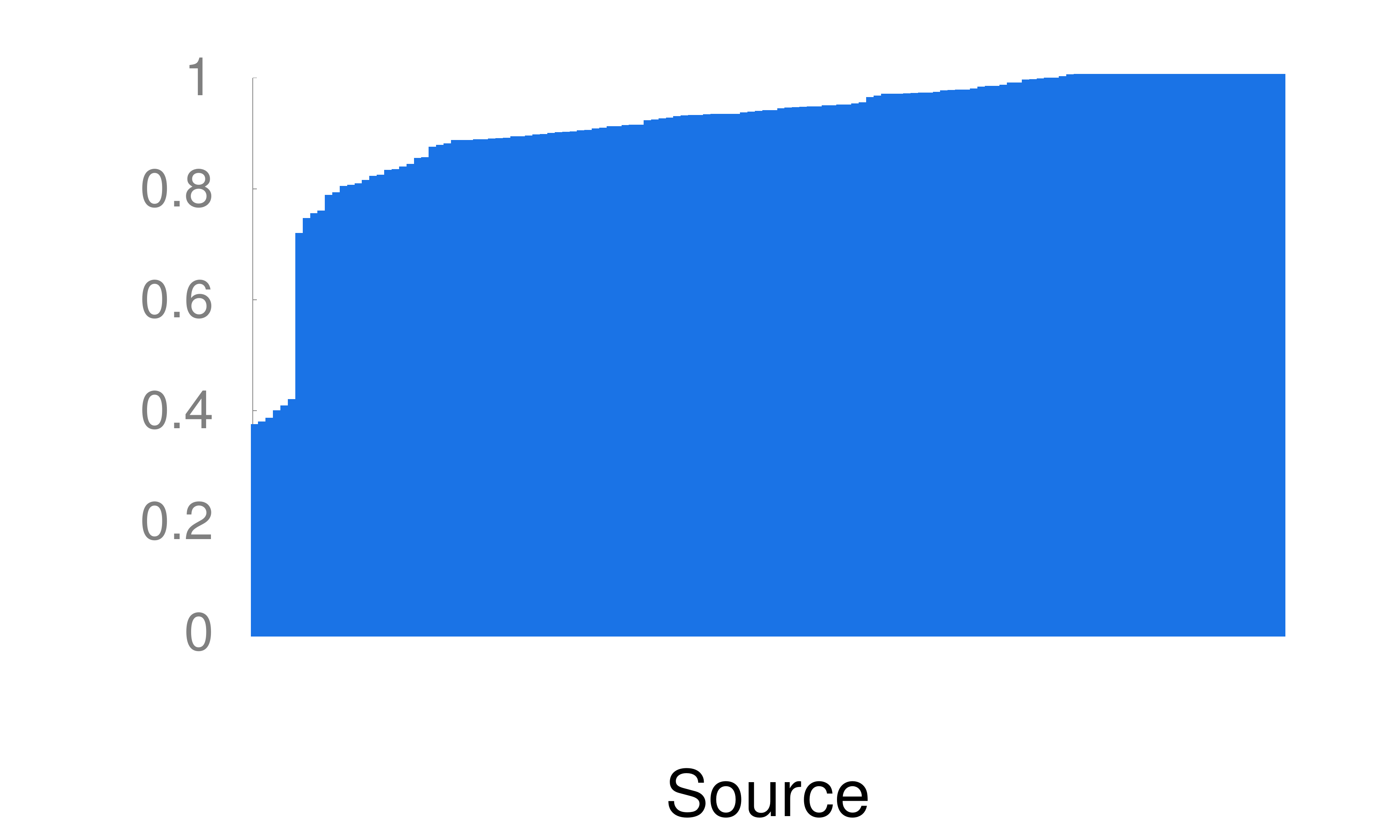}
\caption{\label{fig:channels_trust}Mean trust}
\end{subfigure}


\begin{subfigure}[b]{0.33\textwidth}
\centering
\includegraphics[width=\textwidth]{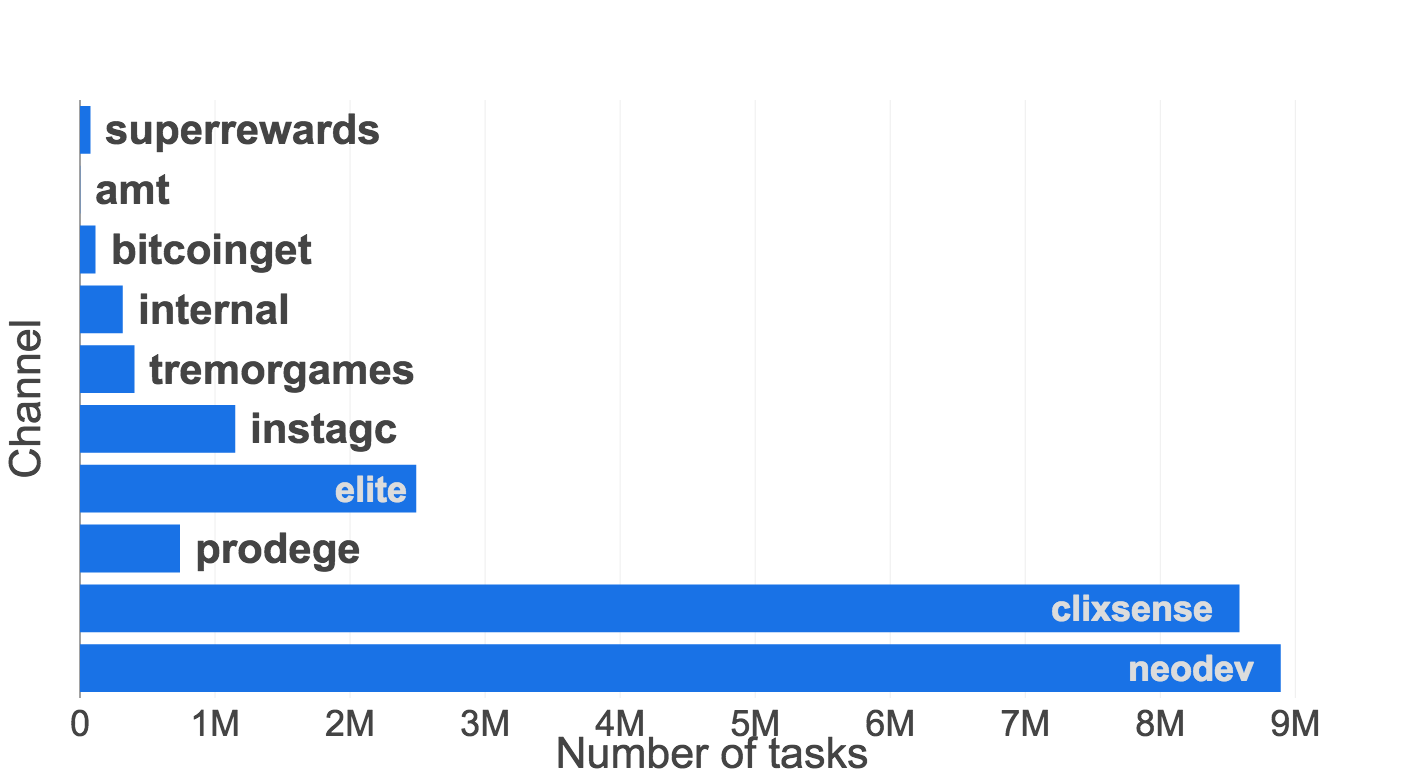}
\caption{\label{fig:channels_tasks}Number of tasks by top sources}
\end{subfigure}
\hfill
\begin{subfigure}[b]{0.33\textwidth}
\centering
\includegraphics[width=\textwidth]{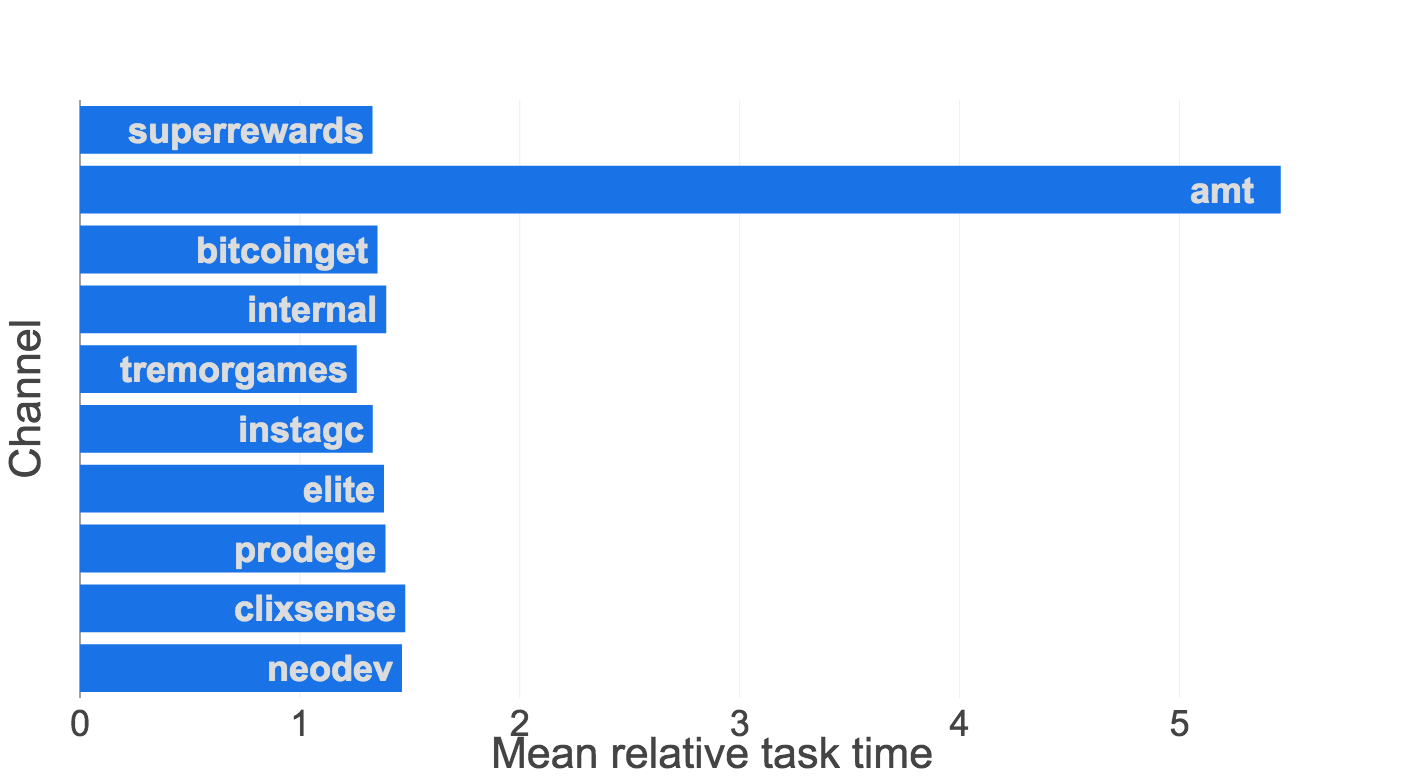}
\caption{\label{fig:major_channels_latency}Mean relative task time of top sources}
\end{subfigure}
\hfill
\begin{subfigure}[b]{0.33\textwidth}
\centering
\includegraphics[width=\textwidth]{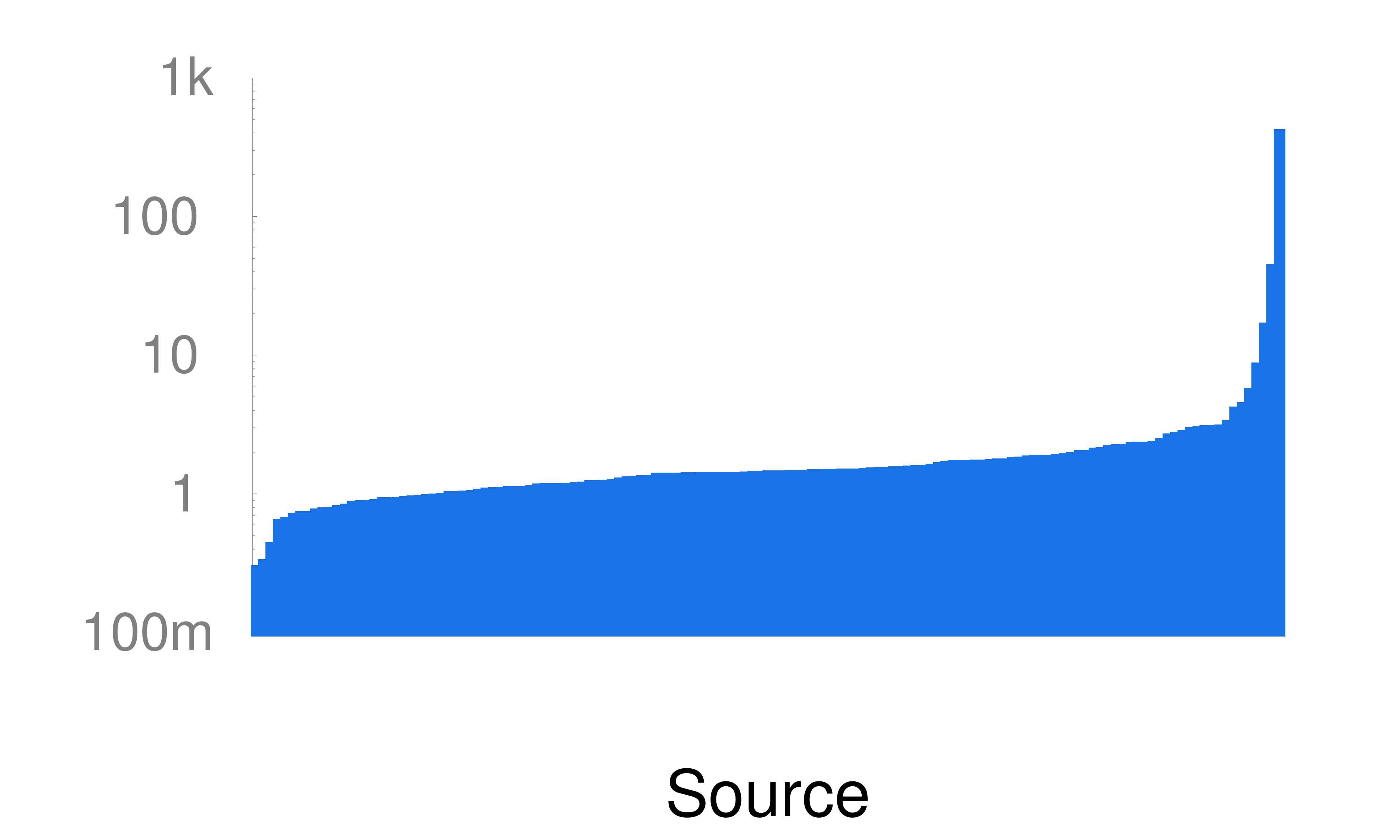}
\caption{\label{fig:channels_latency}Mean relative task time}
\end{subfigure}

\caption{\label{fig:major_minor_channel_stats}Number of tasks, latency and trust distributions of sources}
\end{figure*}

Next, we look at the major contributing sources by (a) number of workers, and (b) number of tasks in Figure~\ref{fig:major_minor_channel_stats}. In Figure~\ref{fig:channels_tasks}, we show the top 10 sources by the number of tasks performed by its workers. 
These 10 sources together account for $\sim$$95\%$ of the tasks, and $\sim$$86\%$ of the workers in the marketplace. Figure~\ref{fig:channels_workers} shows the top 10 sources contributing the most number of workers.
Popular sources include Clixsense~\cite{clixsense} and NeoDev~\cite{neodev} --- companies that provide monetary payment for users taking surveys, and Prodege~\cite{prodege} --- a company that rewards workers in gift cards.
We note some tasks are also routed to Mechanical Turk ({\sf amt}) workers, which accounts for $\sim$$1.5\%$ of all workers. 
While Mechanical Turk has contributed a total of $\sim$$1000$ workers over the period of our evaluation, with a maximum of $\sim$$400$ of their workers being active at any given point of time, by comparison the source NeoDev has contributed a total of $\sim$$27000$ workers in all with as many as $\sim$$2600$ of them being active in a single week.
In addition, the marketplace's own internal workforce ({\sf internal}) accounts for $2.5\%$ of the total workforce and more than $484k$ tasks in our sample during the evaluation period.

\techreport{\ta{The most popular 10 sources account for 95\% of the tasks performed on the marketplace. Furthermore, these 10 include many companies that we, the authors, have never heard of.}}
\sampar{Geographic distribution of workers.}
\techreport{We plot the country-wise distribution of workers in Figure~\ref{fig:worker_countries}. 
The figure shows that crowdsourcing has become a truly global {phenomenon} with workers coming from as many as 148 countries.}
In a study of Amazon Mechanical Turk's workforce~\cite{ipeirotis2010demographics}, the authors noted that more than 60\% of the workers came from USA and India. 
For our marketplace, while these countries continue to contribute a significant number of workers, we also see 17\% of workers coming from the emerging South American and African markets. Close to 50\% of the workers come from 5 countries --- USA (21.3k), Venezuela (5.3k), Great Britain (4.4k), India (4.1k) and Canada (2.8k). \papertext{We observe that crowdsourcing has become a truly global {phenomenon} with workers coming from as many as 148 countries.}

\techreport{
\begin{figure}[h]
\includegraphics[width=\linewidth, height=5cm] {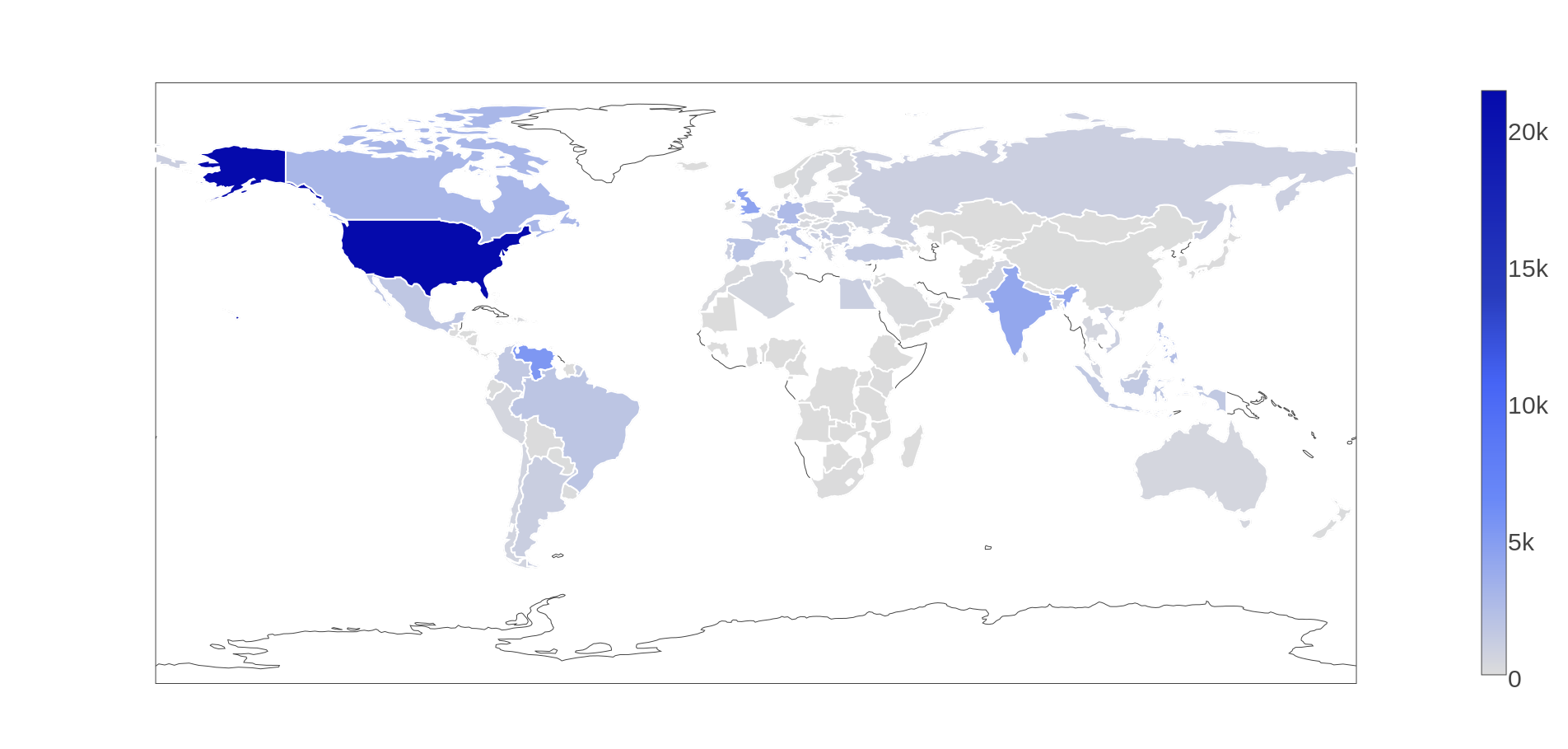}
\caption{\label{fig:worker_countries} Geographical distribution of the crowd workforce}
\end{figure}
}
\sampar{Quality across sources.}
As we noted earlier, different sources bring workers from different locations and specializing in different types of tasks. Furthermore, while some sources have a dedicated workforce performing a large number of tasks regularly, other sources supply an on-demand workforce that performs a small number of tasks occasionally. Given these variations, we investigate if the {\em quality} of workers varies across sources.
We evaluate the quality of different sources on two metrics. Our first metric is the trust score attributed by the marketplace to each completed task performed by a single worker. We compute and report the {\em mean trust} assigned to tasks performed by workers from each source. 
The second metric measures the amount of time taken by workers to complete tasks. To normalize across different tasks, we divide a worker's time by the median time taken by workers to complete that task. We report the average of these {\em relative task times} for tasks performed by each source as the second metric of quality.

The variation in quality (trust and latency) for all sources is shown in Figure~\ref{fig:channels_trust} and Figure~\ref{fig:channels_latency}. 
In terms of mean trust, we observe that close to 10\% of the sources have mean trust $<0.8$. The trust for some sources is even lower than $0.5$. The difference in quality between the sources is more evident when we look at the mean relative task times. While most sources have mean relative task times close to 1, 5\% of the sources have mean relative task times $\geq 3$ ---  the workers from these sources take more than 3$\times$ time to complete the tasks, compared to median task times. Three of these sources even have mean relative task times $\geq 10$.

We further examine the quality of major sources \emph{i.e.}\xspace, sources providing the most number of workers, in more detail in Figure~\ref{fig:major_channels_trust} and Figure~\ref{fig:major_channels_latency}. 
With the exception of Mechanical Turk ({\sf amt}), these sources have high quality --- having mean trust > 0.8 and mean relative task time < 1.5. Mechanical Turk performs poorly on both metrics --- the mean relative task time is more than 5 and mean trust is 0.75.

\techreport{
\ta{Different sources lead to very different qualities. Amazon MTurk, one of the most popular crowdsourcing platforms, shows poor latency and trust. Other lesser known sources show significantly better quality.}
}
\subsection{How do the worker workloads vary?}\label{sec:worker-load}

\begin{figure*}[!t]
\centering
\begin{subfigure}[b]{0.33\textwidth}
\centering
\includegraphics[width=\textwidth]{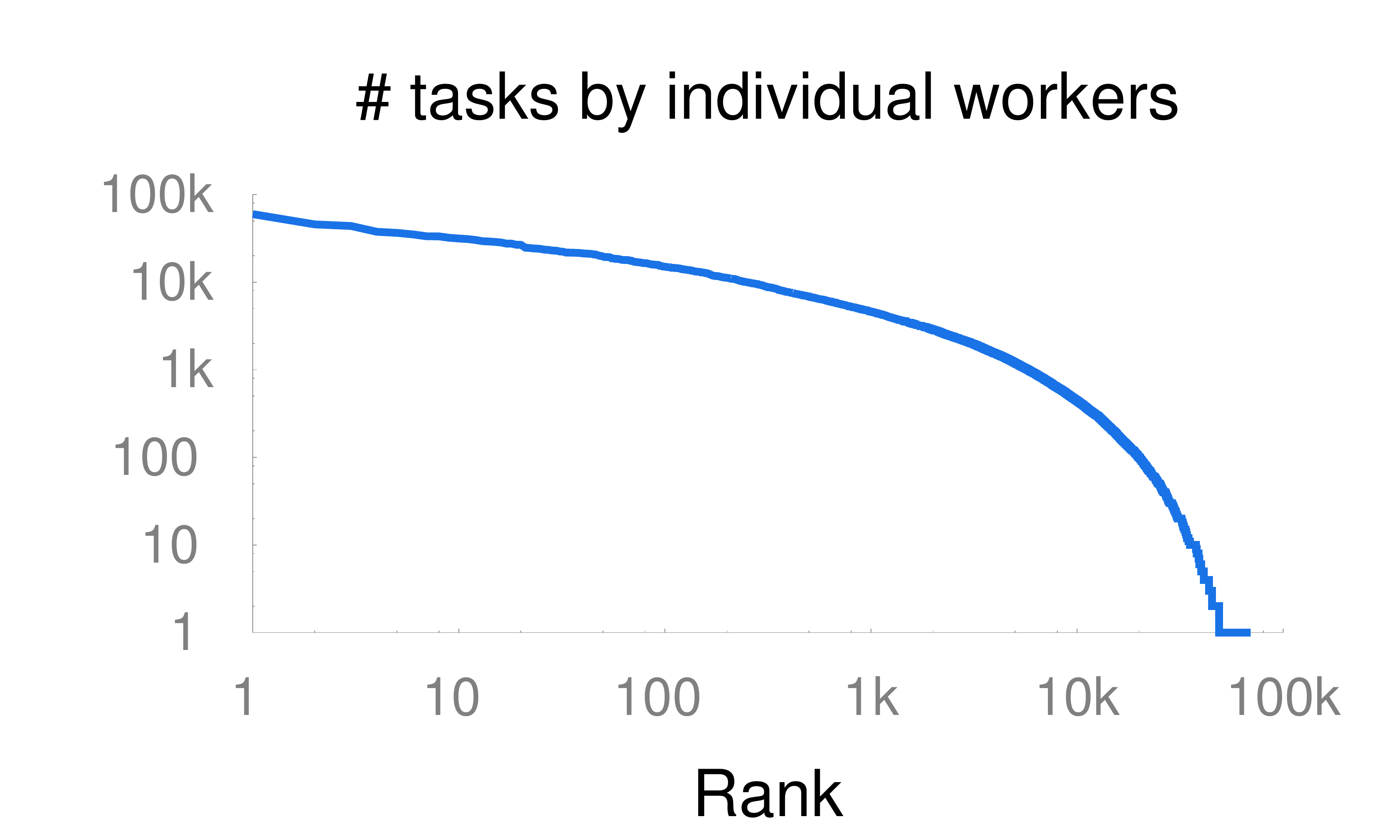}
\caption{\label{fig:worker_tasks}  Workload distribution}
\end{subfigure}%
\hfill
\begin{subfigure}[b]{0.33\textwidth}
\centering
\includegraphics[width=\textwidth]{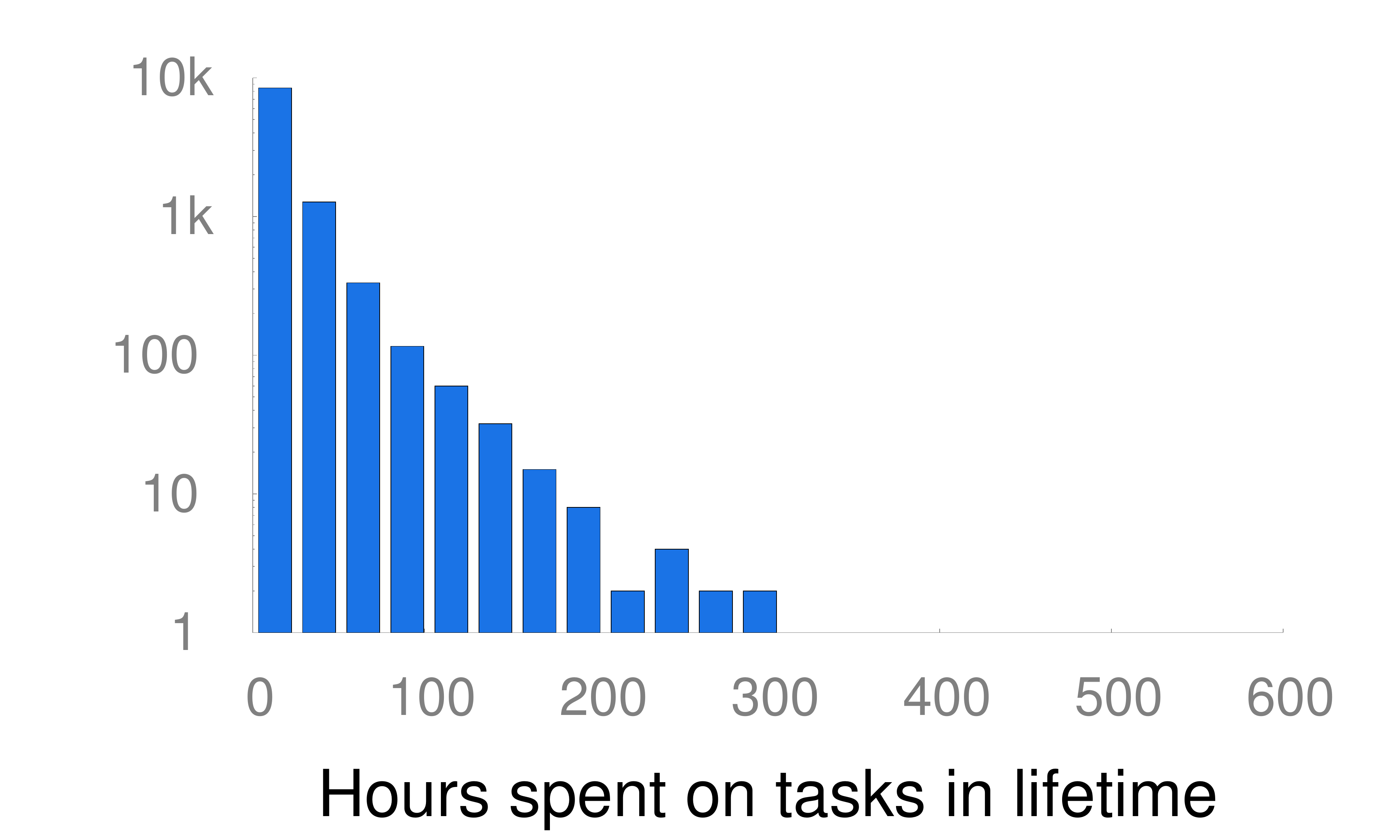}%
\caption{\label{fig:worker_total_hours}Total hours spent on tasks in lifetime}
\end{subfigure}
\hfill
\begin{subfigure}[b]{0.33\textwidth}
\includegraphics[width=\textwidth]{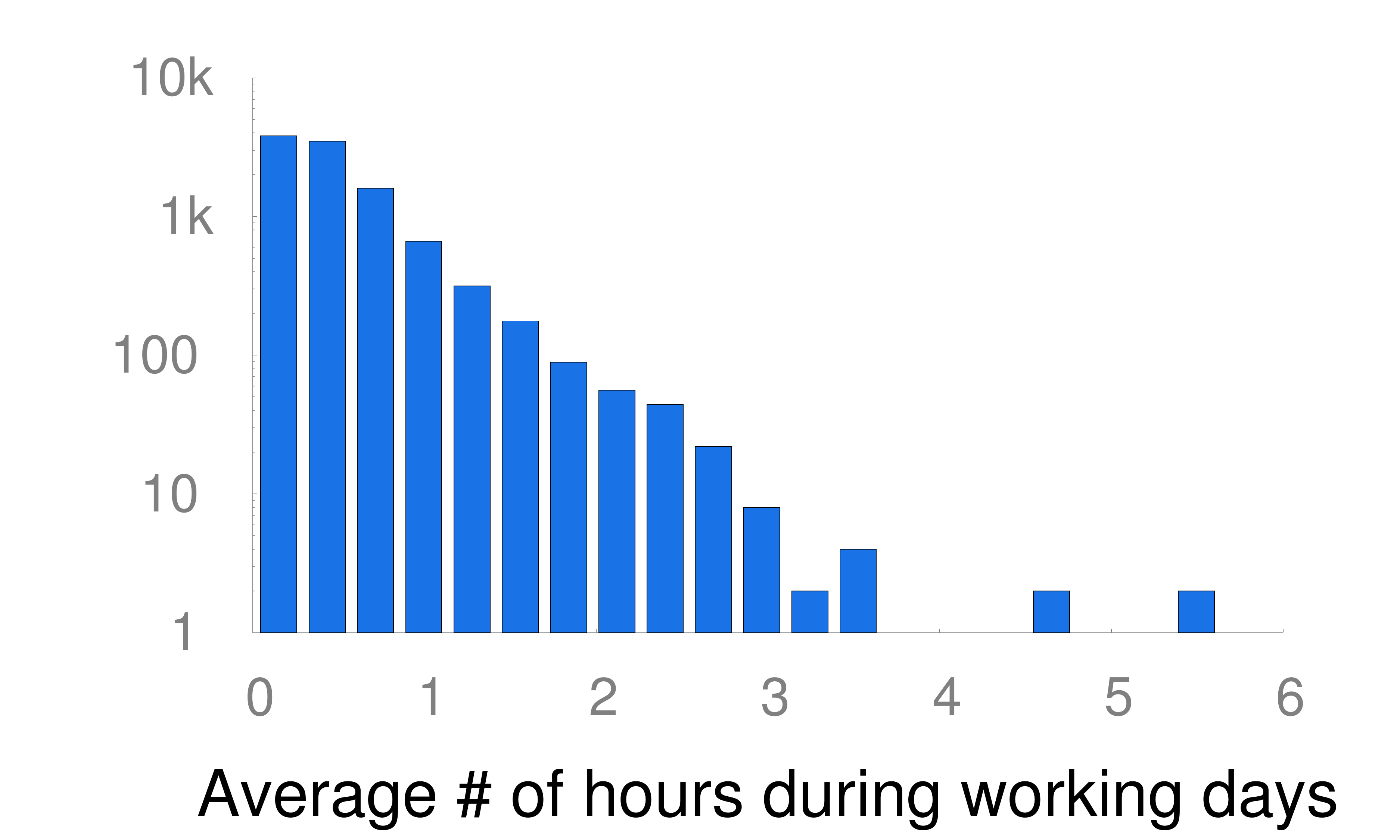}
\caption{\label{fig:worker_avg_hours_per_day}Average number of hours spent on an active day}
\end{subfigure}
\caption{\label{fig:load_and_hours}Workload and Time spent distributions}
\end{figure*}

As our analysis of sources indicated, most of the tasks on the marketplace are completed by a small group of workers. To explore this issue in detail and look at the distribution of worker workloads, we plot the number of tasks performed by each worker in Figure~\ref{fig:worker_tasks}. The x-axis shows the rank of the worker, when workers are sorted in decreasing order of number of tasks completed. The y-axis value shows the number of tasks completed by the worker. From the plot, we note that majority of workers' participation is one-off and the workload is mostly shared by a small group of workers. In fact, more than 80\% of the tasks are completed by just 10\% of the workforce. 
\papertext{Given their experience, it might be worthwhile for marketplaces to collect periodic feedback from these workers}

\techreport{\ta{Most of the tasks on the marketplace are performed by a small group of workers. Given their experience, it might be worthwhile for marketplaces to collect periodic feedback from these workers and direct some of the more difficult tasks to this worker pool.}}

\subsection{How engaged are crowd workers?}\label{sec:worker-engagement}\sampar{Worker Lifetimes.}
We investigate the workers' availabilities on the marketplace through two metrics: (1) the {\em lifetime} of the workers, which is the number of days between their last and first activity on the marketplace during the evaluation period, and (2) the number of {\em working} days (out of their lifetimes) where workers have taken up tasks. 

We first show the distribution of the lifetimes of the workers through a histogram in Figure~\ref{fig:worker_max_lifetime}. Each bar in the x-axis corresponds to a lifetime range, with bar heights denoting the the number of workers having lifetime in corresponding ranges. From these plots, we note that 79\% number of workers are only available over short time frames and hence have lifetimes of less than 100 days. In fact, 52.7\% of the workers have a lifetime of only 1 day in the evaluation period, indicating that a majority of workers are directed to the marketplace through their sources for one-off tasks. However, these workers are not major contributors in terms of number of tasks -- they complete only 2.4\% of the tasks in the marketplace. 

Of the remaining workers who have logged in to the marketplace on more than one day, about one-third have been working on more than 10 days
and have completed 83\% of the tasks in the marketplace. Next, we focus on these {\em active} workers and explore their behavior in greater detail.

\techreport{\ta{More than half of the workers on the marketplace are only active on a single day.
Only 15\% of the entire workforce comes on the marketplace repeatedly to complete tasks. This {\em active} workforce completes more than 80\% of the tasks in the marketplace.}}

\subsection{Active Worker Characteristics\label{sec:active-worker-characteristics}}

\begin{figure*}[!t]
\vspace{-10pt}
\centering
\begin{subfigure}[b]{0.33\textwidth}
\centering
\includegraphics[scale=0.14]{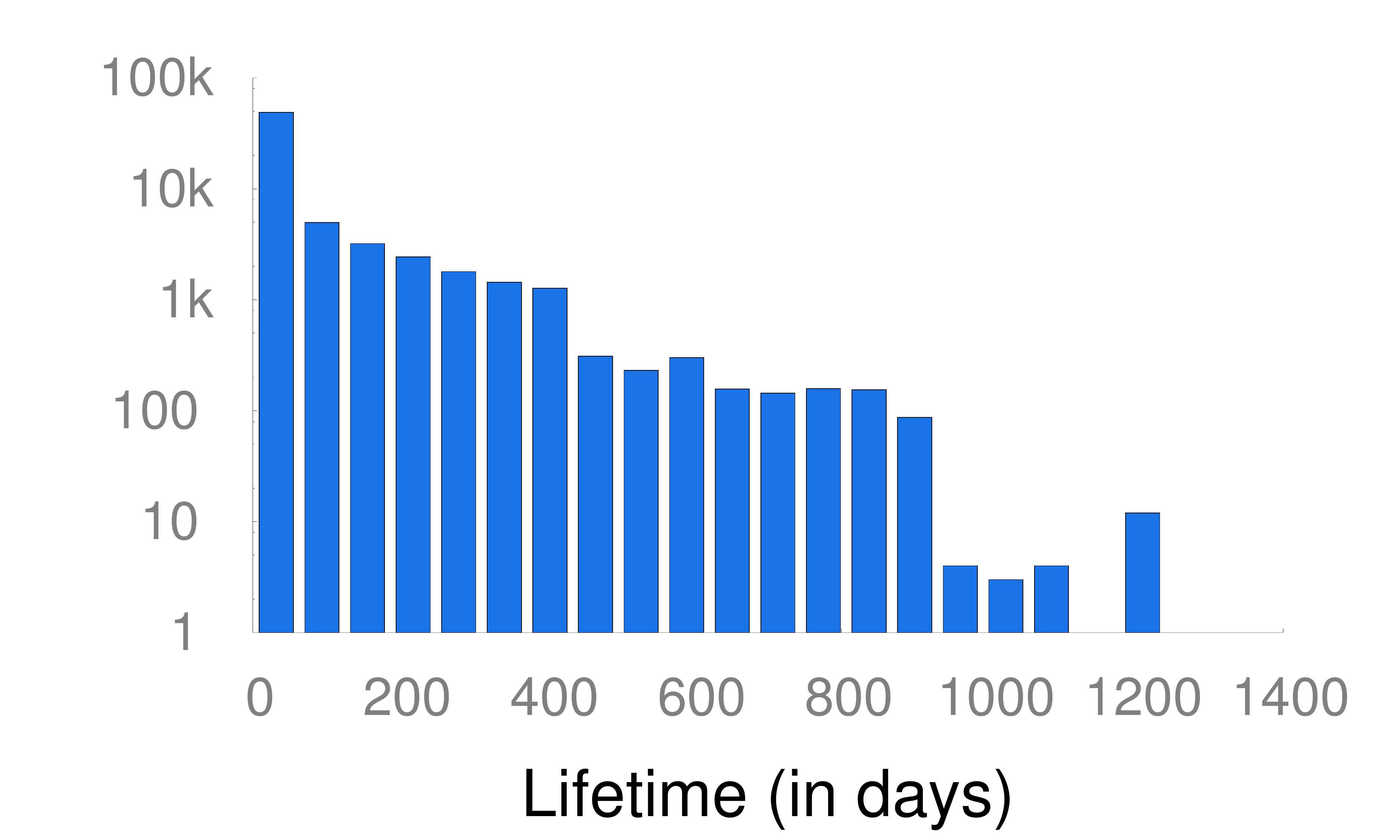}
\vspace{-5pt}
\caption{\label{fig:worker_max_lifetime}Distribution of lifetimes}
\end{subfigure}%
\hfill
\begin{subfigure}[b]{0.33\textwidth}
\centering
\includegraphics[scale=0.14]{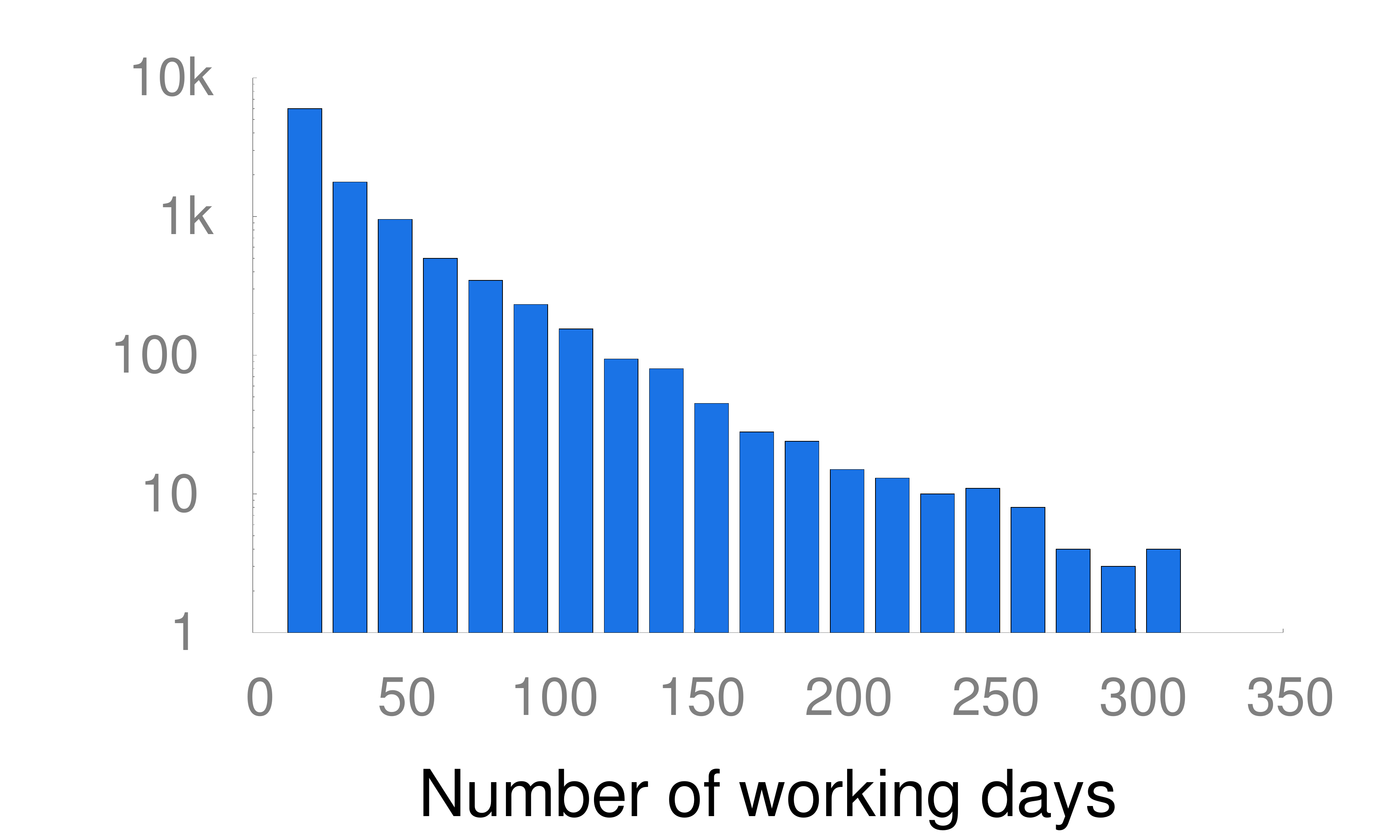}
\vspace{-5pt}
\caption{\label{fig:worker_active_days}Number of working days for active workers}
\end{subfigure}%
\hfill
\begin{subfigure}[b]{0.33\textwidth}
\centering
\includegraphics[scale=0.14]{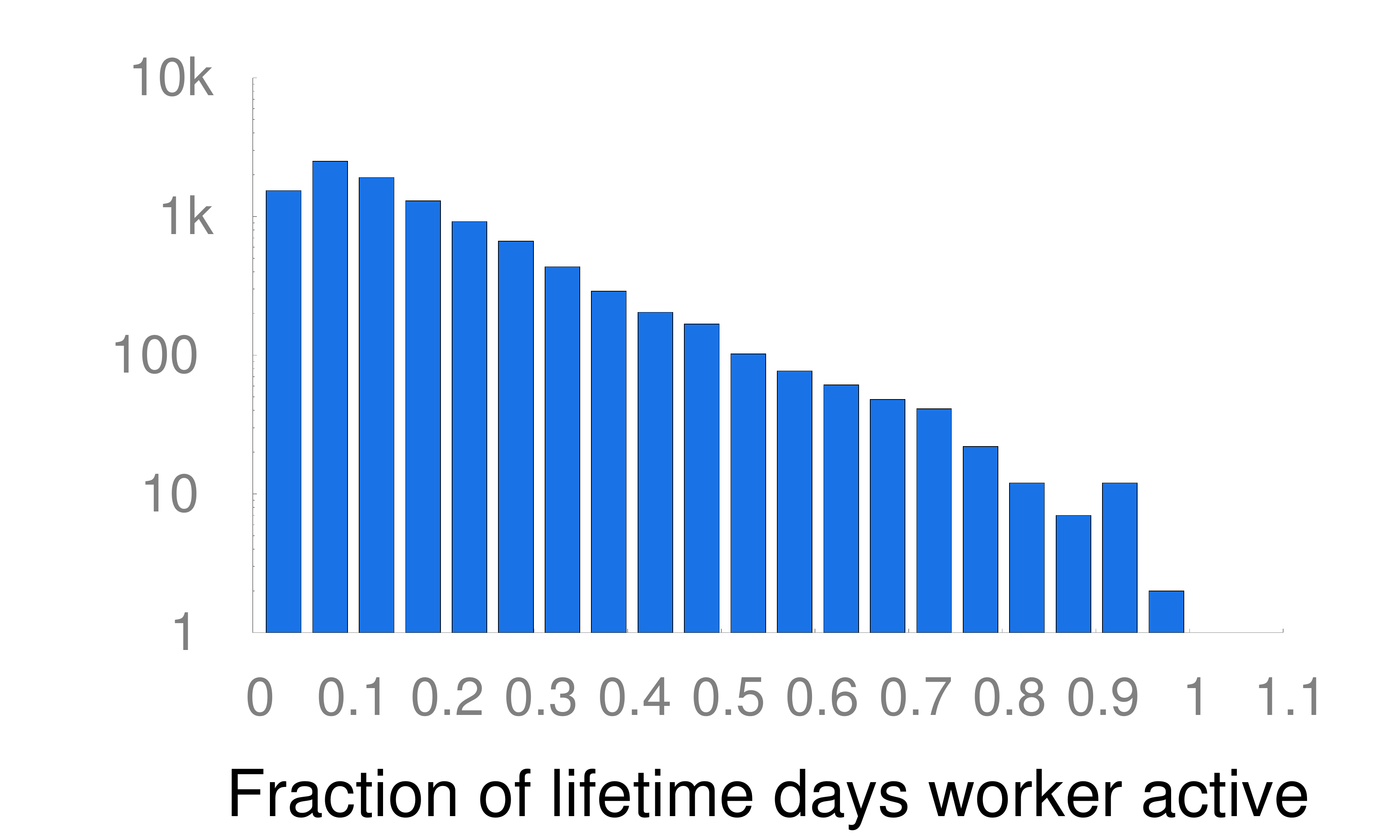}
\vspace{-5pt}
\caption{\label{fig:worker_active_frac}Fraction of lifetime for which workers are active}
\end{subfigure}
\vspace{-5pt}
\caption{\label{fig:worker_lifetime}Worker lifetimes}
\vspace{-5pt}
\end{figure*}

\sampar{Distribution of Working Days.}
Consider the distribution of working days for the active workers 
in Figure~\ref{fig:worker_active_days}. First, on the right side of the plot, 
note the presence of workers who have been working on more than 350 days. 
This is especially remarkable, considering the fact the evaluation period contains regular data for only about 18 months. Second, the bar heights reduce close to linearly (in log scale) with the active days, indicating that the availability of workers decreases exponentially with experience.

For the active workers, we also plot a histogram of the fraction of lifetimes days where they have been working in Figure~\ref{fig:worker_active_frac}. We note that among these active workers, more than 43\% are working at least once a week (on average) during their lifetimes. 

\techreport{\ta{The marketplace has workers who perform tasks regularly -- 5\% of the workforce is working at least one day a week on average. However, such experienced workers are small in number. In fact, the availability of workers decreases exponentially with experience.}}

\sampar{Time Spent.}
We use the amount of time spent by workers on tasks as a proxy for the total time spent working. While this may not be accurate because workers also have to search for tasks, this serves as a good estimate of their productive time. Figure~\ref{fig:worker_total_hours} shows the total number of hours clocked by the active workers during their lifetime. We notice a skewed but long tailed distribution; $\approx$10,000 workers have been working on tasks for less than 25 hours. Nonetheless, there are also a handful of workers who have worked for more than 300 hours during the evaluation period.

Next, in Figure~\ref{fig:worker_avg_hours_per_day}, we plot the average number of hours spent by these active workers on a working day. From this plot, we note that more than 90\% of the workers work for less than 1 hour during their working days. This suggests that crowdsourcing is still not at a scale where it can support many active workers on a full-time basis. However, more than a thousand workers still spend more than an hour a day working on tasks. 

\techreport{\ta{The marketplace only supports a handful of workers on a full-time basis. A majority of the active workers appear to view the marketplace as a supplemental source of income, as is indicated by their daily hours of activity.}}




\sampar{Trust.} 
The mean and median of the average trust of active workers are both above $91\%$, and 90\% of all active workers have average trust higher than $0.84$. Given that the trust scores of workers are all so high, and show such little variation, their distribution does yield any novel insights.

\section{Related Work}\label{sec:related}
One of the first papers in the crowdsourcing literature focused on analyzing the Mechanical Turk 
marketplace for demographic factors~\cite{analyzing-mturk,ipeirotis2010demographics};
a more recent paper studied similar aspects by issuing surveys
to Mechanical Turk workers~\cite{who-are-crowdworkers}.
Other recent papers study the motivations of crowd workers by conducting broad surveys~\cite{brawley2016work, brewer2016would}.
Other papers have evaluated various aspects of marketplaces by interviewing or issuing tasks to workers, 
such as truthfulness~\cite{DBLP:conf/aaai/SuriGM11} and consistency~\cite{sun2016exploring}, 
the efficacy of conducting interface
evaluations via crowdsourcing~\cite{mason2012conducting, turk}, 
limitations in using Mechanical Turk for experimentation~\cite{stewart2015average},
and challenges faced by workers with disabilities~\cite{zyskowski2015accessible}. 
Others attempt to understand worker motivations and behavior using (Turker Nation) forum data~\cite{martin2014being}, and workers' on- and off-network interactions~\cite{gray2016crowd}.
\techreport{Recent work has evaluated the impact of varying price as
well as other economics based concerns~\cite{yihan,faradani2011s,chilton,DBLP:journals/pvldb/HaasW0F15,finnerty2013keep}.
In this paper, since we do not have pricing data, we do not focus on 
evaluating this aspect. }
A recent book~\cite{crowd-book} described the results of interviewing marketplace 
companies (including Samasource~\cite{samasource}, Crowdflower~\cite{crowdflower}) 
for their concerns and problems, but did not conduct a similar
quantitative study based on marketplace data.
Marketplace companies sometimes publish their own reports on demographics,
e.g.,~\papertext{\cite{crowdflower-demographics, samasource-jobs, odesk-stats}}\techreport{\cite{crowdflower-demographics, samasource-jobs, samasource-impact, odesk-stats}}.
In~\cite{kittur2013future}, the authors discuss the challenges faced by crowdsourcing marketplaces, and describe their vision for the future. \techreport{We analyze the state and trends of our crowdsourcing marketplace with a similar goal of improving the crowdsourcing paradigm as a whole---our focus, however, is on making more immediately actionable recommendations through data-driven analyses.}

The paper that is closest to us in adopting a data-driven approach is the one by Difallah et al.~\cite{DBLP:conf/www/DifallahCDIC15}, focused on studying the 
marketplace dynamics of Mechanical Turk,
such as marketplace demand and supply, evolution of task payments over time, as well as other topics.
\techreport{
We describe below, in detail, how our analysis and their analysis differs:

\noindent {\bf What we do that Difallah et al.~doesn't.}
There are several types of analyses that Difallah et al.~does not consider at all, including:
        \begin{denselist}
                \item Effective tasks (Section~\ref{sec:task_analyses}): we quantify the influence of task design parameters on the effectiveness of a task, and make recommendations towards task design based on our observations.
                \item Worker sources (Section~\ref{sec:worker-demo}):  we examine the various sources contributing workers to the marketplace and compare them on contribution and quality. This is not relevant for Difallah et al.~since there is only one ``source'' (Mechanical Turk).
                \item Worker characterization (Sections~\ref{sec:worker-load},~\ref{sec:worker-engagement}): we analyze various aspects of worker demographics and behavior, such as geographic location, lifetime, and engagement.
        \end{denselist}

\noindent {\bf Where we overlap---but how our emphases or granularity is different.}
We overlap with Difallah et al.~on a few topics, but approach
these topics with either a different emphasis or a finer-grained perspective.
        \begin{denselist}
                \item Demand and supply (Section~\ref{sec:marketplace-bursty},~\ref{sec:marketplace-worker-availability}): both papers study the demand and supply interactions in the marketplace, but from slightly different perspectives. Difallah's focus is on making predictions for when batches will be completed. Since they don't have worker data, they do not comment on the supply of workers. Our focus is instead on studying the arrival rate of tasks and availability of workers to complete them, which is more directly relevant for marketplace administrators.
                \item Popular goals (Section~\ref{sec:label-analyses}): Difallah et al.~performs a cursory analysis of goals inferred using keywords within task descriptions; we perform a much more fine-grained analysis using expert-provided labels on all three of \{goals, operators, data types\}, enabling a thorough characterization of the spectrum of crowd work.
        \end{denselist}

\noindent {\bf What Difallah et al.~does that we don't.} This is because we do not have the data to study those aspects.
        \begin{denselist}
                \item Reward analysis: they plot payments assigned to tasks from different topics and workers from different countries over time. We do not have payment information as part of our dataset.
                \item Requester statistics: they plot the number and countries of preferred workers for different requesters over time. We do not have requester IDs as part of our dataset.        
\end{denselist}

} 
\techreport{Thus, w}\papertext{W}e have some overlap with the Difallah et al. paper in terms of marketplace
analysis, but our emphasis and granularity is different; at the same time, the findings in our task design and worker analysis sections are entirely new\papertext{~\cite{techreport}}. 
\papertext{Since their dataset does not have information about individual worker responses, they are unable to study the question of task ``effectiveness'' like we do. }
Also unlike that paper that focuses on Mechanical Turk, our 
crowdsourcing marketplace recruits workers from a collection of labor sources, making it 
a crowdsourcing ``intermediary'' or ``aggregator'', and raising the possibility of a number of interesting additional types of analyses.

Our work in gaining a better understanding of crowd work has broad ramifications
for the database community who has been developing crowd-powered data processing
algorithms~\cite{so-who-won, crowdscreen, DBLP:conf/icdt/DavidsonKMR13, searching, DBLP:conf/sigmod/AmsterdamerGMS13},
and systems, e.g.,~\cite{qurkCIDR, crowddb, cdas, DBLP:conf/www/BozzonBC12, deco-survey}, with dozens of papers published in database conferences each year.
(A recent book surveys this literature~\cite{crowd-book}.)
As examples, understanding the relative importance of various types of processing needs, 
can prioritize the attention of our community to unexplored or underoptimized areas;
understanding how tasks are picked up and worked on can help the community develop
better models of task latency;
understanding the worker perspective and engagement can aid in the design of
better models for worker accuracy and worker behavior in general;
and understanding the impact of task design can help the community adopt ``best practices'' to further
optimize cost, accuracy, and latency.


\section{Conclusions and Future Work}
In this paper, we quantitatively address a number of important open-ended questions aimed towards understanding and improving the entire paradigm of crowdsourcing from three key perspectives --- marketplace dynamics (important to marketplace administrators), task design (important to requesters), and worker characterization (important to labor sources, marketplace administrators, and requesters). 
\techreport{A number of these questions have been speculated upon by practitioners and academics, but there has been little concrete evidence to support these speculations and the so-called ``rules of thumb''. Towards this end, we collect and organize data from a large crowdsourcing marketplace, and additionally enrich this dataset with extensive manual labels.} 
We answer several of what we believe are the most important open questions about crowdsourcing interactions, through quantitative, data-driven experiments on this dataset. Based on our experiments, we come up with a number of valuable insights\techreport{ and takeaways}, that we hope will inform and guide the evolution of crowdsourcing over the coming years. 

There are a number of directions that one could explore, 
following this work.
A natural direction for the database community is to explore the unexplored combinations
of popular task types that have not yet been adequately optimized. 
More broadly, it would be useful to 
pursue a deeper understanding of worker behavior by looking at 
phenomena such as worker anchoring, worker learning, and interactions between various jobs. 
While we have restricted our analyses to a specific set of features and metrics---a 
full analysis of the interplay between various different task 
parameters and notions of job success would be a natural next step. 
Lastly, with full-fledged A/B testing, we may be able to solidify our correlation
and predictive claims with further causation-based evidence.

{\small
\bibliographystyle{abbrv}
\bibliography{ref}
}


\end{document}